\documentclass[fleqn,usenatbib]{mnras}
\usepackage{newtxtext,newtxmath}
\usepackage[T1]{fontenc}

\DeclareRobustCommand{\VAN}[3]{#2}
\let\VANthebibliography\thebibliography
\def\thebibliography{\DeclareRobustCommand{\VAN}[3]{##3}\VANthebibliography}

\usepackage{graphicx}	
\usepackage{amsmath}	

\usepackage{multicol}

\usepackage{xspace} 
\renewcommand{\AA}{\normalfont\r{A}\xspace} 

\usepackage{orcidlink}

\usepackage{color,soul}
\definecolor{lightblue}{rgb}{.70,.95,1}
\sethlcolor{lightblue}

\newcommand{\teff}{\ensuremath{T_{\mathrm{eff}}}\xspace}
\newcommand{\kms}{\ensuremath{\rm{km}\,s^{-1}}\xspace}
\newcommand{\logg}{\ensuremath{\log g}\xspace}
\newcommand{\feh}{\rm{[Fe/H]}\xspace}
\newcommand{\cfe}{\rm{[C/Fe]}\xspace}
\newcommand{\nfe}{\rm{[N/Fe]}\xspace}

\newcommand{\mgfe}{\rm{[Mg/Fe]}\xspace}
\newcommand{\alphafe}{\rm{[\ensuremath{\alpha}/Fe]}\xspace}
\newcommand{\Gaia}{\textit{Gaia}\xspace}

\newcommand{\FERRE}{{\tt FERRE}\xspace}

\defcitealias{lucey23}{L23}
\defcitealias{andrae23}{A23}

\title[C-rich stars from Gaia XP]{Predicting metallicities and carbon abundances from Gaia XP spectra for (carbon-enhanced) metal-poor stars}

\author[A. Ardern-Arentsen et al.]{Anke Ardern-Arentsen$^{1}$\thanks{Email: \url{anke.arentsen@ast.cam.ac.uk}}, 
Sarah G. Kane$^{1}$, 
Vasily Belokurov$^{1}$, 
Tadafumi Matsuno$^{2}$, 
Martin Montelius$^{3}$,
\newauthor
Stephanie Monty$^{1}$, 
Jason~L.~Sanders$^{4}$ 
\\
\\
$^{1}$ Institute of Astronomy, University of Cambridge, Madingley Road, Cambridge CB3 0HA, UK \\
$^{2}$ Astronomisches Rechen-Institut, Zentrum für Astronomie der Universität Heidelberg, Mönchhofstr. 12-14, 69120 Heidelberg,
Germany \\
$^{3}$ Kapteyn Astronomical Institute, University of Groningen, Landleven 12, 9747 AD Groningen, The Netherlands \\
$^{4}$ Department of Physics and Astronomy, University College London, London WC1E 6BT, UK
}

\date{Accepted 2025 January 13. Received 2025 January 09; in original form 2024 October 14}

\pubyear{\the\year{}}

\begin{document}
\label{firstpage}
\pagerange{\pageref{firstpage}--\pageref{lastpage}}
\maketitle

\begin{abstract}
Carbon-rich (C-rich) stars can be found at all metallicities and evolutionary stages. They are often the result of mass-transfer from a companion, but some of the most metal-poor C-rich objects are likely carrying the imprint of the metal-free First Stars from birth. 
In this work, we employ a neural network to predict metallicities and carbon abundances for over 10 million stars with \Gaia low-resolution XP spectra, down to $\feh = -3.0$ and up to $\cfe \approx +2$. 
We identify $\sim2000$ high-confidence bright ($G<16$) carbon-enhanced metal-poor (CEMP) stars with $\feh < -2.0$ and $\cfe > +0.7$. 
The majority of our C-rich candidates have $\feh > -2.0$ and are expected to be binary mass-transfer products, supported by high barium abundances in GALAH and/or their \Gaia RUWE and radial velocity variations. We confirm previous findings of an increase in C-rich stars with decreasing metallicity, adopting a definition of $3\sigma$ outliers from the \cfe distribution, although our frequency appears to flatten for $-3.0 < \feh < -2.0$ at a level of $6-7\%$. We also find that the fraction of C-rich stars is low among globular cluster stars (connected to their lower binary fraction), and that it decreases for field stars more tightly bound to the Milky Way. 
We interpret these last results as evidence that disrupted globular clusters contribute more in the inner Galaxy, supporting previous work. 
Homogeneous samples like these are key to understanding the full population properties of C-rich stars, and this is just the beginning. 
\end{abstract}




\section{Introduction}

The most metal-poor stars are relics from the early Universe, their atmospheres contain traces of the chemical composition of the interstellar medium out of which they were born. These stars can be studied in incredible detail, they can therefore provide unique constraints on early chemical evolution compared to observations at high redshift, with e.g. JWST. Our searches for metal-poor stars have revealed that a large number of them are very rich in carbon \citep[e.g.][]{beers92, beerschristlieb05}, which is easy to recognise from their spectra due to the strong molecular carbon features. 

High-resolution spectroscopic observations of these carbon-enhanced metal-poor (CEMP) stars, currently defined as having \feh\footnote{[X/Y] $ = \log(N_\mathrm{X}/N_\mathrm{Y})_* - \log(N_\mathrm{X}/N_\mathrm{Y})_{\odot}$, where the asterisk subscript refers to the considered star, and N is the number density. Throughout this work, we often refer to \feh as ``metallicity''.} $< -2.0$ and $\cfe > +0.7$ \citep{aoki07}, revealed that there are two types of CEMP stars. 
The ``CEMP-no'' stars are more metal-poor and are typically also enhanced in nitrogen and oxygen, but do not have enhanced heavier element abundances such as iron \citep[e.g.][]{norris13b}. This striking pattern has also been observed in some high-redshift galaxies and absorbers \citep[e.g.][]{saccardi23, deugenio24}. Their detailed abundance patterns are typically well-fit by the yields of so-called faint supernovae \citep{umedanomoto03, nomoto13, tominaga14,ishigaki18} and/or may indicate contributions from rapidly rotating massive stars \citep{meynet06, hegerwoosley10}, both of which can occur for the very dense, metal-free Population III/First Stars. 
The slightly more metal-rich ``CEMP-s'' stars are enhanced in slow neutron-capture (s-) process elements \citep[e.g.][]{lucatello05}, and their radial velocity variations reveal that they are typically in binary systems \citep[][]{hansen16b}. Their abundance patterns are well-fit by the yields of asymptotic giant branch (AGB) stars, and CEMP-s stars are expected to have received carbon and s-process rich material from a former AGB companion \citep{bisterzo10, abate15_detailed}. They are the very metal-poor counterpart of the more metal-rich Ba and CH-stars \citep[e.g.][]{McClureWoodsworth90}. It is worth noting that CH/CEMP stars are not classical carbon stars -- those are very evolved stars that enriched themselves in carbon during the AGB phase, whereas CH/CEMP stars can be found in any evolutionary phase. We refer to stars with carbon over-abundances in ``unexpected'' evolutionary phases as C-rich stars from here on, regardless of their origin. 

To fully understand the role that C-rich stars play in stellar populations, at least two approaches are necessary: 1) large, homogeneous samples/analyses of high-resolution spectroscopic observations of C-rich stars, to use their detailed elemental abundances to constrain chemical processes, and 2) statistical studies of \emph{populations} of C-rich stars, specifically the relative frequency of CEMP-no and CEMP-s (type) stars as a function of metallicity and Galactic environment. Although they have been incredibly valuable so far in teaching us about early chemical evolution and binary interaction, most individual samples of high-resolution spectroscopic observations of CEMP stars are relatively small. The CEMP compilation by \citet{yoon16} has $\sim 300$ stars, but it is an inhomogeneous set: the largest sub-samples from individual papers contain $30-50$ stars each \citep{yong13a, cohen13, roederer14}. To derive the frequencies of CEMP stars with metallicity and Galactic environment, high-resolution spectroscopy metal-poor samples \citep[e.g.][]{lucatello06, yong13b, placco14} and large low/medium-resolution spectroscopic surveys \citep[e.g.][]{frebel06, Lee13, yoon18} have been used, finding increasing CEMP frequencies with decreasing metallicity and increasing distance from the Galactic centre, but each with their own limitations and biases \citep{arentsen22}. For example, high-resolution samples tend to have a human, non-reproducible element to their selection functions, and while low-resolution surveys are likely less biased in that way, the analysis is more challenging and different surveys appear to have inconsistent abundance scales. 

The third data release (DR3) of the \Gaia mission \citep{gaiadr3} includes 200 million very low-resolution ($R \sim 40-150$) BP and RP spectra \citep{carrasco21, deangeli22}, covering $\sim3500-10000$~\AA \citep{montegriffo23}. These are typically referred to as XP spectra. Each of the BP and RP spectra is represented by 55 basis function coefficients (their "continuous" form). In this work, we use the coefficients directly. One of the first mining efforts of the XP spectra was by \citet{lucey23}, using the \texttt{XGBoost} algorithm \citep{xgboost} to identify candidate CEMP stars, but they do not predict \feh or \cfe values. Before DR3, \citet{witten22} had already suggested that it should be possible to identify C-rich stars from the XP spectra. Over the past couple of years, many have used the XP spectra to predict stellar parameters and metallicities \citep[e.g.][]{andrae23, zhang23}, alpha abundances \citep[e.g.][]{ji24, hattori24} and/or carbon and nitrogen abundances (\citealt{fallows24}, \citealt{kane24}) -- most of these works employ machine learning methods to do so. Others have used the XP spectra to identify ``true'' carbon stars in the AGB phase \citep[e.g.][]{sanders23,ye24}. Some have also derived simultaneous metallicities and carbon abundances from narrow-band photometry \citep[e.g.][]{whitten21, huang24}. 

Motivated by these encouraging results, we decided to revisit CEMP/C-rich stars with the goal of estimating their metallicities and carbon abundances from the XP spectra, rather than only identifying whether a star is C-rich or not. Our goals are to create a clean sample of bright C-rich metal-poor stars for spectroscopic follow-up efforts, as well as to provide a largely unbiased sample of metal-poor stars with carbon abundance estimates. To achieve these goals we employ a neural network with a dedicated training set based on LAMOST spectra and supplemented with high-resolution samples of very metal-poor (VMP, $\feh < -2$) and CEMP stars, which we apply to the \citet{lucey23} sample of XP/CEMP candidates and to the \citet{andrae23} XP/giants sample. 

The datasets we use are presented in Section~\ref{sec:data}, and our re-analysis of the LAMOST spectra for the training sample with \FERRE is described in Section~\ref{sec:ferre}. We then proceed to train and test our neural network in Section~\ref{sec:predict} and apply it to the two different XP samples. We use the predictions in Section~\ref{sec:discussion} to investigate the sky distribution of C-rich stars, to derive the C-rich fraction with metallicity and to check the effect of carbon on previous metallicity estimates with XP.

\section{Data}\label{sec:data}

\begin{figure*}
\centering
\includegraphics[width=1.0\hsize,trim={0.0cm 0.0cm 0.0cm 0.0cm}]{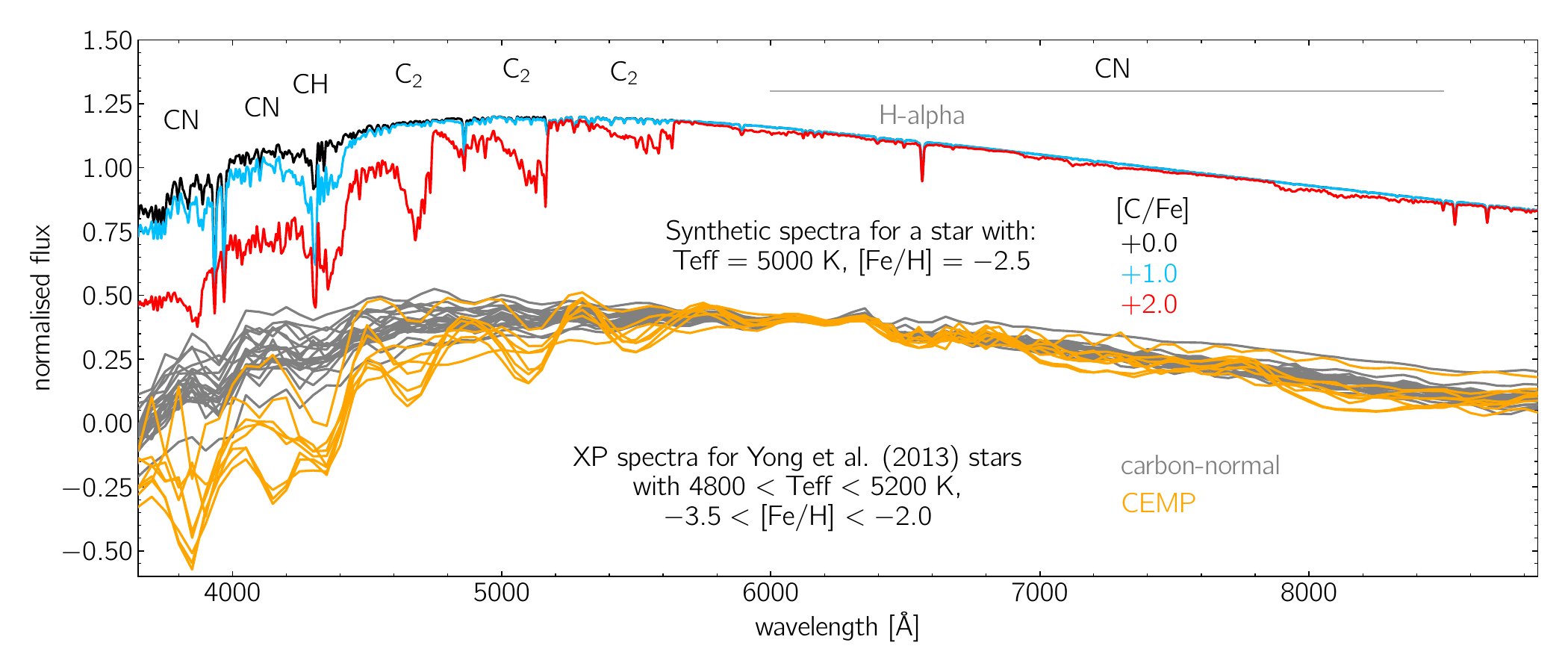}
\caption{Comparison of synthetic spectra (the same as those used in \citealt{arentsen21}) with different \cfe with XP spectra for known C-normal/rich stars from \citet{yong13a}. The synthetic spectra are shown at a higher resolution than that of the XP spectra to display the different spectral features between C-normal and C-rich stars. For some of the observed stars the [C/N] is likely lower than in the models, resulting in stronger CN features.}
\label{fig:xpspec} 
\end{figure*}

First, to motivate the appropriateness of using XP spectra for CEMP stars, in Figure~\ref{fig:xpspec} we show a comparison of synthetic spectra with different carbon abundances (black, blue and red for increasing \cfe, respectively) with XP spectra for normal VMP stars (grey) and for CEMP stars (orange) from the \citet{yong13a} high-resolution spectroscopic study. The carbon-rich stars show clearly different features in their XP spectra compared to the normal VMP stars, matching with known molecular carbon features in the synthetic spectra. There is some variation among the observed CEMP star XP spectra because they cover a range of \teff and \feh, and they also have differences in their carbon and nitrogen abundances. 

Next, we will describe the various datasets we use in this work. These consist of two large \Gaia XP-based samples, high-resolution spectroscopic literature data and LAMOST data.

\subsection{XP vetted giants sample from Andrae et al. (2023)}\label{sec:a23}

To predict \teff, \logg and [M/H] from the XP spectra of \Gaia stars, \citet[][hereafter A23]{andrae23} used the \texttt{XGBoost} algorithm trained on APOGEE data. They augmented it with the \citet{li22} sample of VMP stars to improve the predictions at low metallicity, and they show that their metallicity predictions work well down to [M/H] $\sim -3.0$. This work does not take into account carbon as a variable, which has many molecular features across the XP wavelength range, so for CEMP stars their [M/H] does not correspond tightly to [Fe/H] (see Figure~\ref{fig:a23} in the Appendix). 

We will apply our neural network to the \citetalias{andrae23} vetted giant branch sample with \Gaia radial velocities to derive new \teff, \logg and \feh, as well as adding \cfe. This vetted sample consists of bright stars with high-quality parallaxes (uncertainties $<$ 25\%), \citetalias{andrae23} $\teff < 5200$ and $\logg < 3.5$, and additional cuts on \Gaia and WISE photometry to remove spurious outliers (see \citetalias{andrae23} for details). \citetalias{andrae23} note that these cuts are made to create a \emph{pure} sample, especially at lower metallicity, not necessarily a \emph{complete} one.

\subsection{CEMP candidates from Lucey et al. (2023)}

To identify CEMP stars among the many \Gaia DR3 stars with XP spectra, \citet[][hereafter L23]{lucey23} applied the \texttt{XGBoost} algorithm to all XP spectra for stars with $0.8 <$ (BP$-$RP) $< 2.75$ and M$_G < 7$ -- $\sim 183$ million in total. Bluer stars are likely too hot to produce good estimates for whether a star is carbon-rich or not, because the molecular features become too weak, and redder and fainter absolute magnitude stars are lacking from their training sample. For each star, their \texttt{XGBoost} setup predicts the probability $P_c$ that a star is carbon-enhanced, according to the typical $\cfe > 0.7$ definition. We will refer to these as C-rich candidates. 

The reference sample that \citetalias{lucey23} used to train and test their method comes from the low-resolution part of the Sloan Digital Sky Survey (SDSS, \citealt{york00}), specifically the low-resolution optical side (mostly the Sloan Extension for Galactic Understanding and Exploration (SEGUE) project). The stellar parameters in the reference sample come from the SEGUE Stellar Parameter Pipeline (SSPP, \citealt{lee08}), and include \teff, \logg, \feh and \cfe (with \cfe not corrected for evolutionary effects). The sample had been vetted for defective spectra and other issues and covers main sequence, turn-off and giant stars. Their final reference sample consists of $> 140\,000$ carbon-normal stars (of all metallicities, but dominated by $\feh > -1.0$) and $\sim 1500$ carbon-enhanced stars ($93\%$ of them with $\feh < -1.0$) -- see their figure~3. That figure also shows the magnitude distribution, which is dominated by relatively faint stars (mostly $G>15$, and all the way to the faint limit of XP at $G\sim17.6$). Of the reference sample, $70\%$ is used for training and $30\%$ for testing. 

\begin{figure}
\centering
\includegraphics[width=0.9\hsize,trim={0.0cm 0.0cm 0.0cm 0.0cm}]{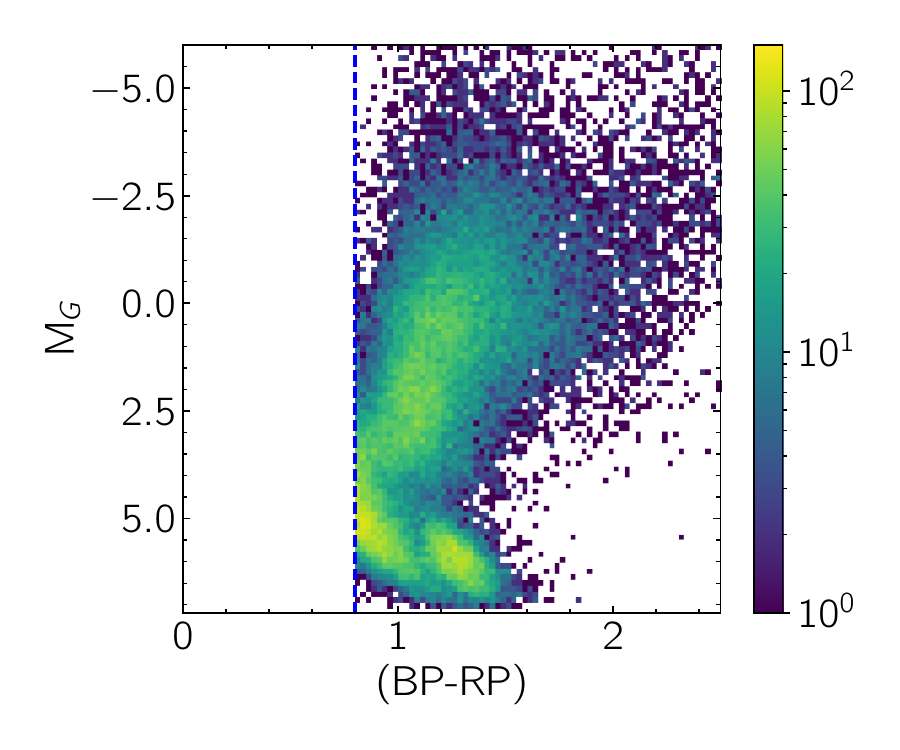}
\includegraphics[width=0.9\hsize,trim={0.0cm 0.0cm 0.0cm 0.0cm}]{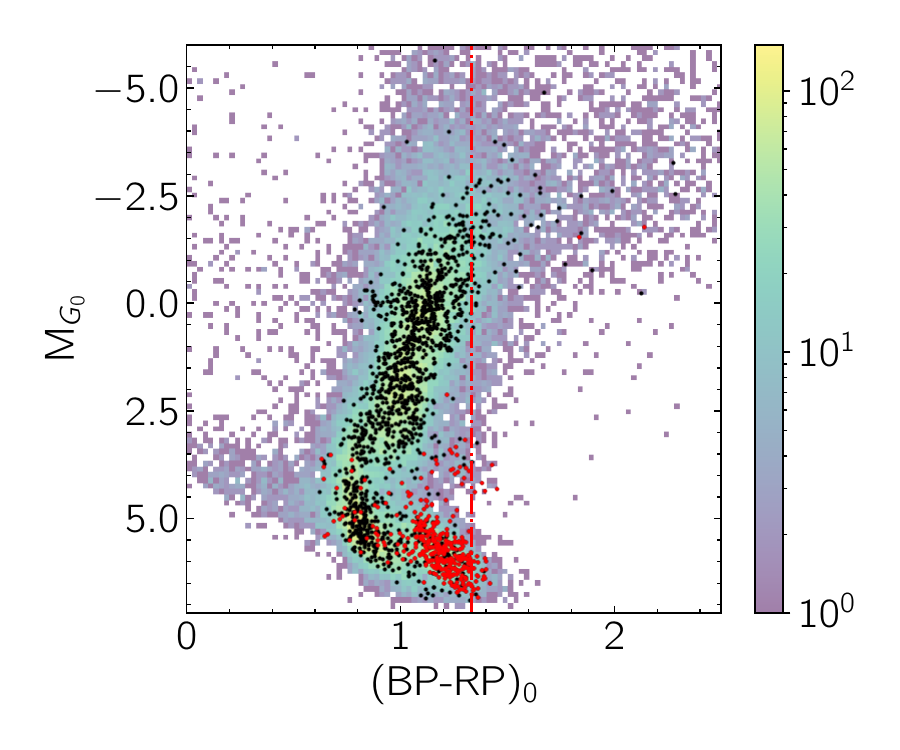}
\caption{Top: CMD of the full \citetalias{lucey23} sample, not corrected for extinction. Bottom: the same sample is shown in the background, now corrected for extinction, with the sub-sample that is in LAMOST shown on top (black and red dots, see the text in Section~\ref{sec:ferre} for the meaning of the red dots). Only stars with E(B$-$V)~$<1.0$ are included. The blue vertical line in top panel is the (BP$-$RP) cut applied by \citetalias{lucey23}, and the red vertical dot-dashed line in the bottom panel corresponds to $\teff \sim 4500$~K. The colour bars show the number of stars in each pixel.}
\label{fig:cmd} 
\end{figure}

\citetalias{lucey23} test the contamination and completeness of their predictions using their test sample. For their $P_c > 0.5$ sample, they find an overall $0.04\%$ false positive rate (number of false positives divided by the sum of the true positives and false positives) and a $26\%$ true positive rate (true positives divided by the sum of the false negatives and true positives) for C-rich stars. The true positive rate is higher for giant stars, for more carbon-rich stars, for stars with lower extinction and for brighter stars (their figures~7 and 8). When cross-matching with an independent high-resolution spectroscopic sample, their completeness rate goes up to $69\%$, missing mostly CEMP stars with the lowest metallicities and "intermediate" carbon-enhancement (Group II stars, \citealt{yoon16}). This is likely due to this sample containing more bright giant stars. \citetalias{lucey23} expect $\sim 9\%$ of their true C-rich candidates to be C-rich and metal-rich ($\feh > -1.0$), assuming the distribution of C-rich stars in their training sample is representative of the underlying population. When comparing their final sample with metallicities from \citetalias{andrae23}, they find that $57\%$ of their candidates has $\feh > -1.0$, which they suggest is likely mostly due to uncertain \citetalias{andrae23} metallicities for C-rich stars and many are likely more metal-poor. 

The final \citetalias{lucey23} catalogue contains 58\,872 C-rich candidates, with $P_c>0.5$ (with the contamination and completeness rates described above). This is the largest, homogeneously identified sample of C-rich stars to date. Of the \citetalias{lucey23} sample, $54\%$ has $G<16$ (we will call this the ``bright sample'') and $46\%$ has $G>16$ (``faint sample''). 
The top panel of Figure~\ref{fig:cmd} shows the colour-magnitude diagram (CMD) of the sample, adopting 1/parallax for the distance and indicating the blue (BP$-$RP) cut at 0.8 from \citetalias{lucey23}. The CMD is divided into three main features: a broad red giant branch (RGB, partly due to extinction), an upper main sequence and a blob on the main sequence with (BP$-$RP) $>1.2$. The bottom panel is corrected for extinction using the \citet{schlegel98} extinction map (re-scaled according to \citealt{schlafly11}) and the \Gaia extinction coefficients depending on \teff, A$_0$ and \feh from \citet{martin24}, fixing \feh to $-1.0$ and iteratively determining the photometric \teff using the \citet{casagrande19} \Gaia infrared flux method. It shows that the sample does contain some hotter stars with (BP$-$RP)$_0 < 0.8$, although it is a small fraction. The black and red points in this diagram will be discussed in more detail in Section~\ref{sec:ferre}, but in short, the red points are likely young, active stars.

\subsection{High-res literature}

We use the high-resolution spectroscopy compilation of CEMP stars by \citet[][hereafter Y16]{yoon16} to enhance the training sample of our machine learning analysis at low metallicity and high carbon abundance. We only use those stars that are included in the \citetalias{lucey23} catalogue. To further complement the sample at low metallicity and "normal" carbon abundances, we use the \citet{li22} sample of high-resolution follow-up of VMP and EMP stars selected from LAMOST. Most of their carbon abundances for giant stars have uncertainties of the order of $0.1-0.15$~dex. The stars we use from Y16 and \citet{li22} are shown with orange dots on the $\feh - \cfe$ plane in Figure~\ref{fig:fcferre}. The other points in this Figure are discussed in Section~\ref{sec:ferre}. 

\subsection{LAMOST DR8}\label{sec:lamost}

We retrieve from the LAMOST spectroscopic archive (DR8) all spectra with SNR $>10$ for stars in the \citetalias{lucey23} catalogue. We remove duplicates and only keep the spectrum with the highest S/N for each star, resulting in 2876 spectra. The main LAMOST pipeline is not appropriate for carbon-enhanced stars, and also misses many of the more metal-poor stars \citep[see e.g.][]{li18, arentsen23} -- we therefore do not make use of the LAMOST DR8 catalogue and derive our own stellar parameters, including \cfe. 

We also create a set of "normal" stars from LAMOST, to be used as a reference and for inclusion in the training sample for the machine learning analysis in a later section. Using stellar parameters from LAMOST DR8, we selected $\sim200$ stars per 0.5~dex metallicity bin between $\feh = 0$ and $-2.5$, roughly equally spread across the Kiel diagram. This results in a sample of $\sim 950$ stars. We re-analyse these spectra in the same way as the LAMOST/\citetalias{lucey23} sample. 

\subsection{Note about evolutionary corrections}

When metal-poor stars ascend the RGB, due to extra mixing, their carbon abundances decrease while their nitrogen abundances increase \citep{gratton00, stancliffe09, placco14}. The strength of this extra mixing depends on the metallicity, \logg, carbon and nitrogen abundances. Corrections are estimated for stars in our sample with $\feh < -1$ only, using the mean prediction of 15 draws of our predicted \logg, \feh and \cfe and their uncertainties, and assuming \nfe = 0 (V. Placco, private communication). The uncertainties we adopt are based on the dispersion in the predictions from the multiple networks (see next section). Given the uncertainties and assumptions, these carbon corrections are quite rudimentary, but it is better than not taking into account the evolutionary effects at all. We apply these evolutionary corrections to the \citetalias{andrae23} sample when using it on its own, for example when showing the distribution of \feh and \cfe or when investigating the C-rich fraction with metallicity and orbital energy. Corrected carbon abundances are indicated as $\cfe_\mathrm{corr}$ or \cfe (predicted) + C cor. Throughout this paper, when discussing or showing \emph{spectroscopic} \cfe values, no corrections for evolutionary effects have been applied.

\section{Spectroscopic analysis of L23 stars in LAMOST}\label{sec:ferre}

\subsection{Approach}\label{sec:ferre1}

We analyse the LAMOST spectra for \citetalias{lucey23} and the "C-normal" sample with the full spectrum fitting \FERRE\footnote{FERRE \citep{allende06} is available from \url{http://github.com/callendeprieto/ferre}} code. In the first \FERRE run, the LAMOST radial velocities are adopted to shift the spectra to their rest wavelength (which is what \FERRE needs as input). We found that these were not always precise for our stars, which is not surprising given the peculiar nature of these stars. We therefore re-derived the radial velocities with \texttt{FXCOR} in IRAF, using the \FERRE output observations and models. After shifting the spectra to the zero-frame we ran \FERRE again to derive final parameters.

We found that \FERRE struggles to estimate good \logg values for metal-poor stars, especially carbon-rich stars. There are not many \logg-sensitive features in low-resolution blue spectra, except for the carbon bands, so if these are not well-fitted, \FERRE appears to compensate by changing \logg. This effect has previously also been seen in some other low/medium-resolution spectroscopic \FERRE analyses \citep{arentsen21, arentsen22}, and biases the carbon abundances. We therefore decided to derive \logg independently for the \citetalias{lucey23}/LAMOST sample and fix it in the \FERRE fit. We limit the sample to stars with good parallaxes ($\varpi/\Delta \varpi > 4$) and derive \logg by matching the absolute RP magnitude to the closest value from a metal-poor PARSEC isochrone \citep{bressan12}. We use RP rather than G because it is expected to be less affected by the molecular carbon bands.


In the \FERRE analysis, we use a synthetic model grid similar to the CRUMP grid presented by \citet{aguado17}. The model grid was computed with the ASSET code \citep{koesterke08}, using model atmospheres computed with the Kurucz code \citep{meszaros12}, assuming a microturbulence of 2~\kms and $\alphafe = +0.4$. The former is appropriate for giants and the latter is appropriate for most metal-poor stars ($\feh < -1.5$). This work focuses on metal-poor giants and therefore these assumptions are justified, but they may bias the results for more metal-rich and/or dwarf stars. The nitrogen abundance is fixed to Solar: $\nfe = 0$. While this is appropriate for carbon-normal VMP stars, it is not ideal for C-rich stars. According to the compilation of CEMP stars in the JINAbase \citep{jinabase}, CEMP-no stars have a wide range in \nfe, from solar to $+3$, while for CEMP-s stars [C/N] is typically around $\sim 0.8$ (corresponding to roughly $+1 < \nfe < +2$). Ideally, nitrogen would be another free parameter in the grid, but this is beyond the scope of this work to explore. 

We use an extended version of the grid, covering the following parameter range \citep[see also e.g.][]{arentsen20b}: 

\begin{itemize}
\item $4500~\mathrm{K} < \teff < 7000~\mathrm{K}$, $\Delta \teff = 250~\mathrm{K}$
\item $1.0  < \logg < 5.0$, $\Delta \logg = 0.5$
\item $-4.0  < \feh < +0.5$, $\Delta \feh = 0.5$
\item $-1.0  < \cfe < +3.0$, $\Delta \cfe = 1.0$
\end{itemize}

The synthetic spectra are smoothed to the LAMOST resolution ($R\sim1800$) and both observed and synthetic spectra are normalised using a running mean with a window of 30 observed-spectrum pixels. We use the cubic Bézier interpolation algorithm and the global search is performed using Powell’s truncated Newton algorithm. We fit the spectra between $3900-5700$~\AA since we want to avoid the CN feature below 3900~\AA and from experience the LAMOST spectra often have artefacts for redder wavelengths.

We found that the resulting \FERRE metallicity estimates were biased high, especially for very C-rich stars. For such stars, most of the fitted wavelength range is dominated by carbon features, as can be appreciated from Figure~\ref{fig:xpspec}, and the main metallicity features are the Ca H\&K lines between $3900-4000$~\AA. To help \FERRE, we decided to significantly increase the weight of the latter region in the fit compared to the rest of the spectrum (1:0.1), which improved our results. The improvement is judged by the quality of the fit in the Ca H\&K region, and supported by improved predictions from our neural network (Section~\ref{sec:predict}) after our adapted methodology. 

We also note that there appears to be a bias in \teff for C-rich stars in the \FERRE analysis: on the Kiel diagram, C-rich giants are offset by a few 100~K towards hotter temperatures compared to C-normal giants (which overlap with the high-resolution samples). This is possibly again related to the mismatch between carbon features in the synthetic and observed spectra, which becomes worse for lower temperatures because the molecular features are larger. One might expect that this \teff bias results in \feh to be somewhat overestimated, as well as the absolute carbon abundance -- however, \cfe is a \emph{ratio} of iron and carbon so its potential bias depends on the relative change of the two. Future improvements might be made in this regard if better external \teff estimates can be found for C-rich stars (e.g. photometric estimates in the infrared or using regions of the XP spectrum that are less affected by molecular features), which is beyond the scope of this work.

\begin{figure}
\centering
\includegraphics[width=1.0\hsize,trim={0.0cm 0.0cm 0.0cm 0.0cm}]{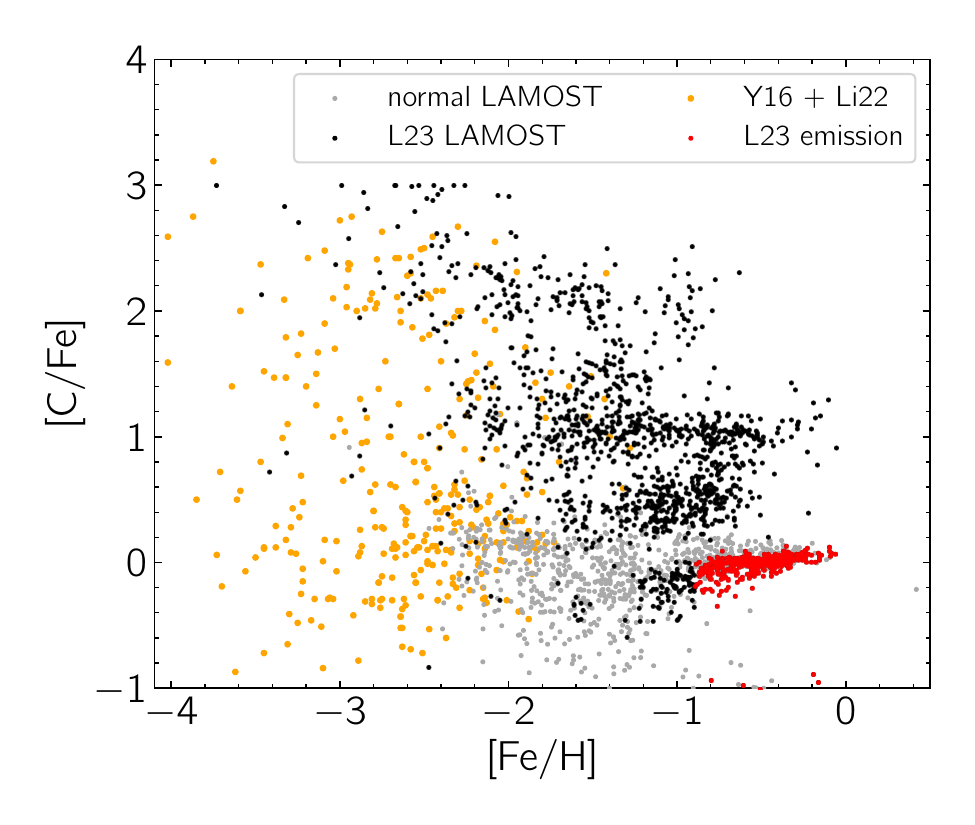}
\caption{Metallicity vs. carbon abundance for the spectroscopic sample used in this work based on LAMOST and high-resolution literature studies. Small orange points are stars from the Y16 CEMP compilation that are among the CEMP candidates of \citetalias{lucey23}, as well as stars from \citet{li22} with XP spectra. The \FERRE analysis of stars with LAMOST spectra in the \citetalias{lucey23} C-rich candidate catalogue is shown in black and red (the red group is selected by metallicity and carbon abundance), and for the ``normal'' LAMOST sample in grey. The red stars turn out to be mostly stars with Ca H\&K in emission, hence they are labelled `\citetalias{lucey23} emission'. The apparent horizontal ``stripes'' are likely due to the fact that the \FERRE grid has steps in \cfe of $1.0$~dex. Note that the results may be slightly biased for the more metal-rich stars due to the grid assumption of $\alphafe = +0.4$.  The carbon abundances in this figure have \emph{not} been corrected for evolutionary effects. 
}
\label{fig:fcferre} 
\end{figure}

\subsection{Results}

The \feh vs. \cfe results for the \citetalias{lucey23}xLAMOST sample analysed with \FERRE are shown in Figure~\ref{fig:fcferre} with black and red dots. The ``normal'' LAMOST sample (see Section~\ref{sec:lamost}) is shown in grey. Only stars with SNR~$>20$, $\feh > -3.8$, $\teff < 6000$~K and $\cfe$ uncertainty $<1.0$ are kept for this figure (and the remainder of the paper). There are a few striking features. 

\begin{figure*}
\centering
\includegraphics[width=0.8\hsize,trim={0.0cm 0.0cm 0.0cm 0.0cm}]{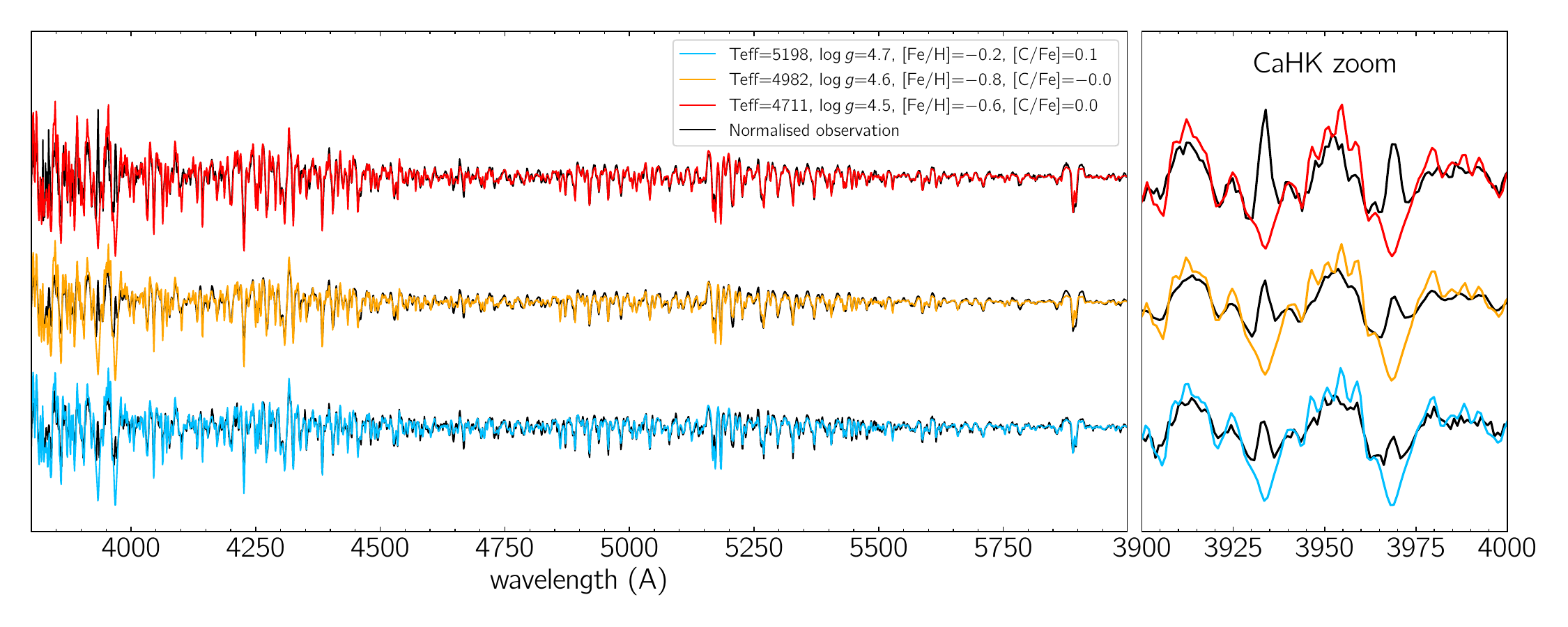}
\caption{\FERRE fits for three stars from the cloud of red points (metal-rich, carbon-normal dwarf stars -- see the legend) in Figures~\ref{fig:cmd} and \ref{fig:fcferre}, where we chose stars with S/N $>50$ to have good quality spectra at blue wavelengths. All show emission in the calcium H \& K lines, which can be seen in the zoomed-in panels on the right. The spectra have been running-mean normalised and have been arbitrarily offset from each other vertically.}
\label{fig:emis} 
\end{figure*}

In the \citetalias{lucey23} sample there is a significant, tight blob at low carbon abundance and $\feh > -1.0$, which is highlighted in red. After inspecting the \FERRE fits, we find that these are almost all stars with Ca H\&K lines \emph{in emission}. We show a few example spectra with fits in Figure~\ref{fig:emis}. We suspect that they were identified as metal-poor C-rich candidates by \citetalias{lucey23} because they have weak Ca H\&K lines but solar carbon features. In the CMD of Figure~\ref{fig:cmd} they are shown in red, occupying a distinct region in the CMD, where young, active dwarfs may be expected to be found. Indeed we find that almost all of these stars are variable in their photometry -- defining Pvar as the probability of being photometrically variable based on the distribution of the number of \Gaia $G$-band observations and the flux/error ratio as a function of $G$ following \citet[][]{martin24}, the bulk of these stars have Pvar $>0.9$. These stars could for example be pre-main sequence stars, magnetically active stars or interacting binaries. 

Another striking feature is the piling up of \cfe values, particularly around $+1.0$ and $+2.0$. This is unlikely to be physical. We experimented with different \FERRE interpolation and global search options, as well as different options for smoothing and the fitted wavelength coverage, but this piling up feature did not go away. It did slightly improve after we applied the higher weighting to the Ca H\&K region. Our suspicion is that it could be due to a mismatch between carbon features in the synthetic and observed spectra, resulting in \FERRE preferring solutions close to the grid nodes. This may be due to non-linearity between grid points and/or a different mix of CNO elements -- the grid assumption of $\nfe = 0.0$ is not really appropriate for most CEMP/C-rich stars, the observed \nfe values are typically higher for both CEMP-s and CEMP-no stars (JINAbase, \citealt{jinabase}). However, it is sufficient for using our predictions to know whether a training sample star is more likely to have $\cfe = +1.0$ or $+2.0$. Future improvements could likely be made for the training sample by for example employing models with more appropriate CNO values (or varying both C and N at the same time), using denser model grids, and/or using different interpolation mechanisms.

The \citetalias{lucey23}/LAMOST sample is shown on the CMD in the bottom panel of Figure~\ref{fig:cmd}. It appears to cover similar regions of the CMD as the full \citetalias{lucey23} sample, suggesting that the LAMOST subset is likely representative of the full \citetalias{lucey23} sample. One region that is underrepresented, and possibly biased, is that of the cool stars with e.g. (BP$-$RP)$_0 > 1.5$. There are not many stars in the training sample in this regime. Also, the \FERRE grid stops at 4500~K and it cannot derive accurate parameters for cooler stars. A photometric temperature of 4500~K for a giant star corresponds to (BP$-$RP)$_0 \sim 1.33$ according to the \citet{casagrande19} method, assuming $\feh = -1.0$ and $\logg = 1.0$. Therefore the results for redder stars are likely somewhat biased and should not be over-interpreted. 

\section{Predicting [Fe/H] and [C/Fe] for XP stars}\label{sec:predict}

\subsection{Methodology}

Next, we proceed to transfer the information from the spectroscopic samples to the full \citetalias{lucey23} XP sample. Our methodology of choice is heteroscedastic regression with an artificial neural network (ANN), following the approach of \citet{kane24} who used it to estimate [N/O] and [Al/Fe]. A very similar approach was used by \citet{fallows24}. We slightly update the ANN to make it more appropriate for our sample. In summary: our ANN relies on PyTorch \citep{pytorch} and consists of an input layer, two hidden layers of 64 nodes each, and an output layer. During training, we use a dropout of $30\%$ (this is a random subset of nodes not used) -- this helps with over-fitting the training data. We use the \texttt{Adam} optimizer \citep{adam} with a learning rate of 0.001. All other \texttt{Adam} hyperparameters are set to the PyTorch defaults. The details of the loss function can be found in \citet{kane24}, but briefly, we use a negative log-likelihood loss function that includes the variances\footnote{These are the variances inferred by the network, they are not learned from including uncertainties in the training labels.} of each of the predictions -- this helps to avoid "regression towards the mean" and provides uncertainty estimates on each of our predictions. The features we input are the XP coefficients and the labels \teff, \logg, \feh and \cfe (the uncertainties on these are not taken into account), and we output the predicted labels, including their variances. We normalise the input XP spectra to their first coefficient, as well as by using the $1$st and $99$th percentiles of the coefficients across the whole sample. We run 50 different networks and average their predicted results to get robust predictions. 

\begin{figure*}
\centering
\includegraphics[width=0.33\hsize,trim={0.2cm 1cm 0.2cm 0.0cm}]{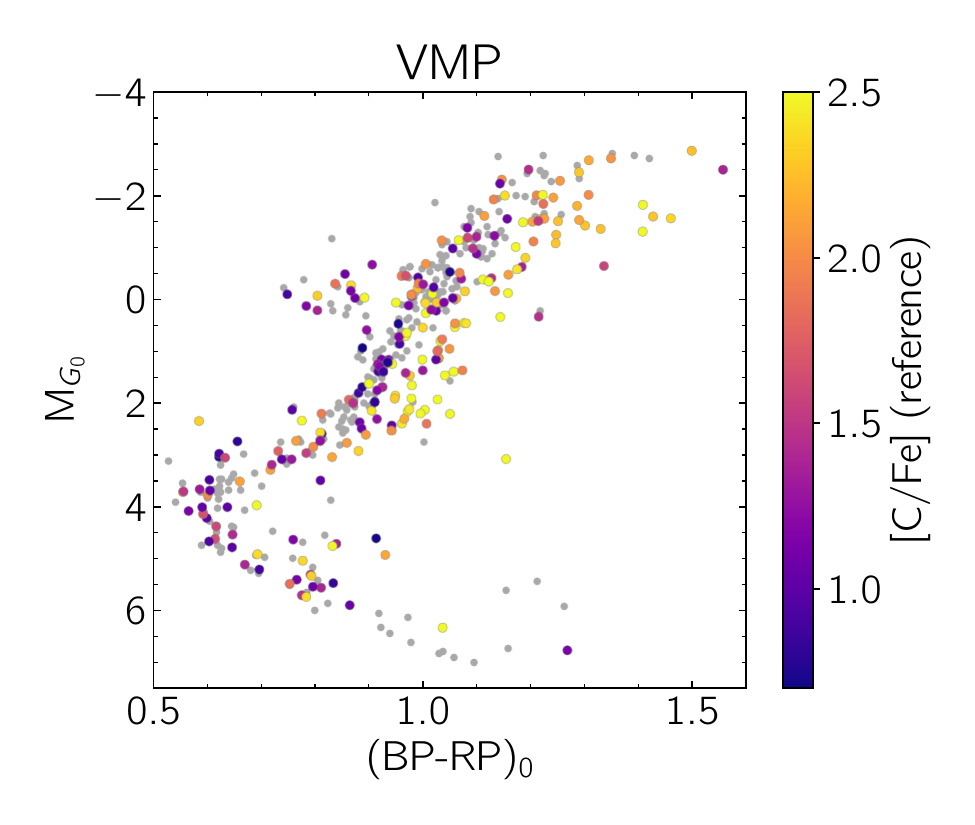}
\includegraphics[width=0.33\hsize,trim={0.2cm 1cm 0.2cm 0.0cm}]{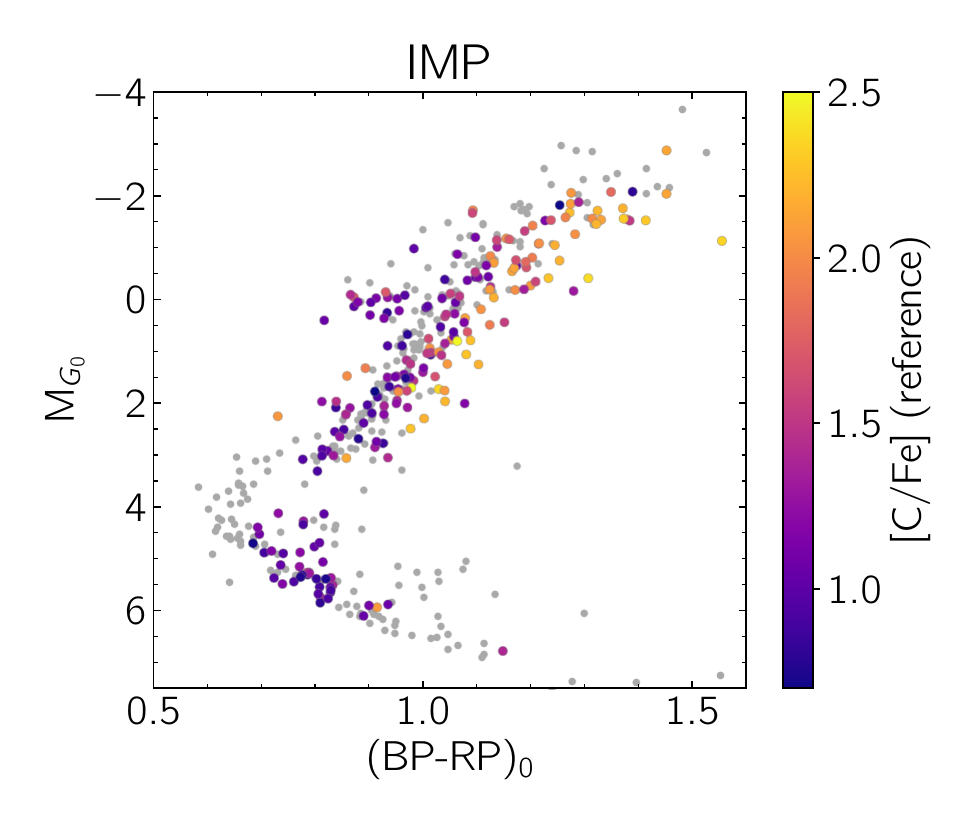}
\includegraphics[width=0.33\hsize,trim={0.2cm 1cm 0.2cm 0.0cm}]{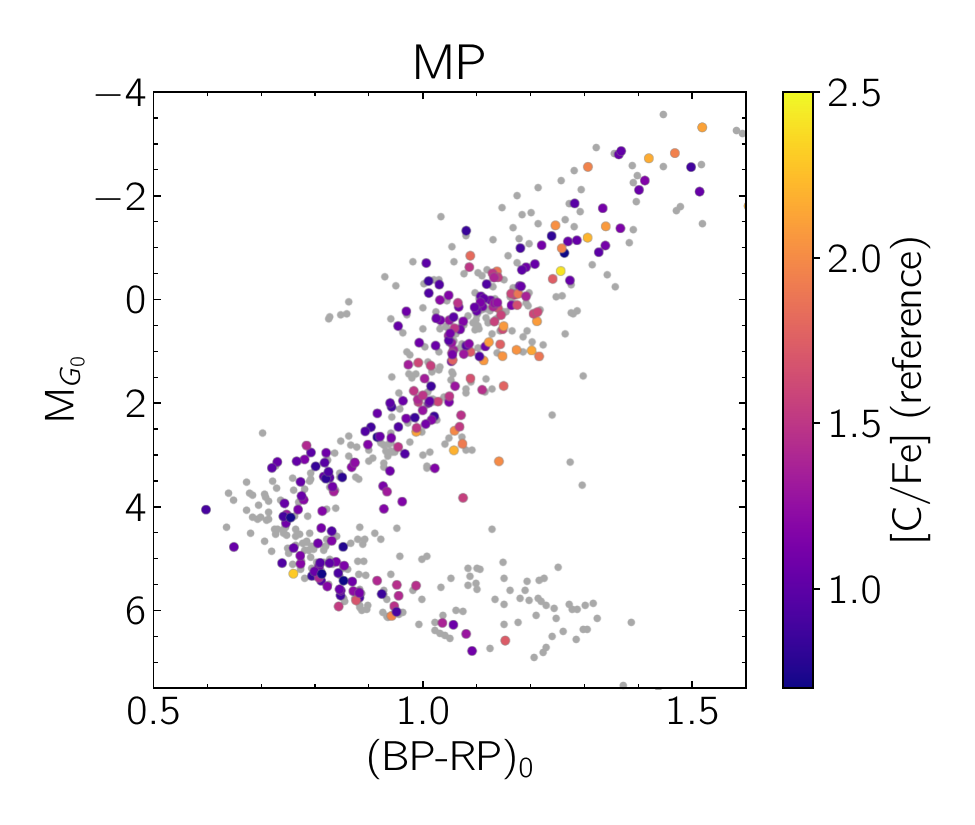}
\caption{CMDs of stars in the reference sample with fractional parallax uncertainty $<25\%$ and E(B$-$V) $<0.3$ in three metallicity bins: $\feh < -2$, $-2 < \feh < -1.5$ and $-1.5 < \feh < -1$, from left to right, respectively. Carbon-normal stars are shown in grey, and stars with $\cfe > 0.7$ are colour-coded by their carbon abundance (not corrected for evolutionary effects). The colour bar is capped at 2.5. Roughly $10\%$ of reference sample is missing in each bin due to the parallax quality cut.}
\label{fig:testcmd} 
\end{figure*}

\begin{figure}
\centering
\includegraphics[width=0.49\hsize,trim={0.0cm 0.0cm 0.0cm 0.0cm}]{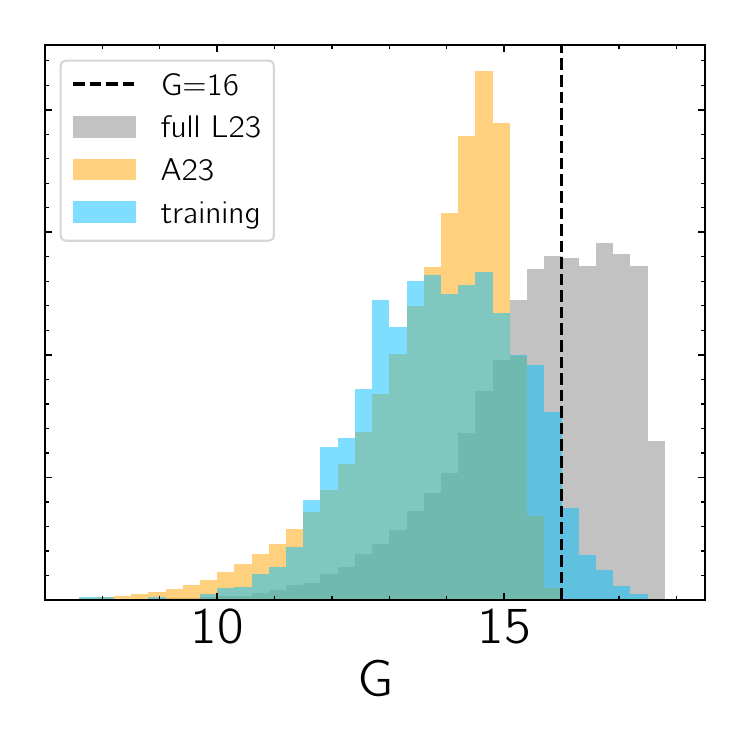}
\includegraphics[width=0.49\hsize,trim={0.0cm 0.0cm 0.0cm 0.0cm}]{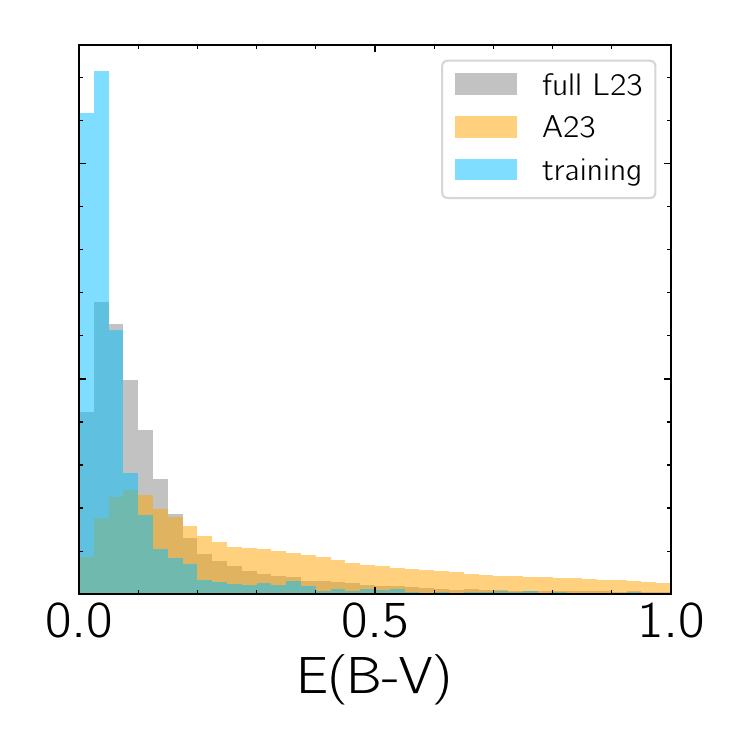}
\caption{Relative distributions of G magnitudes (left) and E(B$-$V) (right) in the \citetalias{lucey23} (grey), \citetalias{andrae23} (orange, after quality cuts and for $\feh_\mathrm{pred} < -0.5$ only) and reference spectroscopy (blue) samples. }
\label{fig:test} 
\end{figure}

\begin{figure*}
\centering
\includegraphics[width=0.495\hsize,trim={0.0cm 1.0cm 0.0cm 0.0cm}]{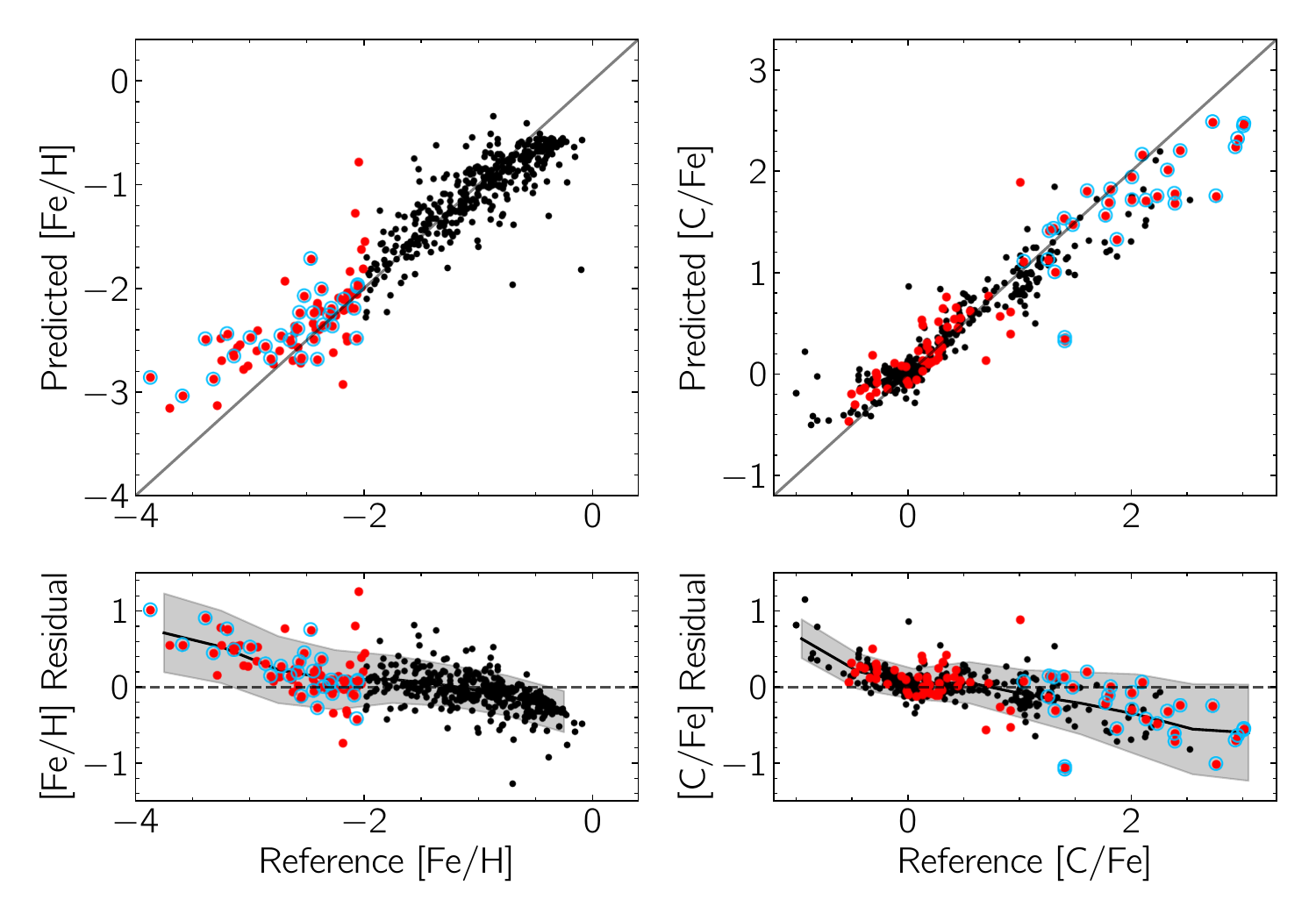}
\includegraphics[width=0.495\hsize,trim={0.0cm 1.0cm 0.0cm 0.0cm}]{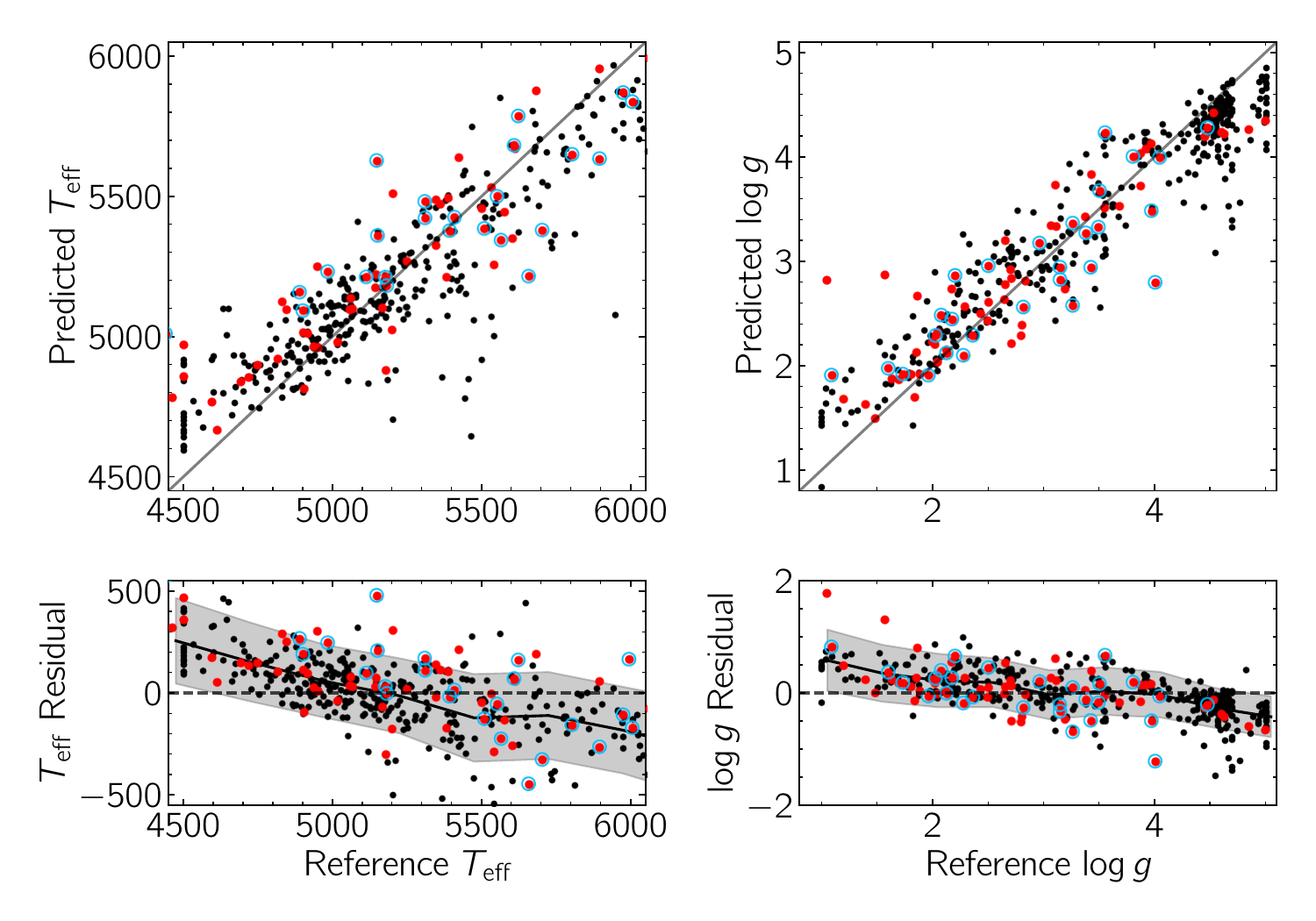}
\caption{Comparison of the reference and predicted [Fe/H] (left), [C/Fe] (middle left), \teff (middle right) and \logg (right) for the test sample (15\% of the available spectroscopic sample, which was not included in the training). The residuals are shown in the bottom panels, with the mean trend line and the shaded region representing the mean predicted standard deviation (in bins of 0.5~dex or 250~K). Throughout the panels, VMP stars with $\feh < -2.0$ are highlighted in red and VMP stars with $\cfe > +1.0$ are circled in light blue. 
}
\label{fig:testpred} 
\end{figure*}

\begin{figure}
\centering
\includegraphics[width=1\hsize,trim={0.0cm 0.0cm 0.0cm 0.5cm}]{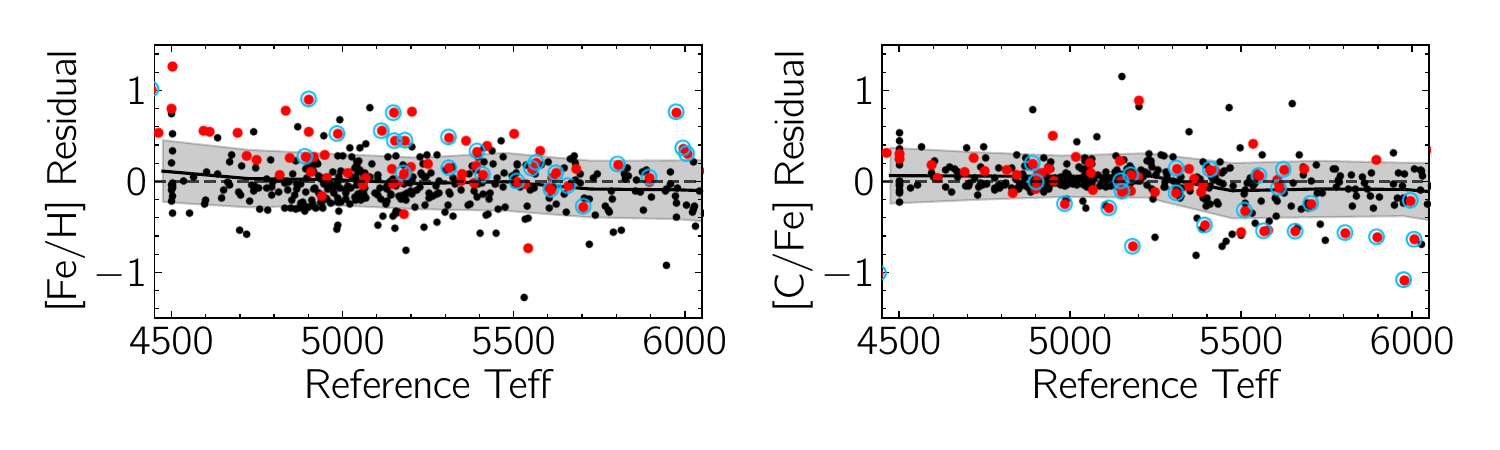}
\includegraphics[width=0.99\hsize,trim={0.0cm 0.0cm 0.0cm 0.0cm}]{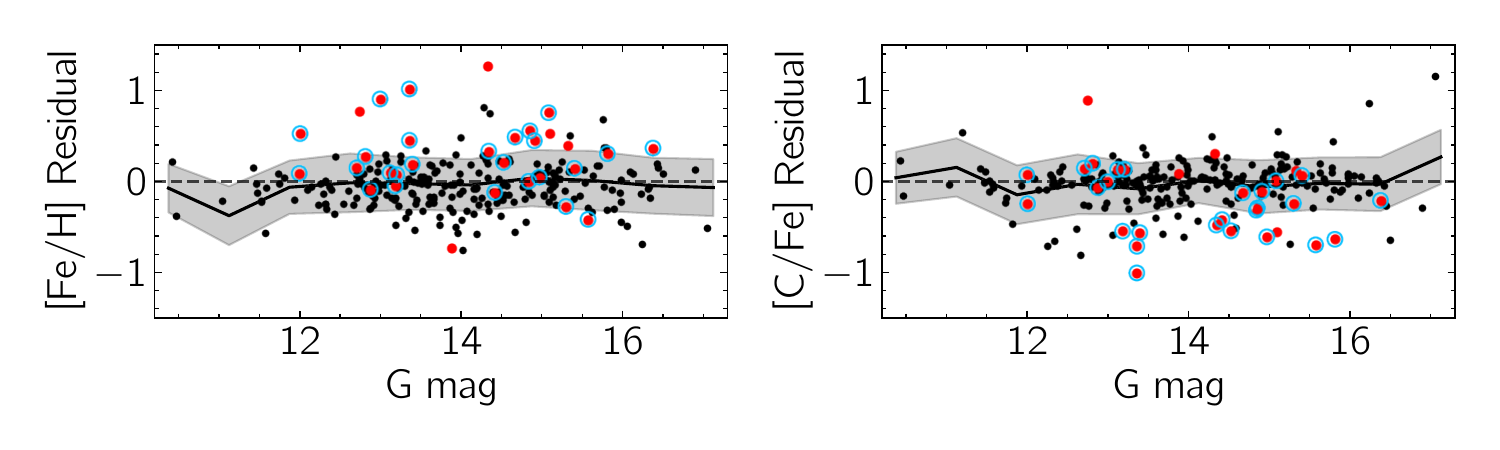}
\includegraphics[width=0.99\hsize,trim={0.0cm 0.0cm 0.0cm 0.0cm}]{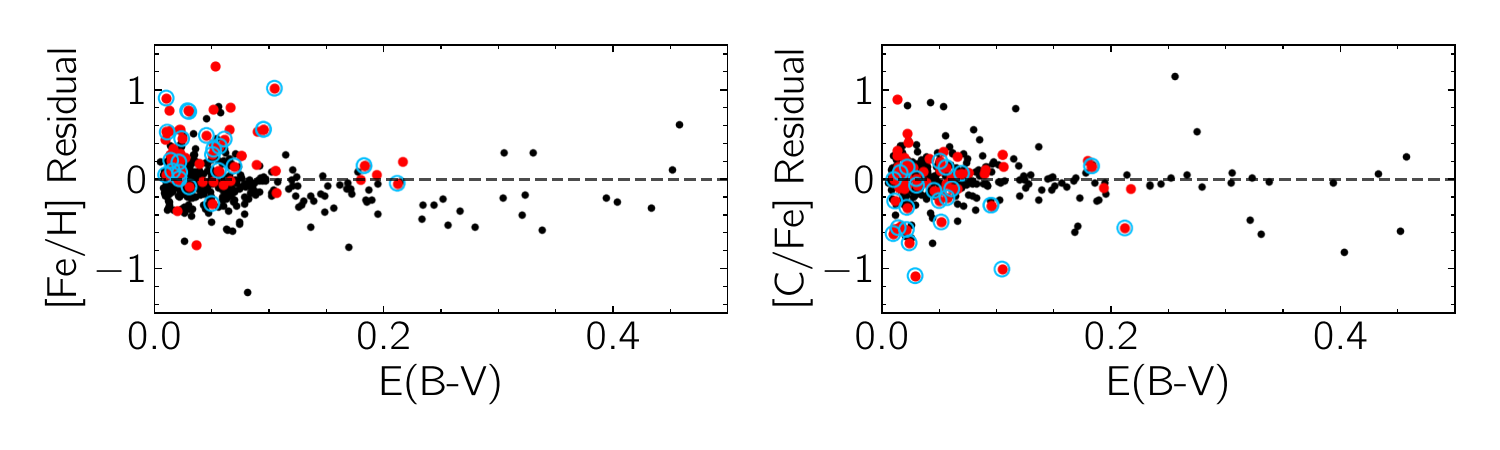}
\caption{Test sample residuals for \feh and \cfe as function of effective temperature (top), \Gaia G magnitude (middle) and E(B$-$V) (bottom). Symbols are the same as in Figure~\ref{fig:testpred}.
}
\label{fig:testpred2} 
\end{figure}

We have the advantage of starting with the pre-selected catalogue from \citetalias{lucey23}, and the possibility of building a training sample that is particularly appropriate for these C-rich candidates. The training sample is based on the \FERRE/LAMOST sample described in the previous section, for which no separation of CEMP-s/-no types is known because they have no measured s-process abundances, supplemented with the Y16 CEMP compilation and metal-poor stars from \citet{li22}. This is an inhomogeneous combination, but it is difficult to homogenise them so it is the best we can do at the moment. The total number of stars in the spectroscopy sample is 2809 (1626 from \citetalias{lucey23}/LAMOST, 862 from "normal" LAMOST, 226 from \citet{li22} and 95 from Y16), which we will call the ``reference sample'' from now on. The $\feh$ and $\cfe$ coverage of the reference sample is shown in Figure~\ref{fig:fcferre}. Of these stars, 909 have $\cfe > 0.7$, with a subset of 244 stars having $\cfe > 0.7$ and $\feh < -2.0$. 

We also present the CMDs of the reference sample in Figure~\ref{fig:testcmd}, in three different metallicity bins. There is a good coverage of the giant branch for the relevant range of [C/Fe] in each metallicity bin. This is especially important for our analysis of the \citetalias{andrae23} giants sample later in this work. The main sequence is less well-covered, but there should be enough stars to help distinguish between dwarfs and giants -- mostly relevant for the \citetalias{lucey23} sample in this work. We note that C-rich stars are typically redder than their C-normal counterparts at the same metallicity in these CMDs. This is expected, due to the molecular carbon features being stronger in the blue compared to the red (see Figure~\ref{fig:xpspec}), which therefore changes the apparent colour of C-rich stars. This kind of effect of carbon on photometry is well known, and has plagued various photometric metal-poor surveys in their target selection resulting in biases against C-rich stars \citep[e.g.][]{dacosta19, martin24}.

We use $85\%$ of the reference sample for training and $15\%$ for testing. We note that most stars in the reference sample have $G<16$, as shown in the left-hand panel of Figure~\ref{fig:test}, whereas the full \citetalias{lucey23} sample has a significant fraction of fainter stars ($>45\%$). Due to this mismatch between the training and science samples, and also because the uncertainties on the XP spectra increase for fainter stars, we expect the quality of the predictions to decrease for fainter stars. We focus mostly on bright stars ($G<16$) in this work. A potential future improvement could be made by including more faint stars in the reference sample (e.g. from SDSS). The extinction coverage of the reference and \citetalias{lucey23} samples are similar (see the right panel of Figure~\ref{fig:test}), although the full \citetalias{lucey23} sample has slightly more stars with intermediate E(B$-$V) values ($0.1-0.2$). Most stars are in relatively low extinction regions -- $80\%$ of the bright stars have E(B$-$V) $< 0.3$, and $93\%$ of the faint stars. Of the stars with higher E(B$-$V), most have $\feh > -1$.

Compared to the training sample in \citetalias{lucey23}, ours is significantly brighter. This is not surprising since LAMOST contains more bright stars than SDSS. The LAMOST and SDSS spectra were also analysed with different methods (FERRE here and SSPP for \citetalias{lucey23}), which have different systematics \citep[see e.g.][who cross-analysed various low-resolution spectroscopic samples with these two pipelines]{arentsen22}. Finally, we supplement our training sample with high-resolution spectroscopic samples (both carbon-normal and CEMP), which \citetalias{lucey23} do not do.

\subsection{Results}

\subsubsection{Test sample}

The results for the test set (the $15\%$ that was kept out of training) are shown in Figure~\ref{fig:testpred}, where the top panels show the reference versus the predicted values and the bottom panels show the residual with the mean trend and predicted standard deviation as a function of the reference values. VMP and CEMP stars are shown in red and circled in blue, respectively. 
In general the predictions are good, especially for \feh (even down to $-3.0$!) and \cfe\, -- our main parameters of interest. The measured standard deviation between reference and predicted values is $\sim 0.30$~dex for \feh and $\sim 0.25$ for \cfe, averaging over the full parameter range. The predicted standard deviations from the network range from $[0.43,0.25]$ for $\feh = [-3.0, -0.5]$, and from $[0.24,0.45]$ for $\cfe = [-1.0,3.0]$. 

Our main focus is on \feh and \cfe, but we predict \teff and \logg as well. We find that the predictions for the chemistry improve when including these, which makes sense since the strength of the absorption features also depends on these parameters. As can be seen in Figure~\ref{fig:testpred}, there is a residual trend for both \teff and \logg. However, for the bulk of the stars, with $\teff = 4800-5500$~K, the residual is less than $\sim 80$~K (with $\sigma \sim 200$~K) for \teff and 0.4~dex for \logg (with $\sigma \sim 0.6$). We do not use these parameters in the remainder of this work, apart from comparing the $\teff$ with \citetalias{andrae23}. 

We also show the dependence of the \feh and \cfe residuals on effective temperature, magnitude and E(B$-$V) in Figure~\ref{fig:testpred2}. This test suggests that there are no clear systematics with temperature and magnitude. It appears that there might be a small bias in [Fe/H] for E(B$-$V) $> 0.1$ for more metal-rich stars, although not for the cluster of VMP stars around 0.2. Unfortunately the coverage of high E(B$-$V) in the test sample is too limited to do a more detailed investigation. We make an additional test with higher extinction PIGS stars in Section~\ref{sec:pigs}, finding some bias in the metallicity. For most of our analyses later in this work, we only use stars with E(B$-$V) $< 0.3$.

\subsubsection{Predicted uncertainties}

As can be appreciated from Figure~\ref{fig:testpred}, the predicted uncertainties (shaded bands in the bottom panels) have some dependence on the predicted parameters. Most notably, the uncertainties increase with decreasing metallicity and with increasing carbon abundances. We find that similar trends are present for the predicted uncertainties in the full \citetalias{lucey23} sample, focusing on the brightest stars for which the uncertainties are not dominated by the quality of the XP spectra. When limiting the sample to a fixed \feh range, there is still an increasing \cfe uncertainty with increasing \cfe\,-- this is not just a metallicity effect. It is the same case the other way around, when only considering a fixed \cfe range, the \feh uncertainties still rise with decreasing \feh. There is also still a range of uncertainties for a fixed value of each parameter. These appear to be mainly due to the brightness of a star -- most clearly varying with the BP magnitude. 

These observations are encouraging in that the uncertainties are likely representing true physical limitations and/or limitations in the training set (e.g. it is harder to measure \feh at low metallicity and for C-rich stars, and there are larger uncertainties in the spectroscopic sample for high \cfe stars) as well as uncertainties due to the quality of the XP spectra (probed by the brightness). 

\subsubsection{Full \citetalias{lucey23} sample predictions}

\begin{figure}
\centering
\includegraphics[width=1.0\hsize,trim={0.5cm 0.0cm 1.0cm 0.0cm}]{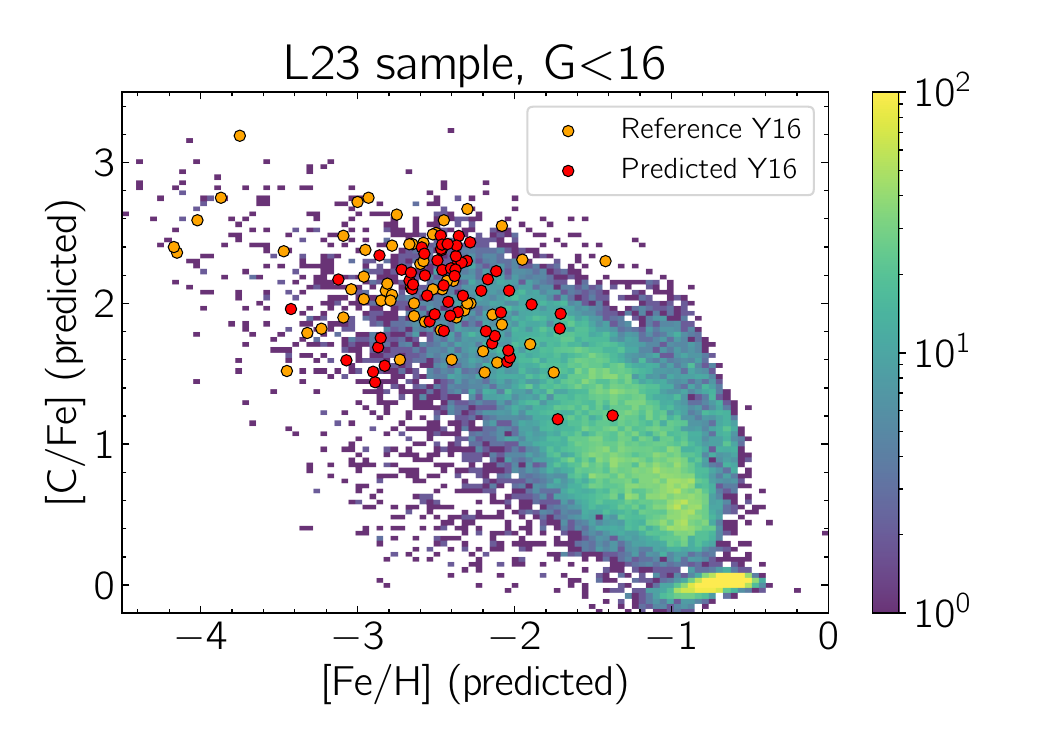}
\includegraphics[width=1.0\hsize,trim={0.5cm 0.0cm 1.0cm 0.0cm}]{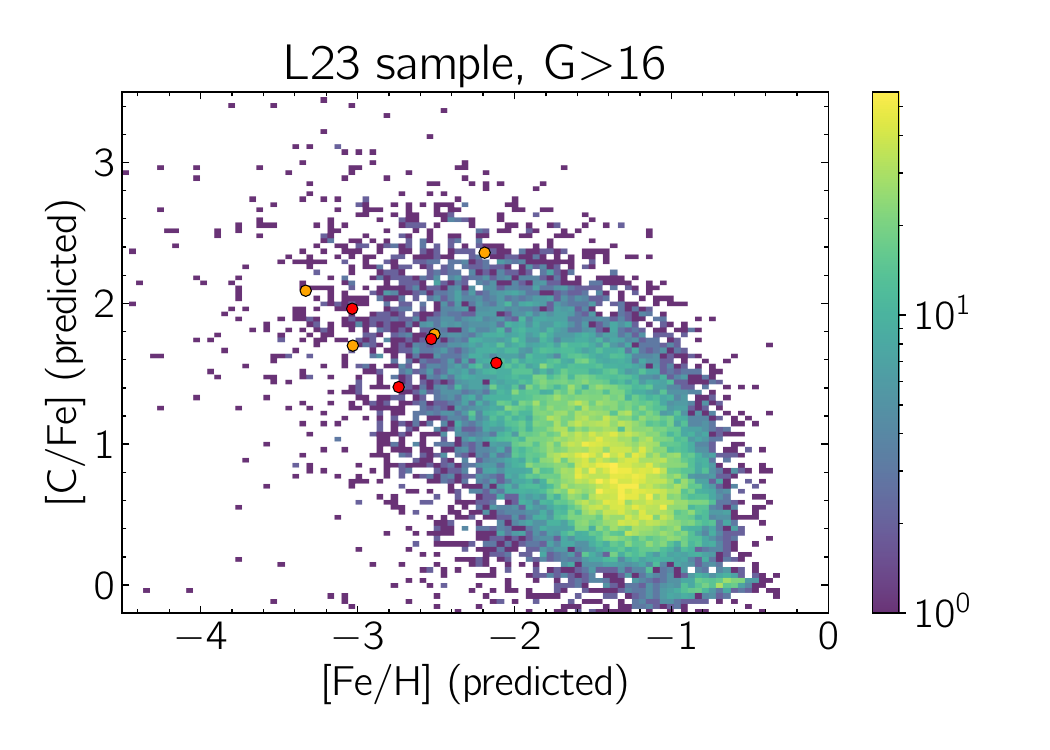}
\caption{Predicted [Fe/H] vs. [C/Fe] for the full \citetalias{lucey23} catalogue. Top: bright stars with $G<16$, bottom: fainter stars with $G>16$. CEMP stars from Y16 are highlighted in orange (high-resolution spectroscopic values) and red (predicted values). The colour bar shows the number of stars in each pixel, and the maximum in the top panel has been fixed to $10^2$ to showcase the distribution of C-rich stars better. No quality cuts or evolutionary carbon corrections have been applied.}
\label{fig:pred} 
\end{figure}

The predicted $\feh - \cfe$ diagrams are shown in Figure~\ref{fig:pred} for the full \citetalias{lucey23} sample, split into bright (top) and faint (bottom) stars. Focusing on the bright stars, three over-densities can be noticed: the pronounced metal-rich, carbon-poor blob in the same location as the Ca H\&K emission stars in the LAMOST sample (red points in Figure~\ref{fig:fcferre}), an almost vertical branch around $\feh \sim -0.8$, and a large blob with some slight sub-structure for the rest of the stars. The substructure in the main blob may potentially be connected to the training sample piling up around $\cfe = +1.0/+2.0$. They are unlikely connected to the two different types of CEMP stars, as most C-rich stars with $\feh > -2.5$ are expected to come from the binary mass-transfer scenario. Also shown are the Y16 stars with reference $\cfe > +1.5$ (orange), to highlight where they end up in the predictions (magenta). Note that most of these were used in the training, so this is not an independent test -- just a sanity check on the stars we are particularly interested in. 

In the reference sample, $\sim 25\%$ of the LAMOST/\citetalias{lucey23} stars were found to be impostor carbon-rich stars that are actually stars with Ca H\&K in emission. This percentage turns out to be very similar in the full \citetalias{lucey23} sample: among the bright stars, $\sim20\%$ lie in this same region of the parameter space ($\cfe < +0.25$ and $\feh > -1.0$). The blob is less well-populated for the faint stars (only $3\%$), although it is still present. The CMD for the faint stars is also not well-populated in the emission star region (red blob region in the bottom panel of Figure~\ref{fig:cmd}) supporting their absence. The bulk of these metal-rich, carbon-poor stars are variable in their photometry (Pvar $>0.9$), consistent with what we found for the emission line stars in the training sample.

\begin{table}
\centering
\caption{\label{table:L23summary}Summary of bright ($G<16$) stars in \citetalias{lucey23} passing our quality cuts}
\begin{tabular}{lcl}
 \hline
  & $\feh < -1.5$ & $\feh < -2.0$ \\
 \hline
 Fraction of sample & 39\% & 13\% \\
 $\cfe > +0.7$ & 91\% & 95\% (N=1982) \\
 $\cfe > +1.5$ & 40\% & 65\% (N=1356) \\ 
 \hline
\end{tabular}
\end{table}

It is curious that such a large number of these metal-rich stars made it into the \citetalias{lucey23} C-rich candidates. As mentioned before, our suggestion is that \texttt{XGBoost} predicts these stars to be CEMP from the XP spectra because they have weak calcium lines and solar-level carbon features. Furthermore, potentially these stars were not present in the fainter SDSS training sample compared to our brighter LAMOST training sample, and/or they did not have appropriate labels from the SSPP, but the SSPP results for \cfe are not publicly available so we cannot check this. 

Of the full sample, $25\%$ of the stars has predicted $\feh > -1.0$, going up to $35\%$ for the bright sample. These numbers are somewhat smaller than the $\sim 60\%$ based on \citetalias{andrae23} metallicities reported by \citetalias{lucey23}, but still quite significant. This might partly be due to the \citetalias{andrae23} metallicities being strongly biased towards higher metallicities for C-rich stars. For the C-rich stars ($\cfe > 0.7$), we find that $10/14\%$ (all/bright) have $\feh > -1.0$, roughly double compared to the $7\%$ \citetalias{lucey23} predicted. 

Next, we summarise the properties of our cleaned bright sample $(G<16)$, after applying the following quality cuts: E(B$-$V) $< 0.3$, Pvar $<0.5$ and (BP$-$RP)$_0 < 1.35$ (not shown in the figures). Only $1\%$ of the stars now lies in the region of the emission line stars, and the branch around $\feh \sim -0.8$ almost completely disappears. Most of the stars with predicted $\feh < -3.0$ disappear as well, largely due to the variability cut -- these were likely spurious predictions. The numbers of IMP, VMP and CEMP stars are summarised in Table~\ref{table:L23summary}. There are almost 2000 promising VMP C-rich candidates, an exciting sample that deserves to be followed up spectroscopically in the future. We note that this is a conservative sample -- there are $\sim 1500$ CEMP candidates that are being removed by our quality cuts, which have less certain estimates but many of them are still likely to be good candidates. Furthermore, while we focus here on the most metal-poor stars, the more metal-rich stars are still very valuable -- they are likely to be CH stars, the slightly more metal-rich counterpart of CEMP-s stars.

\begin{figure}
\centering
\includegraphics[width=1.0\hsize,trim={0.5cm 0.65cm 0.0cm 1.0cm}]{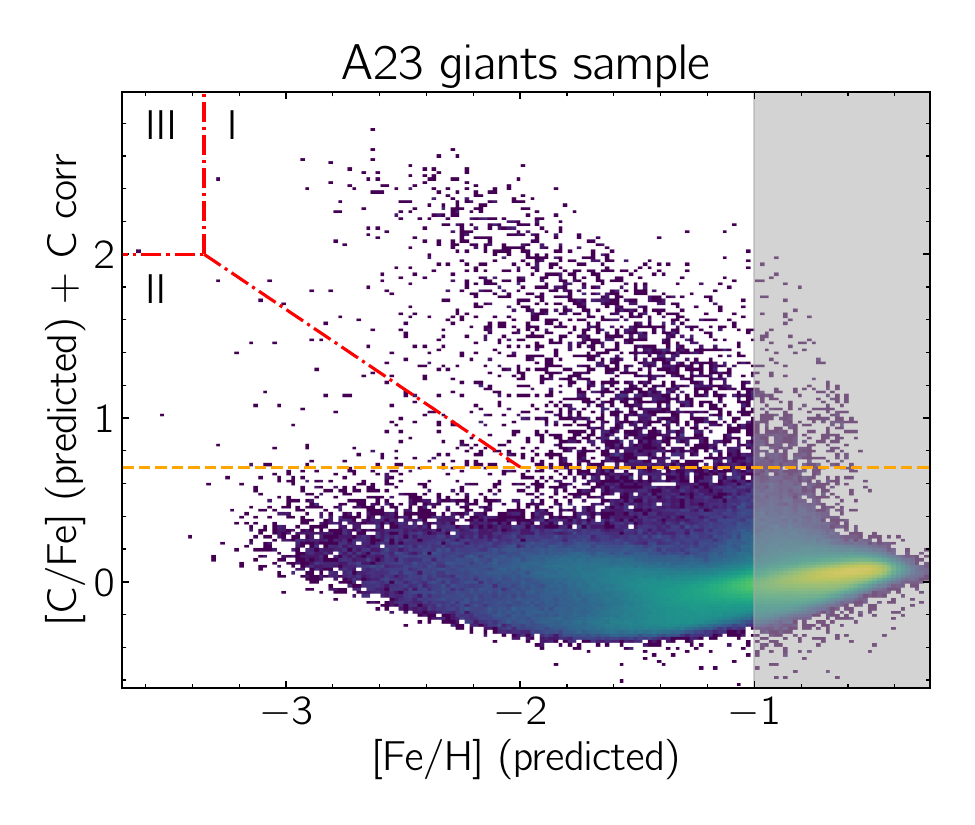}
\caption{Distribution of predicted \feh and $\cfe_\mathrm{corr}$ (corrected for evolutionary effects) for giants from the \citetalias{andrae23} sample with \Gaia RVs. The orange dashed horizontal line is at $\cfe = 0.7$. The red dashed-dotted lines indicate a rough separation between the Group I, II and III CEMP stars according to \citet{yoon16}. Note that the x- and y-axes have slightly different limits compared to Figure~\ref{fig:pred}.
}
\label{fig:cfe_a23} 
\end{figure}

\subsubsection{Application to the \citetalias{andrae23} giants sample}\label{sec:a23appl}

To get a better sense of the frequency of C-rich objects in a high-quality sample, and to potentially identify other promising bright CEMP candidates that were missed in the \citetalias{lucey23} sample, we proceeded to apply the network to the vetted \texttt{XGBoost} giant sample of \citet{andrae23}, limiting to stars with \Gaia RVS radial velocities (mostly stars with $G<15$). There are 10 million stars in this sample. The regime of bright giants is exactly where our methodology is expected to work very well, given the high quality XP spectra and the relatively strong carbon features for cool, evolved stars. It is, however, a much stricter sample (see Section~\ref{sec:a23}) and will lack many of the \citetalias{lucey23} candidates. 

The resulting $\feh - \cfe_\mathrm{corr}$ diagram is shown in Figure~\ref{fig:cfe_a23}, for stars with E(B$-$V) $<0.3$, bp\_chi\_squared $< 10.5 \, \times$ bp\_degrees\_of\_freedom (indicating good quality XP spectra), Pvar $<0.5$ (to remove photometrically variable stars, of many different kinds) and (BP$-$RP)$_0 < 1.35$ (because of the limitations of the training sample for cooler stars). The quality cuts are important to clean up the sample, for example the Pvar cut removes a significant number of EMP candidates that are clearly spurious upon further investigation. There is a clear "main" sequence and a cloud of C-rich objects, with $\sim 3200$ objects having predicted $\cfe > 0.7$ and $-4.0 < \feh < -0.5$. The main sequence appears to be split into two. We find that, at low metallicity, the lower sequence has lower \logg\,-- the double sequence is therefore likely the result of evolutionary effects. 

The rough division lines between the Groups I, II and III CEMP classes of \citet{yoon16} have been indicated in the figure, where type~I corresponds predominantly to CEMP-s and type~II and III predominantly to CEMP-no stars (with a small number of exceptions). The overwhelming majority of our C-rich sample falls in the Group I region, and is likely of the binary mass-transfer type (CEMP-s/CH stars). There is also a significant population with slightly higher metallicity ($-1.6 < \feh < -0.6$) and intermediate enhanced carbon abundances ($ 0.3 < \cfe < 0.7$). We show in Section~\ref{sec:galah} that many of these stars appear to be enhanced in barium, consistent with them being CH-stars too.

\begin{figure}
\centering
\includegraphics[width=1.0\hsize,trim={0.0cm 0.8cm 0.0cm 0.5cm}]{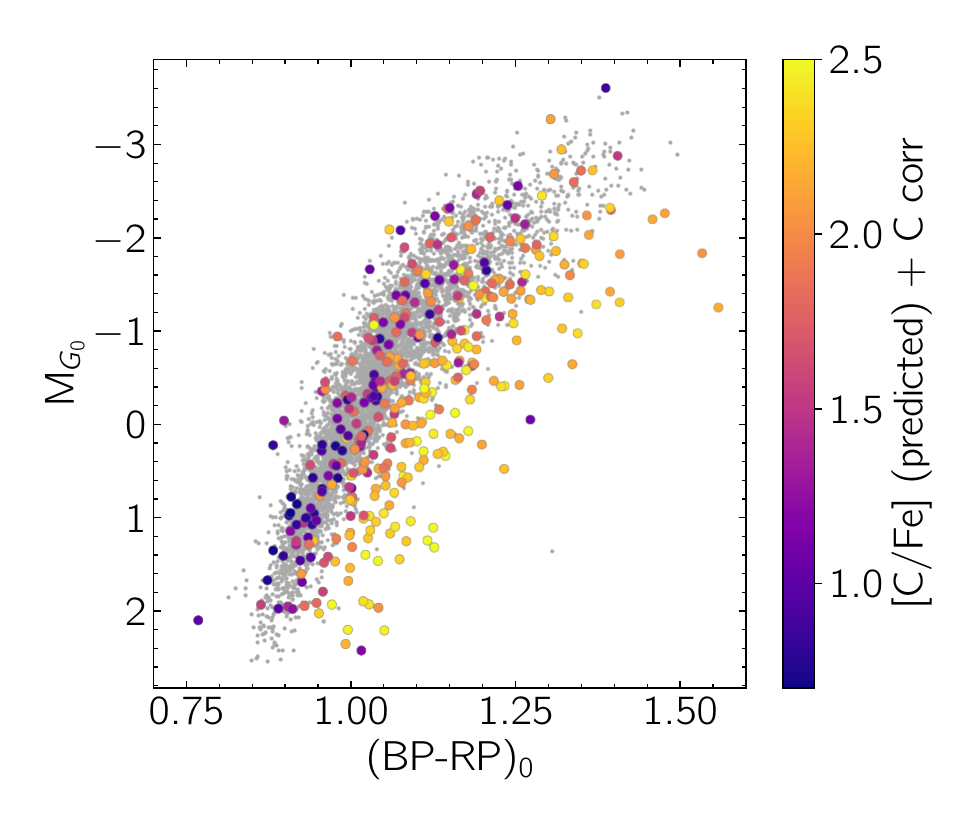}
\caption{CMD of VMP stars ($\feh < -2.0$) in our \citetalias{andrae23} sample (grey), with C-rich VMP stars colour-coded by their carbon abundance (the colour bar is capped at 2.5). Only stars with E(B$-$V) $< 0.1$ have been included to avoid potential shifts due to extinction, and the red (BP-RP)$_0$ colour-cut has been slightly relaxed compared to other analyses in this work to show most of the CMD.}
\label{fig:cmd_crich} 
\end{figure}

Finally, we note that, as in the \FERRE training sample, C-rich stars among our \citetalias{andrae23} predictions appear to be offset towards warmer \teff compared to C-normal stars of the same metallicity. As discussed in Section~\ref{sec:ferre1}, this is a limitation of the training sample analysis and likely not physical. The expectation is that C-rich stars have similar \teff to C-normal VMP stars.\footnote{Mass-transfer could make a star slightly more massive and therefore hotter, but this is expected to be an increase on the order of only $\sim 0.05$~M$_{\odot}$ \citep[e.g.][]{abate15_binarypop}, which cannot explain the discrepancy found here.} 
When looking at the CMD, the C-rich stars are redder than C-normal VMP stars, see Figure~\ref{fig:cmd_crich}. This is likely not connected to their \teff, but due to the molecular carbon features -- we also see this effect in the CMD of the reference sample (Figure~\ref{fig:testcmd}).

\subsubsection{Completeness of \citetalias{lucey23} sample}

A test of completeness of the sample of C-rich candidates from \citetalias{lucey23} for bright stars can be done by comparing with our predictions for the \citetalias{andrae23} giants sample. For example, we would like to check whether there are any CEMP stars missing from \citetalias{lucey23} that we are picking up with the giants sample. Using the same quality cuts on extinction, XP quality and Pvar as in the previous section and a definition of $\cfe > 0.7$ as C-rich, we find that of the 3200 predicted C-rich metal-poor stars in \citetalias{andrae23}, $90\%$ are already present in the \citetalias{lucey23} catalogue. Of the missing stars, $\sim 80\%$ has predicted $\cfe < +1.0$ and mostly $\feh < -1.5$, these have the weakest \cfe features so are most challenging to pick up. Overall, the completeness of C-rich, bright, cool giant stars in the \citetalias{lucey23} sample appears to be very good, especially for $\cfe > +1.0$.

\subsection{Spectroscopic comparisons}\label{sec:speccomp}

Next, we compare our \citetalias{andrae23} predictions to various high-resolution spectroscopic samples. In all comparisons we only include stars with E(B$-$V) $< 0.3$, Pvar $<0.5$, and spectroscopic $\teff > 4400$~K (for cooler stars our predictions are less solid). It should be noted that all these comparisons are for bright stars (mostly $10 < G < 14$), where our predictions may be expected to be better than for fainter stars.  

Comparisons with other \feh and/or \cfe predictions from \Gaia XP (\citetalias{andrae23}, \citealt{fallows24}) and photometry \citep{huang24} are shown in the Appendix.

\begin{figure*}
\centering
\includegraphics[width=0.45\hsize,trim={0.0cm 0.0cm 0.0cm 0.0cm}]{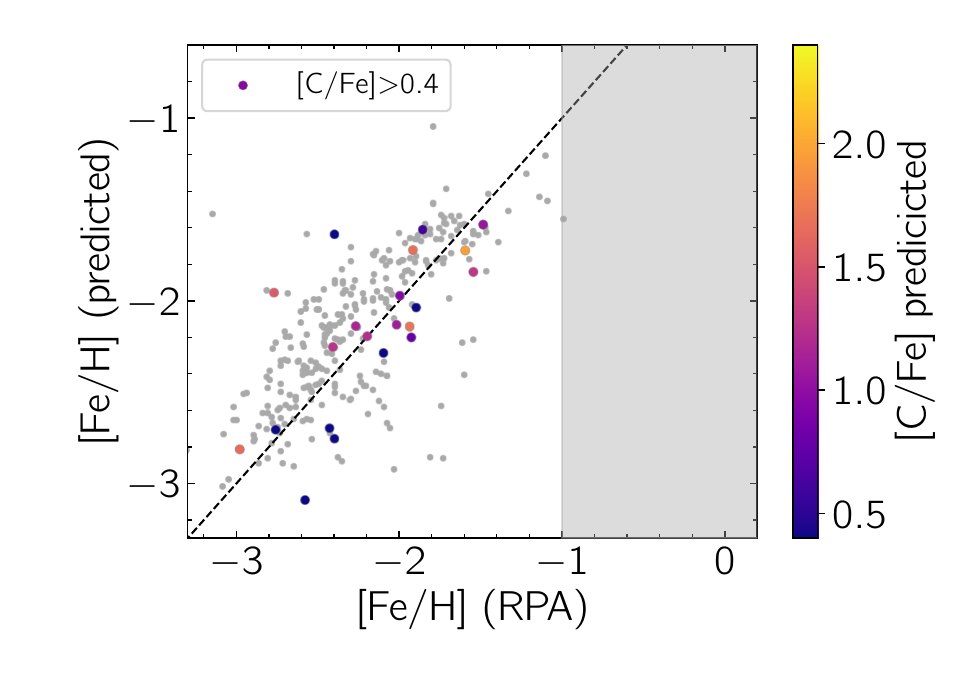}
\includegraphics[width=0.35\hsize,trim={0.0cm 0.0cm 0.0cm 0.5cm}]{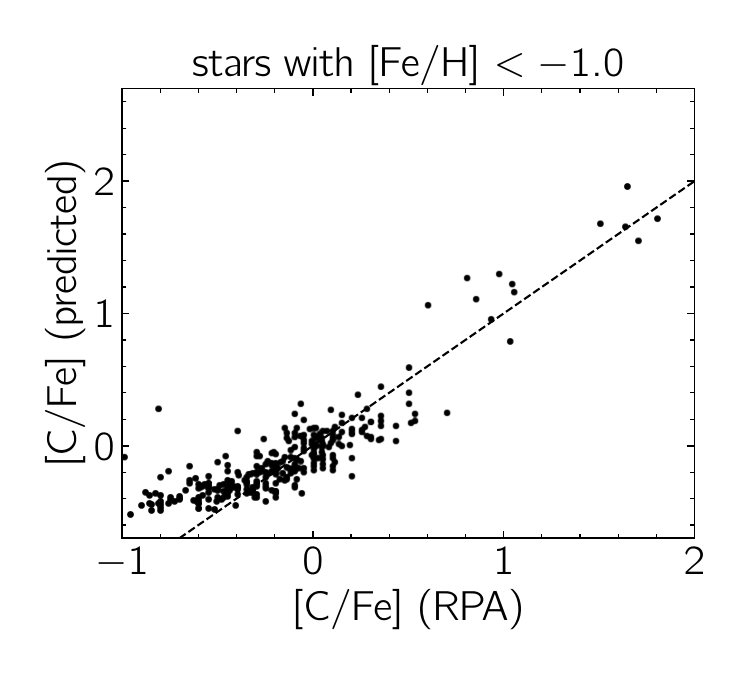}
\includegraphics[width=0.45\hsize,trim={0.0cm 0.0cm 0.0cm 0.0cm}]{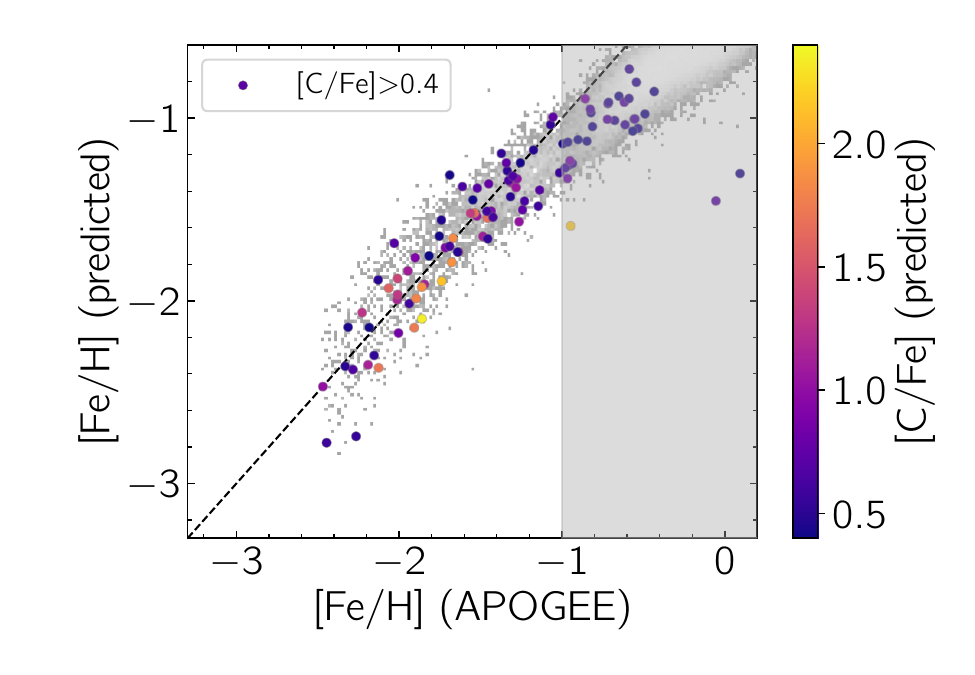}
\includegraphics[width=0.35\hsize,trim={0.0cm 0.0cm 0.0cm 0.5cm}]{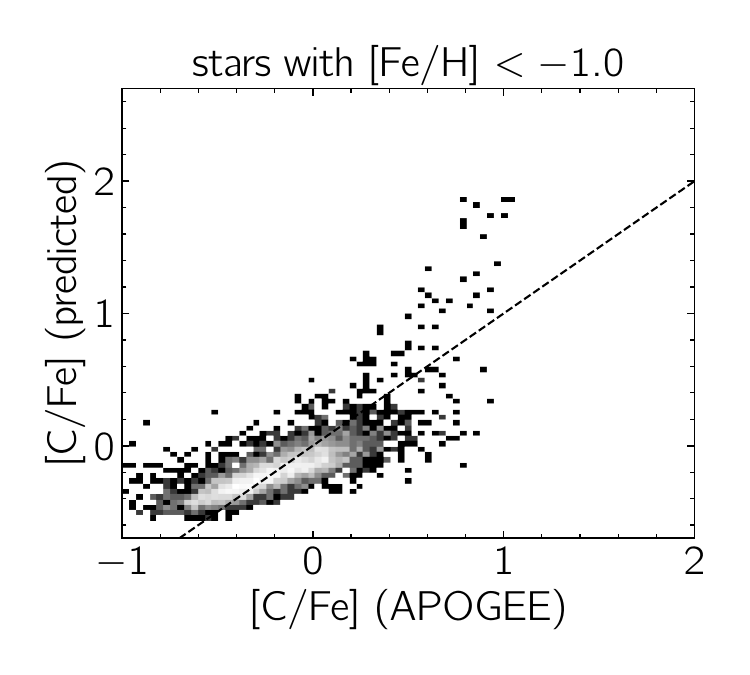}
\includegraphics[width=0.45\hsize,trim={0.0cm 0.0cm 0.0cm 0.0cm}]{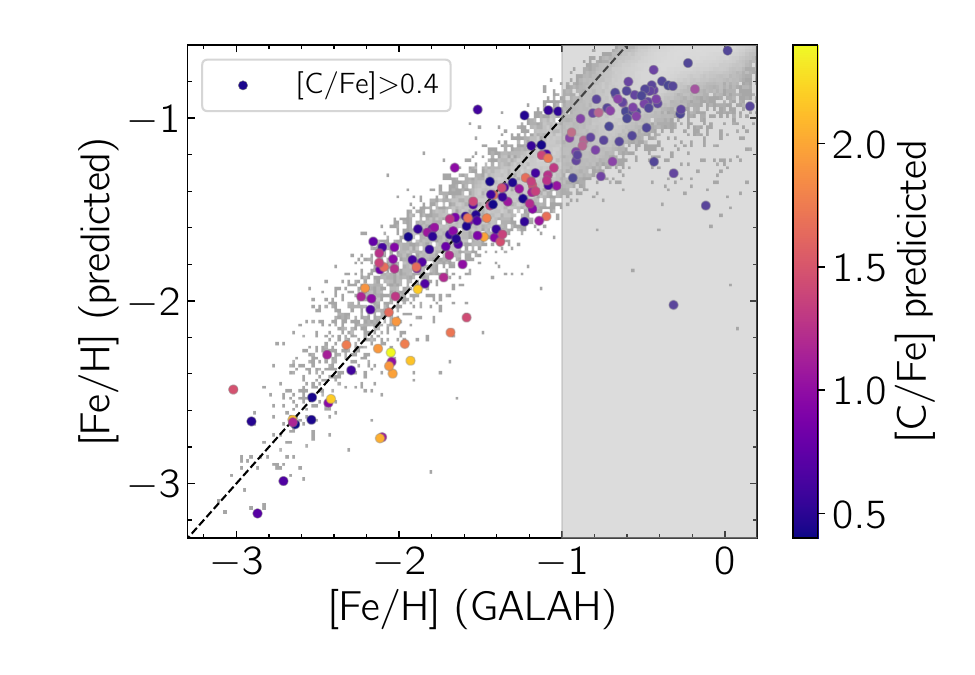}
\includegraphics[width=0.35\hsize,trim={0.0cm 0.0cm 0.0cm 0.5cm}]{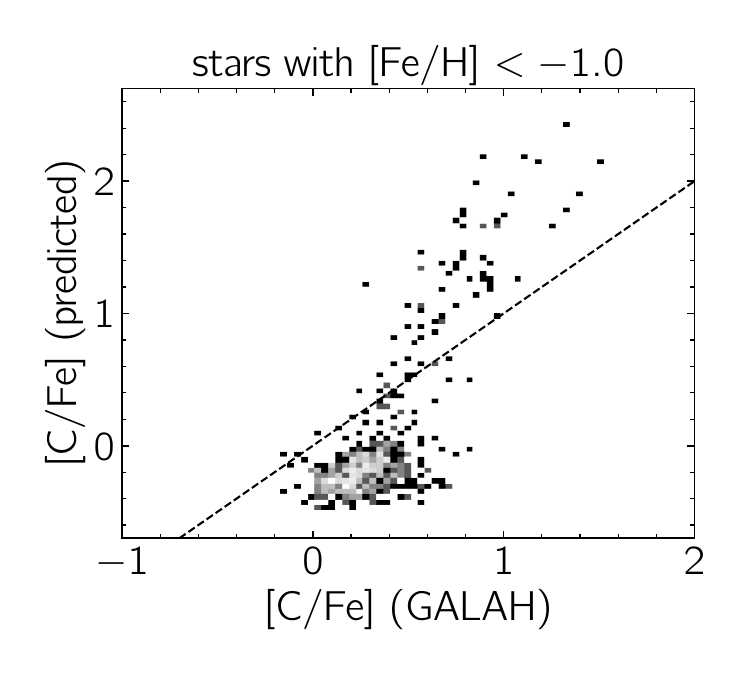}
\caption{Comparisons of our predicted \feh (left) and \cfe (right) to the high-resolution spectroscopy values from RPA (top), APOGEE (middle) and GALAH (bottom). In the left-hand panels, colour-coding is by predicted \cfe for stars to be predicted relatively C-rich ($\cfe > +0.4$). The right-hand panels only show metal-poor stars with spectroscopic $\feh < -1.0$. 
}
\label{fig:apogalahcomp_a23} 
\end{figure*}

\subsubsection{RPA}

The R-Process Alliance (RPA) has followed up many bright metal-poor stars with dedicated high-resolution spectroscopy, and carbon abundances are included in various works \citep{hansen18, ezzeddine20, holmbeck20}. It is an excellent independent sample to check the validity of our predictions. The cross-match between our \citetalias{andrae23} predictions and these samples is shown in the top row of Figure~\ref{fig:apogalahcomp_a23}. We note that we use the evolutionary uncorrected \cfe abundances from RPA to be consistent with our predictions.

We find that our metallicity predictions are slightly biased towards higher metallicities. For this dedicated VMP high-resolution spectroscopy sample, the median offset is $\sim 0.15$~dex with a dispersion of $\sim 0.26$~dex. For the C-rich stars, however, the metallicity predictions show little offset. The carbon abundances show excellent agreement, for both C-normal and C-rich stars. Only for the very carbon-poor RPA stars with $\cfe < -0.6$ there a clear systematic, where our predictions flatten. Selecting the stars with $\cfe > -0.6$, we find a median difference of $0.03$~dex with a dispersion of $0.16$~dex. This independent comparison supports the quality of our predictions for bright stars.

\subsubsection{APOGEE}\label{sec:apogee}

We also compare our predictions to DR17 of the spectroscopic APOGEE survey \citep{apogeedr17}, which has mostly stars between $9 < G < 15$, with a mean of $G=12$. APOGEE is not ideal for metal-poor and C-rich stars, because the lowest metallicity in their grid is $\feh = -2.5$ and the maximum value of \cfe in the grid is $+1.0$ \citep[see also e.g.][]{Kielty17}. However, it is still useful to check the metallicities of intermediate metal-poor stars, as well as to highlight objects more C-rich than normal. The comparisons of \feh and \cfe for stars with APOGEE SNR $> 30$ are shown in the middle row of Figure~\ref{fig:apogalahcomp_a23}. 

In the \feh panel, stars with predicted $\cfe > +0.4$ are colour-coded by the predicted \cfe. Overall, we find that the metallicities agree well between our predictions and APOGEE for $\feh \lesssim -1.0$, for both C-normal and C-rich stars. The \feh scatter between APOGEE and our predictions is $0.18$~dex, with APOGEE being higher by 0.04~dex on average for $-1.5 < \feh < -1.2$ and the difference going to zero for $\feh < -1.5$. We do not expect our metallicities to be good in the more metal-rich regime, given the assumption of $\alphafe = +0.4$ in our spectroscopic sample and our focus on creating a good training set for metal-poor stars. 

In the \cfe panel, only stars with spectroscopic $\feh < -1.0$ are shown. There is a good correlation between our predictions and APOGEE values for C-normal stars, although it is not along the 1-1 line.  All C-rich candidates from our predictions are indeed also relatively C-rich in APOGEE, although less so due to the $\cfe = 1$ limit of the APOGEE grid. Many of our most metal-poor C-rich candidates have a raised \texttt{FE\_H\_FLAG} or \texttt{C\_FE\_FLAG}. 

\subsubsection{GALAH}\label{sec:galah}

We also cross-match our \citetalias{andrae23} predictions with GALAH DR4 \citep{galahdr4}, another high-resolution spectroscopic survey targeting bright stars (the metal-poor stars are mostly between $12 < G < 14$). In this latest GALAH data release, CNO abundances are included for the first time. We remove stars with raised `low S/N' or `not converged' flags, \texttt{snr\_px\_ccd3} $< 30$ and for the carbon comparison we also remove stars with \texttt{flag\_c\_fe}~$!= 0$. The comparisons for \feh and \cfe are shown in the bottom row of Figure~\ref{fig:apogalahcomp_a23}. 

For \feh, the comparison looks similar as for APOGEE. Some of the most C-rich stars have higher metallicities in GALAH than from our predictions. It is not clear whether this is an issue with our metallicities or those from GALAH -- both are possible for such extreme stars. For \cfe, there are much fewer stars due to the quality cut on the C flag. For the stars remaining, there is an offset for C-normal stars between our predicted \cfe and that from GALAH. A similar offset was noticed by \citet{galahdr4} between GALAH and APOGEE for metal-poor stars. For higher carbon abundances stars, the offset goes in the opposite direction. We note that stars identified as C-rich from our predictions are also on the high side of the carbon distribution in GALAH.  

\begin{figure}
\centering
\includegraphics[width=1.0\hsize,trim={0.5cm 0.0cm 0.5cm 0.5cm}]{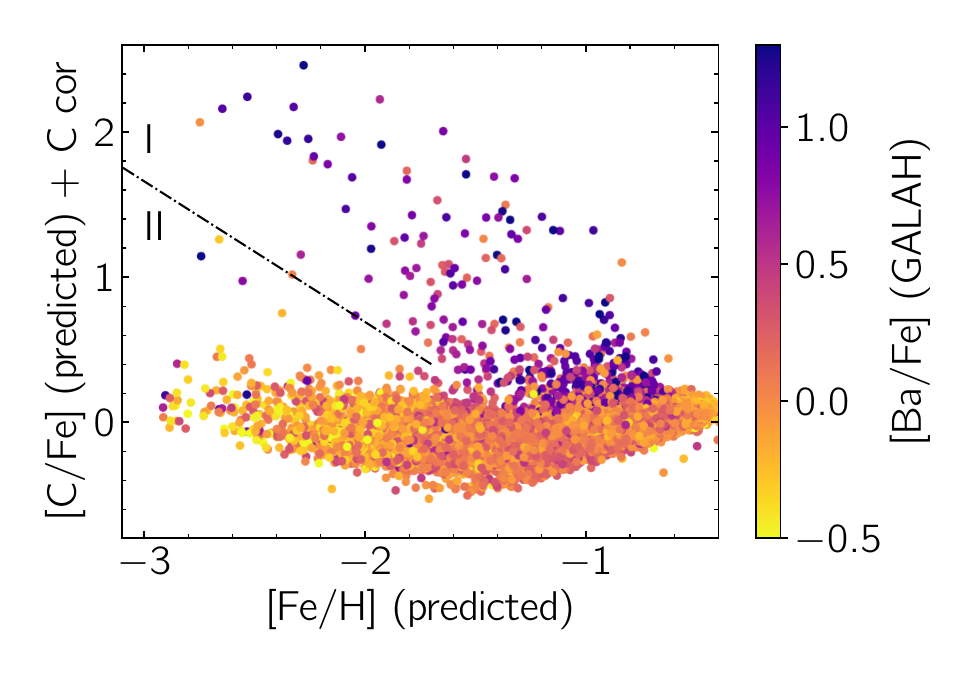}
\caption{The predicted \feh and $\cfe_\mathrm{ccor}$ from our \citetalias{andrae23} sample colour-coded by [Ba/Fe] from GALAH DR4 \citep{galahdr4}. Most of our C-rich candidates are enhanced in Ba, as expected in an AGB binary mass-transfer scenario. The rough separation between Group I and II CEMP stars from \citet{yoon16} is shown with the dashed-dotted line. 
}
\label{fig:galah_a23} 
\end{figure}

Alongside many other elements, GALAH includes barium abundances for many stars. Barium is of interest because it is an s-process element, and CH/CEMP-s stars are typically enhanced in these. In Figure~\ref{fig:galah_a23}, we colour-code our $\feh - \cfe_\mathrm{corr}$ predictions by the [Ba/Fe] from GALAH, only keeping stars with \texttt{flag\_ba\_fe} = 0. Barium has strong lines and is therefore well-measured in many stars. Almost all our C-rich candidates show enhanced Ba, consistent with these stars being AGB binary mass-transfer products. At the lowest metallicities, there appear to be several Ba-normal/poor C-rich stars, which are likely to be CEMP-no stars.

\subsubsection{PIGS CEMP stars}\label{sec:pigs}

Most of the tests we have done are for stars in low extinction regions, which is where most of our training sample is located (see Figure~\ref{fig:test}). To test the results at slightly higher extinctions, and for an independent test, we cross-match our \citetalias{lucey23} sample with the Pristine Inner Galaxy Survey (PIGS) spectroscopic follow-up sample \citep{arentsen20b, ardernarentsen24}. We find 25 stars in common, which have $0.2<$ E(B$-$V) $<0.6$ and $14 < G < 16.5$. There are not many stars in common between PIGS and the \citetalias{andrae23} sample, because PIGS stars are typically fainter and too distant to have good parallaxes.

All PIGS stars with a match in \citetalias{lucey23} are indeed C-rich according to the PIGS spectroscopy. There are 13 stars passing all recommended PIGS spectroscopic quality criteria (good SNR, not double-lined and good \FERRE $\chi^2$). For these, the carbon abundances are fairly consistent between the PIGS spectroscopy and our predicted values, the mean difference $\Delta \cfe$ (PIGS $-$ predicted) is $-0.10$~dex with a standard deviation $\sigma = 0.34$ for VMP stars and $-0.13, \sigma = 0.16$ for $\feh > -2.0$. The metallicities are consistent for $\feh > -2$ ($\Delta \feh = -0.07, \sigma = 0.24$) but significantly offset for VMP stars ($\Delta \feh = -0.59, \sigma = 0.34$). This could potentially be due to extinction affecting the bluer spectral regions more (where the most of the metallicity information is present for C-rich stars), and/or due to the lack of VMP, high-extinction training sample stars. 

This limited test suggests that CEMP stars can still be recognised from the XP spectra in higher extinction regions, but some caution should be taken with their exact parameter predictions.

\section{Discussion}\label{sec:discussion}

\subsection{What kind of C-rich objects do we find?}

As discussed in Section~\ref{sec:a23appl}, given the metallicities and carbon abundances of our C-rich candidates (located in the \citet{yoon16} Group I region), most stars in our sample are likely to be CH/CEMP-s stars -- the result of mass-transfer from a former AGB companion. This is supported by the GALAH results in the previous section showing that many of our candidates are barium-rich. But what about the stars more metal-poor than $\feh < -2.0$? In the high-resolution spectroscopic CEMP compilation of \citet{yoon16}, at $\feh = -2.5$, roughly $85\%$ of the CEMP stars are CEMP-s, going down to $50\%$ for $\feh = -3.0$. In this metallicity range, stars with higher absolute carbon abundances are more likely to be CEMP-s. While this compilation might not be entirely unbiased with regards to different CEMP types (e.g. it might contain an overabundance of \textit{very} C-rich stars due to selection choices in early high-resolution spectroscopy follow-up programs), it suggests that even between $-3.0 < \feh < -2.0$, most of our candidates will be of the binary mass-transfer type. Around $\feh \sim -2.5$ and below, some fraction of CEMP-no stars is expected. 

Binary indicators from \Gaia also show that many of our C-rich candidates are likely to be binaries. We select a nearby high-quality subset (parallax uncertainties $<20\%$, 1/parallax $<1.5$~kpc) from the \citetalias{andrae23} sample, where RUWE can be used as indicator of binarity due to the astrometric wobble (see e.g. \citealt{belokurov20}). We find that for stars with predicted $-2.5 < \feh < -1.5$, the fraction of stars with RUWE $>1.4$ increases from $\sim 10\%$ for C-normal stars to $\sim 40\%$ for $\cfe = +1$ and $\sim 80\%$ for $\cfe = +2$. There are only two C-rich candidates for $\feh < -2.5$, of which one has a high RUWE. Many of the stars with high RUWE also have large radial velocity amplitudes, supporting their binary nature. More detailed investigations of the binary properties are left for a future work. A final note: the binary companions of C-rich stars are currently faint, unseen white dwarfs, which do not contribute directly to the apparent brightness and/or the spectra of the system.

\subsection{What is frequency of C-rich stars and how does it change with metallicity?}\label{sec:freq}

\begin{figure*}
\centering
\includegraphics[width=0.49\hsize,trim={0.0cm 0.0cm 0.0cm 1.0cm}]{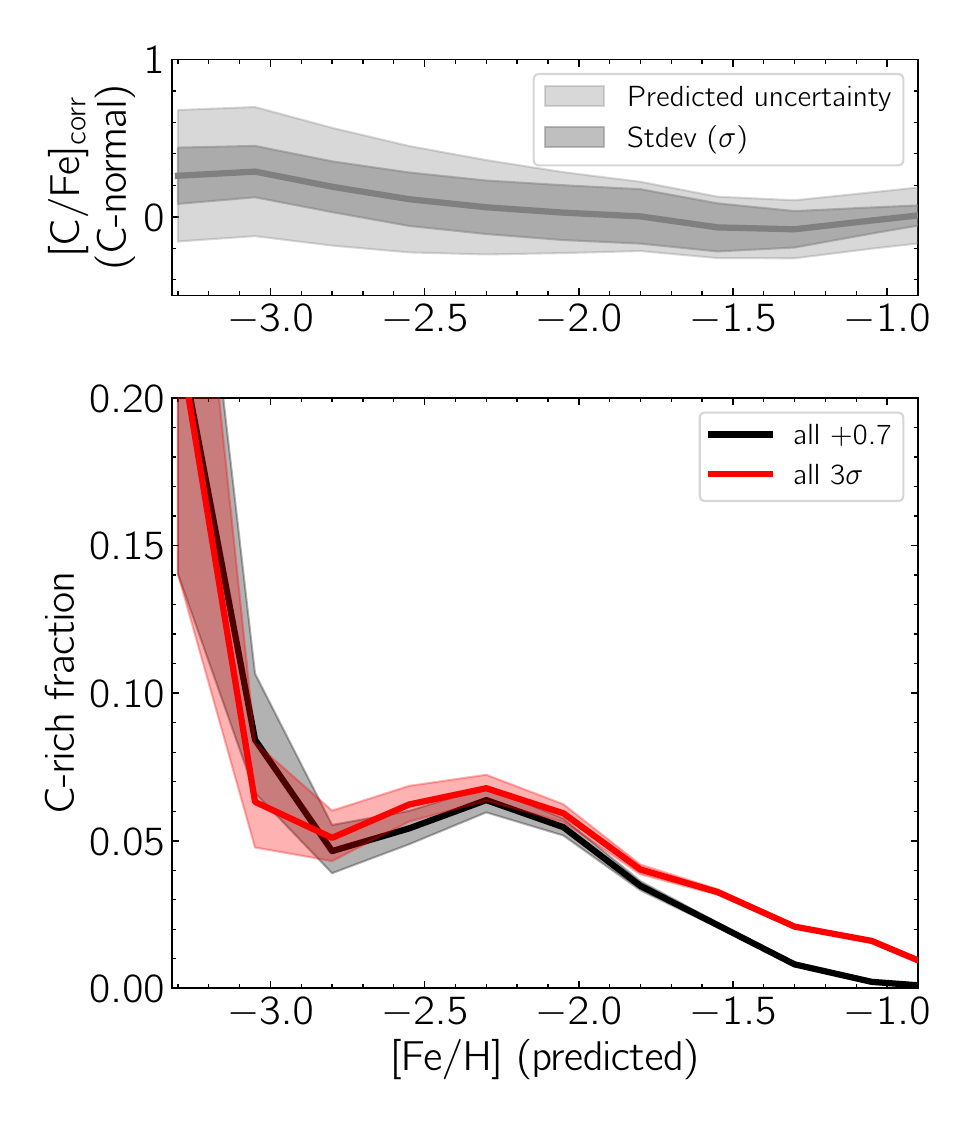}
\includegraphics[width=0.49\hsize,trim={0.0cm 0.0cm 0.0cm 1.0cm}]{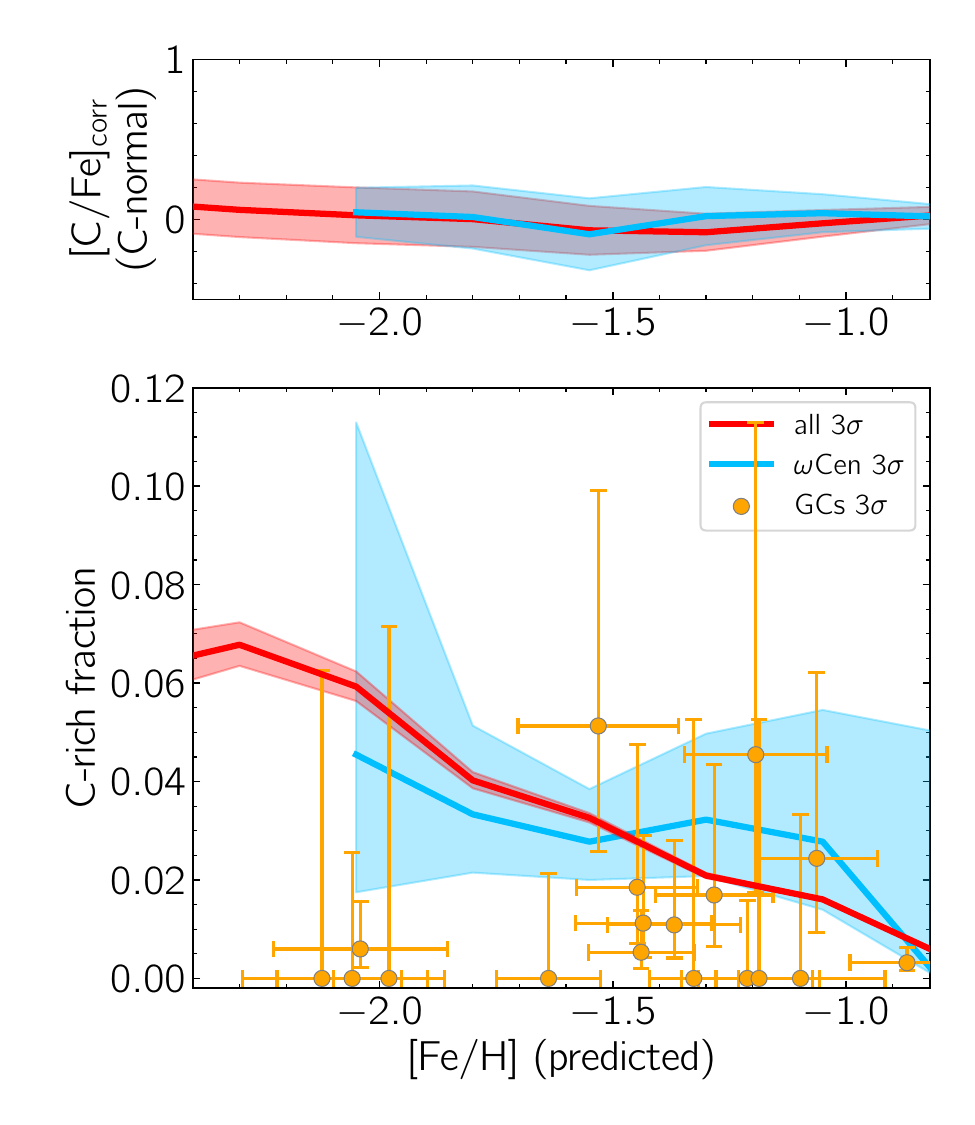}
\caption{Fraction of C-rich objects in the \citetalias{andrae23} sample as a function of metallicity. The \textbf{top left panel} shows the median predicted $\cfe_\mathrm{corr}$ for carbon-normal stars, with the predicted uncertainties and the standard deviation indicated in shaded bands. The \textbf{lower left panel} shows the fraction of stars in the same sample with $\cfe_\mathrm{corr} > +0.7$ (black) or $\cfe_\mathrm{corr} > \mu + 3\sigma$ in each metallicity bin (red), where the shaded bands indicate the binomial proportion confidence interval. The \textbf{right-hand panels} are similar, repeating the lines for the $3\sigma$ sample (red) and adding in results for globular clusters with 10 or more stars (orange points) and $\omega$ Centauri (light blue line). For both we use their respective $\cfe_\mathrm{corr} > \mu + 3\sigma$ C-rich definitions. Bin sizes in all panels are 0.25~dex. The horizontal error bars for the GCs represents the standard deviation of our metallicity predictions for each cluster. 
}
\label{fig:cfe_a23_frac} 
\end{figure*}

\subsubsection{For field stars}

Next, we investigate the \emph{frequency} of C-rich objects, using our predictions for the \citetalias{andrae23} giants sample. We again use a sample of stars with E(B$-$V) $<0.3$, bp\_chi\_squared $< 10.5 \, \times$ bp\_degrees\_of\_freedom, Pvar $<0.5$ and (BP$-$RP)$_0 < 1.35$. First we derive the median $\cfe_\mathrm{corr}$ in bins of metallicity for carbon-normal stars with $\cfe < 0.7$, as well as their predicted uncertainties and standard deviation of the distribution -- this is shown in the top left panel of Figure~\ref{fig:cfe_a23_frac}. A cut of $\cfe_\mathrm{corr} > 0.7$ selects stars that are typically $2-3 \sigma$ outliers from the distribution \citep[see also][for a discussion on a relative selection of CEMP stars]{sestito24_sgrcarbon}, although at lower metallicity there may be more outliers due to larger measurement uncertainties.

\begin{table}
\centering
\caption{\label{tab:cfrac}C-rich fraction as function of metallicity for the classical definition ($\cfe > +0.7$) and the $3\sigma$ definition, in bins of 0.25~dex (see Figure~\ref{fig:cfe_a23_frac}). Confidence intervals are given in square brackets. Note that we provide differential fractions, as opposed to cumulative fractions in much of the literature.}
\begin{tabular}{cccc}
\hline
   [Fe/H] & classical  & $3\sigma$ [C/Fe] &  $3\sigma$ \\
   (bin centre)    & definition & level & definition \\
\hline
$-3.30$ & 0.22 &   0.797 &   0.22 \\
        & [0.14, 0.33] &   &   [0.14, 0.33] \\
$-3.05$ & 0.084 &   0.777 &   0.063  \\
        &    [0.066, 0.107] & & [0.048, 0.083] \\
$-2.80$ & 0.046  &   0.674 &   0.051  \\
        & [0.039, 0.055]      &       &  [0.043, 0.060]     \\
$-2.55$ & 0.054  &   0.621 &   0.062 \\
        & [0.049, 0.060]      &       &   [0.057, 0.069]     \\
$-2.30$ & 0.064  &   0.570 &   0.068  \\
        & [0.060, 0.068]      &       &  [0.063, 0.072]     \\
$-2.05$ & 0.055  &   0.549 &   0.059 \\
        &  [0.052, 0.058]     &       &  [0.056, 0.062]      \\
$-1.80$ & 0.035 &   0.521 &   0.040  \\
        &  [0.033, 0.036]      &       &  [0.039, 0.042]     \\
$-1.55$ & 0.0214  &   0.391 &   0.0326 \\
        & [0.0207, 0.0222]      &       &   [0.0317, 0.0336]     \\
$-1.30$ & 0.0081 &   0.268 &   0.0209  \\
        & [0.0078, 0.0084]       &       &  [0.0204, 0.0214]     \\
$-1.05$ & 0.0022  &   0.225 &   0.0161  \\
        & [0.0021, 0.0023]      &       &  [0.0158, 0.0163]     \\
\hline
\end{tabular}
\end{table}

The fraction of C-rich objects per metallicity bin is shown in black in the lower left panel of Figure~\ref{fig:cfe_a23_frac}, using a definition of $\cfe_\mathrm{corr} > 0.7$. In red, we also show the fraction of C-rich stars based on them being $>3\sigma$ away from the median $\cfe_\mathrm{corr}$ in each metallicity bin. The results are summarised in Table~\ref{tab:cfrac}. These two definitions work very similarly, although the outlier definition gives a larger fraction for higher metallicities. 
A C-rich frequency of $6-7\%$ for giant VMP stars is much lower than previously reported frequencies based on spectroscopic samples -- \citet{arentsen22} compiled and discussed these, finding they are of the order of $10-40\%$ for $-2.5 < \feh < -2.0$ \citep[e.g.][]{frebel06,Lee13, placco14,yoon18}. It is important to note that fractions reported in the literature are typically \emph{cumulative}, whereas we are presenting \emph{differential} fractions. Cumulative fractions can be biased by the metallicity distribution function of a sample (e.g. the relative number of stars at $ \feh = -3.0$ vs. $-2.0$), and this bias is removed when taking the fractions in metallicity bins. Literature fractions are also given for giants in general, whereas our A23 sample is limited to the upper RGB. There could potentially be different fractions of C-rich stars along the RGB (in reality or due to analysis biases). These two points are discussed in further detail in \citet{arentsen22} and are worth exploring more in the future. It is also possible that our derived frequency is a lower limit, given the precision of the XP spectra for C-normal VMP stars and/or the limited precision of our evolutionary correction estimates, although spectroscopic samples have limitations as well. 

The trends we see are roughly in agreement with previous trends seen in spectroscopic samples of a rising carbon abundance with decreasing metallicity for C-normal stars, as well as an increase in the C-rich fraction \citep{Lee13,placco14,arentsen22}. These two might be connected to some degree, with more of the ``normal'' distribution having $\cfe > 0.7$ at lower metallicity (this is why we also try adopting an outlier definition of C-rich stars), but the number of very C-rich stars that are clear outliers appears to be rising too. \citetalias{lucey23} also identified a rising frequency of their C-rich candidates with metallicity, adopting the \citetalias{andrae23} metallicities (although they are biased for C-rich stars, see Appendix~\ref{sec:photcomp}). 

The rising trend for our sample does \emph{not} continue for $\feh < -2.0$, but stays roughly constant down to $\feh = -3$, with a small decrease around $\feh = -2.8$. This might partly be related to our metallicity estimates being somewhat biased (higher than they should be) for $\feh < -2.5$ (see Figure~\ref{fig:testpred} and the discussion at the end of Section~\ref{sec:ferre1}), which would reduce the CEMP frequency for very/extremely metal-poor stars and slightly increase it between $-2.5 < \feh < -2.0$. Our test sample also shows a couple of CEMP candidates with $\cfe \sim 0.7-1.0$ that are assigned ``normal'' carbon abundances, suggesting we might miss some of these slightly C-rich stars, but these two specific ones are warmer (around 5500~K) than our RGB sample used here. Our results could also hint at a real C-rich frequency plateau in the VMP range. In this regime, the CEMP-no stars are not yet dominant, so the main driver of the trend is likely to be the CEMP-s stars and the properties of their orbits and/or AGB companions as function of metallicity, potentially stabilising for VMP stars. A detailed investigation of this is left for a future work.

\subsubsection{For globular cluster stars}

We show the fraction of C-rich objects for stars associated with globular clusters (GCs) in orange in the right-hand panels of Figure~\ref{fig:cfe_a23_frac}. We cross-match our sample with the GC members from \citet{vasiliev21}, keeping stars with member probabilities $>0.5$ that are not in Omega Centauri because of its complexity (see below), resulting in $\sim 1800$ stars. Of these stars, only two have $\cfe > 0.7$. For the figure, we only include clusters with 10 or more stars. The mean carbon abundance for the cluster stars can be slightly different compared to the rest of the \citetalias{andrae23} sample, so we recalculate the $3\sigma$ definition of carbon-enhancement rather than using the fixed cut at $\cfe = +0.7$ or using the cut for field stars. We find that at all metallicities, the C-rich fraction for GC stars is typically lower than in the full field sample (compared to the red line, to be consistent with the $3\sigma$ definition), although the uncertainties are large and there are a few exceptions. 

GCs have been found to have relatively low binary fractions, especially in their outer regions \citep[][using the photometric binary sequence for main sequence stars]{milone12}. Barium-rich objects across different metallicities, all associated with mass-transfer from a former AGB companion (through Roche-lobe overflow, wind-assisted Roche-lob overflow and/or wind transfer), have also been found to be rare in GCs \citep{dorazi10}. \citet{kirby15} also did not find any C-rich stars among $\sim 150$ stars in three VMP GCs, but they note that their selection may have been biased. Since we expect most of the C-rich stars in this metallicity range to be associated with mass-transfer, we can confirm the result of a lower fraction of AGB binary-interaction stars in GCs with our large, homogeneous and hopefully unbiased XP sample of carbon abundances. A possible explanation for this is that most of the binaries that typically would have formed Ba/C-rich stars are disrupted in the high concentration GC environments, before intermediate low-mass stars can evolve into AGB stars \citep{gratton04} -- hence reducing the frequency of Ba/C-rich stars. Known Ba/CH/CEMP-s stars have periods of several hundred to several thousand days \citep[e.g.][]{hansen16a, jorissen19}, so these are typically not very close binaries, and could be destroyed relatively easily. We note that there are still some open questions as to why they have the periods that we observe \citep[see e.g.][]{abate18}. Other binaries might also be hardened by the cluster dynamics, with orbits becoming tighter and potentially into the regime where the system goes through a common envelope phase and/or interacts before the companion reaches the AGB, which would not produce the typical AGB mass-transfer Ba/CH/CEMP-s signature. 

All Milky Way globular clusters display two populations of stars separated by their light element abundances, with the first having abundances similar to field stars and the second showing peculiar abundances \citep[see e.g.][for a review on the subject]{gratton04}. \citet{dorazi10} find that 4 out of 5 of their Ba stars are in first population GC stars, suggesting that potentially the low binary fraction is mostly among second generation GC stars, which are thought to have formed in a higher density environment.  \citet{lucatello15} confirmed this, finding a four times lower binary frequency among second generation GC stars compared to first generation GC stars, from homogeneous radial velocity monitoring of cluster stars at roughly the half-light radius of several GCs. With our current sample, it is challenging to say anything about differences in C-rich frequency between first and second generation stars. Additionally, \citet{milone12} found that higher mass clusters appear to have lower binary fractions. We checked whether there is any trend between C-rich fraction and $M_V$ of the GCs, but we find no clear correlation -- larger samples per cluster may be necessary to investigate this further.

Next, we look at $\omega$ Centauri, not a normal GC but likely a nuclear star cluster that is the remnant core of an accreted dwarf galaxy \citep[e.g.][]{bekki03}. There are around 700 stars associated with $\omega$ Centauri in our catalogue of predictions after quality cuts. Among them, we find 13 stars with $\cfe > 0.7$ -- significantly more than in all the other globular clusters summed together. Within the uncertainties, the fraction of C-rich stars with metallicity according to the $>3\sigma$ definition is similar to that in the main sample, and even slightly higher (within $\sim 2 \sigma$) than that of the main sample for $\feh \sim -1.1$, see the light blue line in the lower right panel of Figure~\ref{fig:cfe_a23_frac}. This suggest that the conditions in $\omega$ Centauri were permissible to the right kind of binaries surviving --  indeed $\omega$ Centauri has a relatively low density and long relaxation time \citep{baumgardt18}, suggesting it may have had a lower number of disrupting encounters than in denser systems. 

One might worry about contamination among the XP spectra of GC stars, since they are densely populated. We checked the \texttt{phot\_bp\_rp\_excess\_flux} for the GC sample, for which high values would indicate possible blends/contamination in the photometry, but after our already applied quality cuts there are none with bad values. Most of our stars are not in the central dense regions of the clusters. Additionally, \citet{mehta24} recently successfully used the XP spectra of GC stars to separate the multiple stellar populations inside the GCs, showing that they have good quality.

\subsection{What are the spatial and dynamical properties of the C-rich candidates?}

\begin{figure}
\centering
\includegraphics[width=1.0\hsize,trim={0.5cm 0.0cm 0.0cm 0.0cm}]{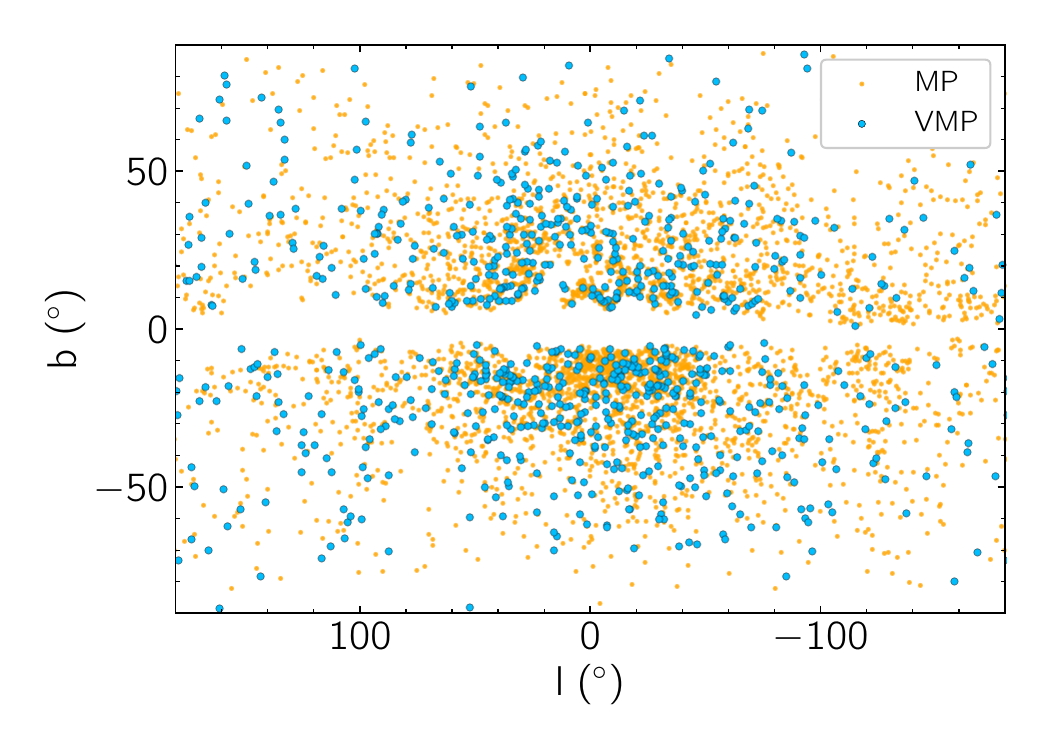}
\caption{Galactic longitude and latitude distribution of metal-poor, C-rich candidates (with $\cfe_\mathrm{corr} > +0.5$) in our cleaned \citetalias{andrae23} sample. MP stars have $\feh < -1$ (small yellow dots) and VMP stars have $\feh < -2$ (larger blue dots). 
}
\label{fig:sky} 
\end{figure}

\subsubsection{Sky distribution}

We present the longitude and latitude distribution of our metal-poor (MP), C-rich candidates with $\feh < -1.0$ and $\cfe > 0.5$ in Figure~\ref{fig:sky}, highlighting those with $\feh < -2.0$ in blue (VMP). The following quality cuts have been applied to our \citetalias{andrae23} sample: E(B$-$V) $< 0.5$, Pvar $<0.5$ and (BP$-$RP)$_0 < 1.35$. The density of both MP and VMP C-rich stars is highest towards the central region of the Milky Way, where the density of metal-poor stars is also the highest \citep[e.g.][]{rix22}. The MP candidates are more concentrated than the VMP stars, and also have a slight disc contribution (particularly visible beyond $l=-100^{\circ}$). The VMP candidates have quite an extended distribution. These things are consistent with the C-normal distributions of MP and VMP stars. Next, we investigate the dynamical properties of our sample. 

\begin{figure}
\centering
\includegraphics[width=0.8\hsize,trim={0.0cm 0.0cm 0.0cm 0.5cm}]{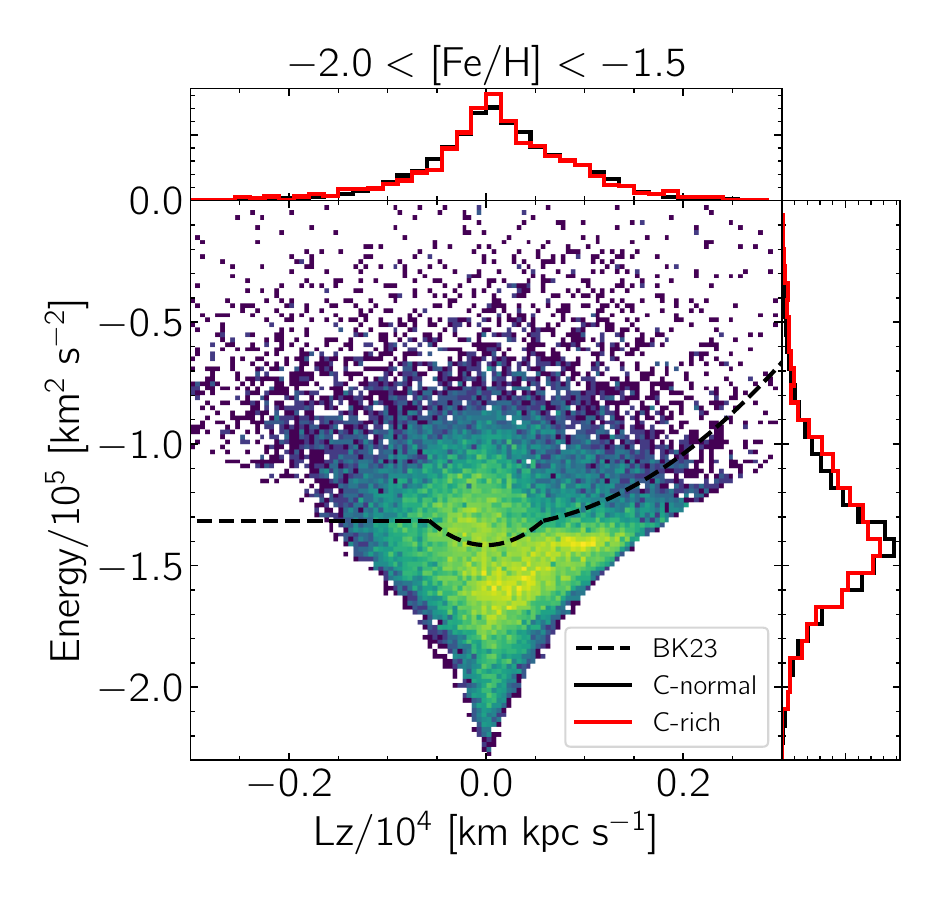}
\includegraphics[width=0.8\hsize,trim={0.0cm 0.0cm 0.0cm 0.0cm}]{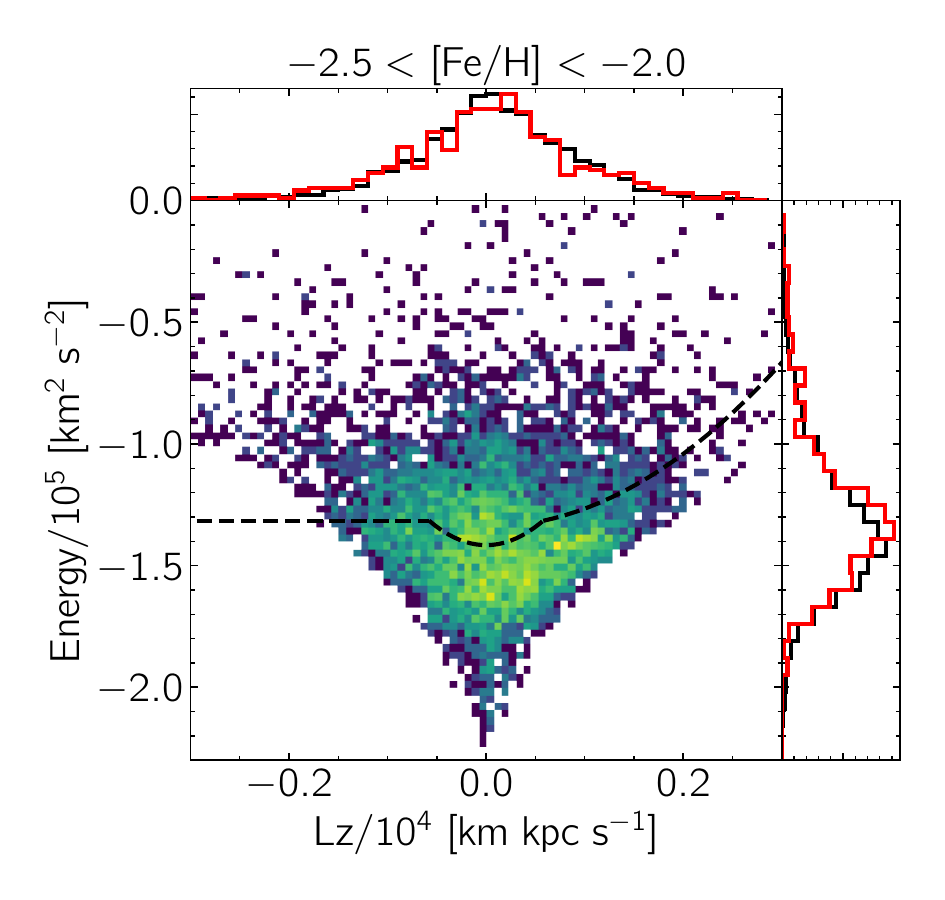}
\caption{Energy and angular momentum distributions for stars in two metallicity bins (top and bottom less and more metal-poor, respectively, see titles), with 1D histograms indicating the distributions of C-normal (black, $\cfe_\mathrm{corr} < +0.5$) and C-rich (red, $\cfe_\mathrm{corr} > +0.7$) stars. The dashed lines indicate the accreted/in-situ division proposed by \citet{belokurov23_Nrich}, shifted for the adopted potential as in \citet{kane24}. }
\label{fig:ELz} 
\end{figure}

\begin{figure}
\centering
\includegraphics[width=1.0\hsize,trim={0.0cm 0.8cm 0.0cm 0.0cm}]{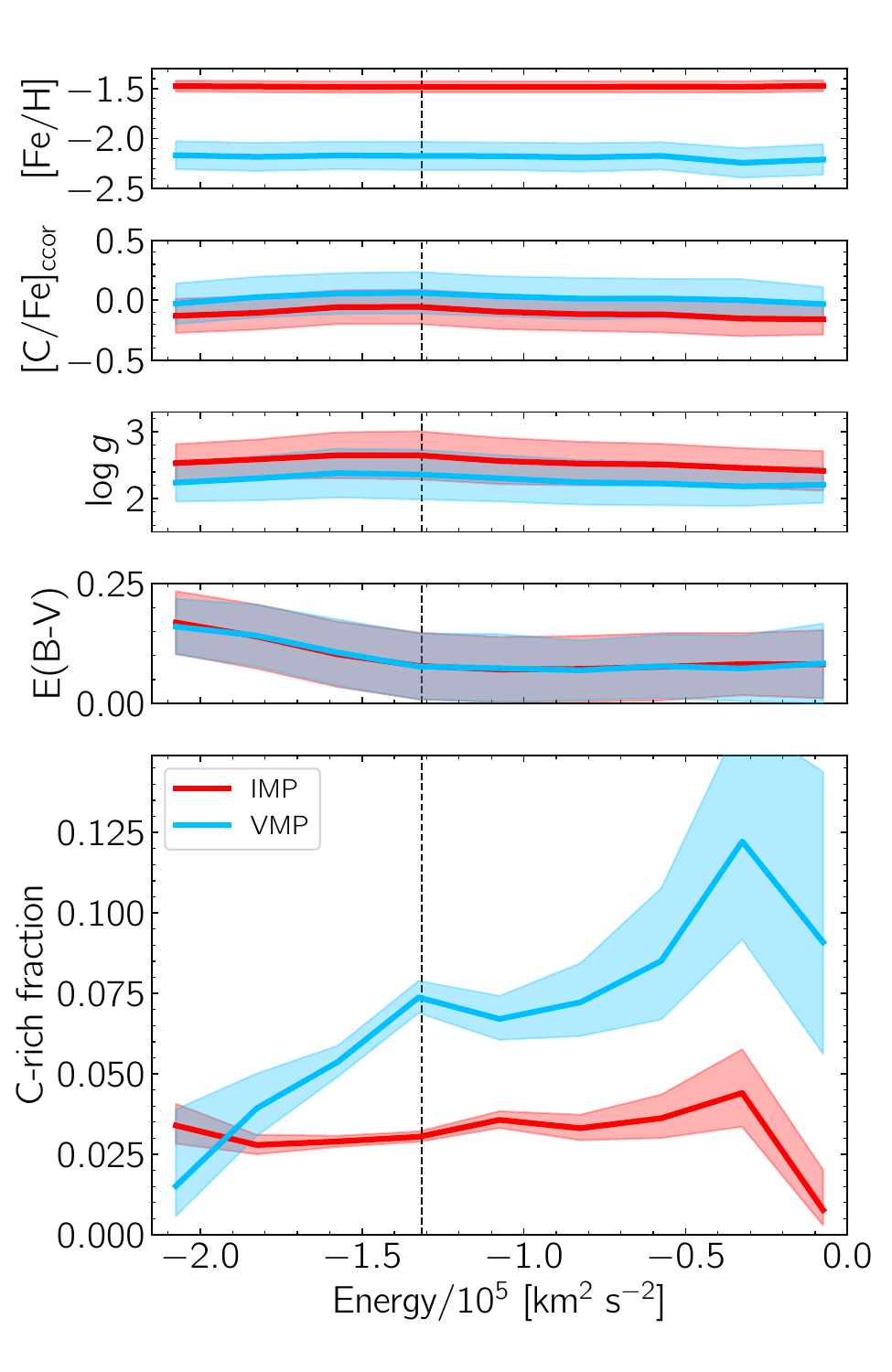}
\caption{For an intermediate metal-poor sample with $-1.6 < \feh < -1.4$ (red) and a VMP sample with $-2.5 < \feh < -2.0$ (blue), the median \feh, $\cfe_\mathrm{ccor}$, \logg and E(B$-$V) for C-normal stars as a function of energy is shown in the top four panels, and the fraction of C-rich stars ($3\sigma$ outliers) in the bottom panel. The dashed vertical line in the bottom panel is the rough separation between in-situ and accreted according to \citet{belokurov23_Nrich} (see Figure~\ref{fig:ELz}). 
}
\label{fig:energyfrac} 
\end{figure}

\subsubsection{Energy/angular momentum distributions}

We derive energy and angular momentum for all the stars in the \citetalias{andrae23} sample, using distance = 1/parallax and the \Gaia radial velocities and proper motions and the \texttt{gala} package with the \texttt{MilkyWayPotential} \citep{gala, pricewhelan20_gala, galpy}. For this exercise, we remove globular cluster stars from the sample. We present the E/$L_z$ distributions for $-2.0 < \feh < -1.5$ and $-2.5 < \feh < -2.0$ in Figure~\ref{fig:ELz}, with 1D distributions of the two parameters in top and side panels for C-normal (black) and C-rich (red) stars. 

We quantify the difference between C-normal and C-rich stars by applying Kolmogorov-Smirnov (KS) tests to the 1D $L_z$ and energy distributions. For both metallicity regimes, there is no statistical difference in the $L_z$ distributions for C-rich and C-normal stars, but there is a difference (p-value = $4 \times 10^{-4}$ for IMP and $3 \times 10^{-3}$ for VMP) in the energy distributions, with C-rich stars having slightly higher energies on average. This could point to a difference in C-rich fraction with energy, although for the IMP bin it could also be related to a slight metallicity gradient with energy within this bin and an increased frequency of C-rich stars with lower metallicity (see Section~\ref{sec:freq}). 

Since most of our C-rich candidates are likely in binary systems, one may wonder whether their radial velocity variations and/or high RUWE could affect their E$/L_z$ distribution. We suggest that this is not likely to have a significant impact on our analysis. The radial velocity reported by \Gaia is an average of the multiple observations, which should be reasonably close to the systemic velocity, and high RUWE only occurs for the most nearby stars, which is a small subset of our \citetalias{andrae23} sample. Additionally, we only find a difference in the energy distribution and not the angular momentum -- if the uncertainties are significantly increased, one might expect a broadening of both distributions. Further investigation of this might be interesting, but is beyond the scope of this work.

\subsubsection{C-rich frequency with orbital energy}

Next, we investigate the fraction of C-rich stars as a function of energy to put further constraints on the difference hinted at in the previous section. The energy is closely connected to the apocentre of a star, so this investigation probes the radial dependence of the C-rich fraction in the Galactic halo. Using energy rather than the distance allows covering a larger range, less biased by selection effects. 

To isolate the effect of the orbital properties from the metallicity dependence, we select two samples: one intermediate metal-poor with $-1.6 < \feh = -1.4$ and one very metal-poor with $-2.5 < \feh < -2.0$. We choose a smaller bin for the intermediate metallicity sample compared to the previous section because of the steep increase in C-rich fraction between $-2.0 < \feh < -1.5$ (Figure~\ref{fig:cfe_a23_frac}), and we hope to separate the trends between \feh, energy and C-rich fraction. The average metallicity, (corrected) carbon abundance, \logg and E(B$-$V) with energy for C-normal stars in these samples is presented in the top four panels of Figure~\ref{fig:energyfrac}, showing that they have very little gradient and are unlikely to cause any spurious trends for the C-rich fraction with energy. The exception is potentially E(B$-$V), which rises left-wards of the in-situ/accreted boundary. It is expected that the extinction increases for more bound stars, as they have to be concentrated towards the inner Galaxy. The bottom panel of that figure presents the fraction of C-rich stars ($3\sigma$ outliers) with energy, where especially the VMP sample shows a strong increase with increasing energy. The change is less obvious for $\feh = -1.5$, although a slight trend is potentially still present. When separating stars with $\feh = -1.5$ in high ($< -1.5 \cdot 10^5$ km$^2$ s$^{-2}$) and low ($> -1.3 \cdot 10^5$ km$^2$ s$^{-2}$) energy groups, there is a $\sim2\sigma$ difference in their C-rich fractions. 

While extinction might potentially affect our identification of C-rich stars, its effect is unlikely to explain away the trend with energy that we observe for the VMP stars. There is still a trend when only considering stars with $\mathrm{energy} > -1.5 \cdot 10^5$~km$^2$ s$^{-2}$, and in this regime there is little change of E(B$-$V).

Similar trends for the C-rich fraction with distance and/or Galactocentric radius, namely more C-rich stars further out, have previously been found with spectroscopic samples \citep[e.g.][]{frebel06,yoon18}. They were interpreted as the result of different types of building blocks, larger and smaller dwarf galaxies, dominating in different halo regions -- surviving ultrafaint dwarf galaxies appear to have a higher frequency of CEMP-no stars than dwarf spheroidal galaxies \citep[e.g.][]{salvadori15,lucchesi24}.

However, we expect most of the C-rich stars in these metallicity ranges to be of the binary-transfer type (CH/CEMP-s). Our result therefore hints at differences in the binary populations with energy. Potentially this could be connected to the contribution of stars from disrupted globular clusters, since we showed above that the frequency of C-rich stars is lower in globular clusters. It has been suggested that there are relatively more stars from disrupted globular cluster in the inner regions of the Milky Way compared to further out into the halo \citep{schiavon17,horta21, belokurov23_Nrich, kane24}. \citet{belokurov23_Nrich} suggest this is related to vigorous star formation in clusters in the early Milky Way, in what is now the \emph{in-situ} halo. 
The previous works focused on stars with $\feh > -1.5$ because they relied on chemistry from the APOGEE data. Our results are consistent with the literature interpretation, and for the first time using a large unbiased dataset, we can now cautiously extend the suggestion of an increased globular cluster star contribution in the inner Galaxy to stars with $\feh < -2.0$. \citet{arentsen21} suggested a similar interpretation for the lower CEMP fraction among VMP stars in the PIGS inner Galaxy dataset, although they were limited by potential selection effects against very C-rich stars. \citet{kim23} also speculated about a possible connection between a higher contribution of disrupted globular clusters and a lower fraction of CEMP stars in a population.

Others have suggested that there could be a connection between the frequency of C-rich binary products and the initial mass function (IMF) of a stellar population \citep{lucatello05_imf, tumlinson07}. If there are more massive stars in a population, the fraction of Ba/CH/CEMP stars should increase due to the larger number of available intermediate mass AGB stars that can donate their carbon and s-process elements to a companion. If we take our results at face value, they would correspond to an IMF with fewer massive stars in the inner Galaxy compared to the outer regions. \citet{tumlinson07} predicted the opposite -- given that we expect the inner Galaxy to have the oldest stars at a given metallicity, they would have formed at a time when the cosmic microwave background radiation was hotter, which would have raised the characteristic mass of a stellar population IMF, increasing the C-rich fraction. This idea is not supported by our results.

\subsection{Note on selection biases in the A23 giants sample}

Finally, we note that it is likely that the \citetalias{andrae23}+\Gaia RV sample is not unbiased with respect to C-rich and/or VMP stars. The \citetalias{andrae23} giants sample is based on cuts in their predicted \teff ($< 5200$~K) and \logg ($<3.5$), as well as some photometric cuts based on \Gaia and WISE photometry -- the latter are aimed to remove stars on the hotter end of the RGB that might be contamination from young hot stars among the metal-poor giants. These cuts may also remove some true very/extremely metal-poor stars, which might be slightly different for C-normal and C-rich stars. 

There may also be other subtle effects due to cuts in magnitude and/or the use of absolute magnitudes and its effect on \logg, because C-rich stars are slightly fainter than their C-normal counterparts. Furthermore, for cool, very C-rich objects, there will be molecular features in the calcium triplet region, and some quality cuts applied to the \Gaia RVs may remove such objects. 

We leave the detailed investigation of selection effects and/or estimates for less biased samples to future work. This work is showing that this analysis works as a proof of concept, and it is sufficient to be aware that there are some biases -- in particular the frequency of C-rich objects as a function of metallicity (and/or distance) may be affected.

\section{Conclusions}

In this work, we parameterised carbon-rich (C-rich) stars from the \Gaia low-resolution XP spectra using a heteroscedastic regression neural network, labelling them with \teff, \logg, \feh, \cfe and their associated uncertainties. Our training sample consists of re-analysed LAMOST spectra for C-normal and C-rich stars, supplemented with high-resolution spectroscopic samples of CEMP stars \citep{yoon16} and V/EMP stars \citep{li22}. We applied our network to two \Gaia XP samples: the largest sample of CEMP candidates from \citet[][L23]{lucey23} and the \citet[][A23]{andrae23} vetted RGB sample with \Gaia radial velocities to study population properties. We summarise our results: 

\begin{enumerate}
    \item The predictions for the test sample (Figure~\ref{fig:testpred}) show that we are able to predict \feh down to $-3.0$ and \cfe from sub-solar to $\cfe \gtrsim +2.0$. We tend to predict slightly lower \cfe and/or higher \feh for very metal-poor, very C-rich stars, but CEMP candidates can still be identified confidently. Our predictions are best for bright stars ($G<16$). Comparisons with high-resolution spectroscopy from the RPA, APOGEE and GALAH datasets show good overall agreement, and although APOGEE and GALAH struggle with very C-rich stars, our C-rich candidates are also relatively C-rich in these surveys (Figure~\ref{fig:apogalahcomp_a23}). As has been found previously, C-rich stars are redder than their C-normal counterparts at the same metallicity (Figure~\ref{fig:cmd_crich}). 
    
    \item We found that $\sim 40\%$ of bright \citetalias{lucey23} C-rich candidates have predicted $\feh < -1.5$, of which $90\%$ has $\cfe > +0.7$. We discovered that the \citetalias{lucey23} sample is contaminated by stars with Ca H\&K in emission ($\sim 20\%$ of the sample), which can easily be removed with our predictions and a variability cut. There are $\sim 2000$ bright CEMP candidates with predicted $\feh < -2.0$ and $\cfe > +0.7$ in our sample after quality cuts, which deserve further study. 

    \item Using our predictions for the vetted \citetalias{andrae23} RGB sample, we confirmed previous findings regarding an increase of C-rich stars with decreasing metallicity (Figure~\ref{fig:cfe_a23_frac} and Table~\ref{tab:cfrac}), potentially flattening for $-3.0 < \feh < -2.0$ to a plateau at $6-7\%$. The reality of this plateau needs to be tested in the future, as well as our much lower frequencies compared to spectroscopic samples ($10-40\%$). We found that for $\feh > -2.0$, a more appropriate definition of C-rich stars is whether they are $3\sigma$ outliers from the carbon distribution rather than selecting $\cfe > +0.7$. For $\feh < -2.0$ the two definitions are very similar (within 1$\sigma$). 
    
    \item Most of the C-rich objects we have identified are expected to be of the AGB binary mass-transfer type, with current (unseen) white dwarf companions. The bulk of candidates lies in the Group I region from \citet{yoon16} and indeed we found a large fraction of barium-enhanced stars among our C-rich candidates in a cross-match with GALAH (Figure~\ref{fig:galah_a23}), as well as a large fraction of high \Gaia RUWE stars in a nearby sub-sample -- an indication of binarity. Some of the more metal-poor C-rich candidates may instead carry the imprint of the metal-free First Stars from their birth.
        
    \item In our sample, the fraction of C-rich candidates ($3\sigma$ outliers) in globular clusters appears to be lower than in the field, except for the likely nuclear star cluster $\omega$ Centauri (Figure~\ref{fig:cfe_a23_frac}). The fraction in GCs is potentially lower because binaries tend to get destroyed in GC environments, specifically those with orbital properties that might allow AGB mass-transfer. 

    \item We found that the fraction of C-rich candidates increases with decreasing orbital energy (Figure~\ref{fig:energyfrac}), implying that fewer AGB binary mass-transfers have happened among stellar populations in the inner Galaxy compared to further out. We suggest that this might be connected to previous studies finding increased contributions to inner Galaxy field stars from disrupted GCs or GC-like systems, which had lower binary fractions. 
 
\end{enumerate}

We have shown that it is feasible to parameterise metal-poor C-rich stars using the \Gaia XP spectra. This bright giants sample is a perfect input catalogue for spectroscopic follow-up observations, for example for WEAVE \citep{weave} or dedicated efforts. In \Gaia DR4, the XP data will go even deeper to fainter magnitudes, allowing for example to probe deeper into the Galactic halo and to target distinct dynamical substructures and their C-rich populations.

\section*{Acknowledgements}

We thank Elisabeth Sola, Else Starkenburg, Zhen Yuan, Terese Hansen and David Aguado for useful discussions or for comments on a draft of this work. We thank Terese Hansen for sharing a table with uncorrected carbon abundances for the RPA data releases. We thank Vinicius Placco for computing evolutionary carbon corrections for our sample. We thank the reviewer for useful comments.

AAA acknowledges support from the Herchel Smith Fellowship at the University of Cambridge and a Fitzwilliam College research fellowship supported by the Isaac Newton Trust. SGK acknowledges PhD support from the Marshall Scholarship. JLS acknowledges support from the Royal Society (URF\textbackslash R1\textbackslash191555). 

This work has made use of data from the European Space Agency (ESA) mission {\it Gaia} (\url{https://www.cosmos.esa.int/gaia}), processed by the {\it Gaia} Data Processing and Analysis Consortium (DPAC, \url{https://www.cosmos.esa.int/web/gaia/dpac/consortium}). Funding for the DPAC has been provided by national institutions, in particular the institutions participating in the {\it Gaia} Multilateral Agreement.

Guoshoujing Telescope (the Large Sky Area Multi-Object Fiber Spectroscopic Telescope LAMOST) is a National Major Scientific Project built by the Chinese Academy of Sciences. Funding for the project has been provided by the National Development and Reform Commission. LAMOST is operated and managed by the National Astronomical Observatories, Chinese Academy of Sciences.

Funding for the Sloan Digital Sky Survey V has been provided by the Alfred P. Sloan Foundation, the Heising-Simons Foundation, the National Science Foundation, and the Participating Institutions. SDSS acknowledges support and resources from the Center for High-Performance Computing at the University of Utah. SDSS telescopes are located at Apache Point Observatory, funded by the Astrophysical Research Consortium and operated by New Mexico State University, and at Las Campanas Observatory, operated by the Carnegie Institution for Science. The SDSS web site is \url{www.sdss.org}.

SDSS is managed by the Astrophysical Research Consortium for the Participating Institutions of the SDSS Collaboration, including Caltech, The Carnegie Institution for Science, Chilean National Time Allocation Committee (CNTAC) ratified researchers, The Flatiron Institute, the Gotham Participation Group, Harvard University, Heidelberg University, The Johns Hopkins University, L’Ecole polytechnique f\'{e}d\'{e}rale de Lausanne (EPFL), Leibniz-Institut f\"{u}r Astrophysik Potsdam (AIP), Max-Planck-Institut f\"{u}r Astronomie (MPIA Heidelberg), Max-Planck-Institut f\"{u}r Extraterrestrische Physik (MPE), Nanjing University, National Astronomical Observatories of China (NAOC), New Mexico State University, The Ohio State University, Pennsylvania State University, Smithsonian Astrophysical Observatory, Space Telescope Science Institute (STScI), the Stellar Astrophysics Participation Group, Universidad Nacional Aut\'{o}noma de M\'{e}xico, University of Arizona, University of Colorado Boulder, University of Illinois at Urbana-Champaign, University of Toronto, University of Utah, University of Virginia, Yale University, and Yunnan University.

This work made use of the Fourth Data Release of the GALAH Survey \citep{galahdr4}. The GALAH Survey is based on data acquired through the Australian Astronomical Observatory, under programs: A/2013B/13 (The GALAH pilot survey); A/2014A/25, A/2015A/19, A2017A/18 (The GALAH survey phase 1); A2018A/18 (Open clusters with HERMES); A2019A/1 (Hierarchical star formation in Ori OB1); A2019A/15, A/2020B/23, R/2022B/5, R/2023A/4, R2023B/5 (The GALAH survey phase 2); A/2015B/19, A/2016A/22, A/2016B/10, A/2017B/16, A/2018B/15 (The HERMES-TESS program); A/2015A/3, A/2015B/1, A/2015B/19, A/2016A/22, A/2016B/12, A/2017A/14, A/2020B/14 (The HERMES K2-follow-up program); R/2022B/02 and A/2023A/09 (Combining asteroseismology and spectroscopy in K2); A/2023A/8 (Resolving the chemical fingerprints of Milky Way mergers); and A/2023B/4 (s-process variations in southern globular clusters). We acknowledge the traditional owners of the land on which the AAT stands, the Gamilaraay people, and pay our respects to elders past and present. This paper includes data that has been provided by AAO Data Central (datacentral.org.au).

This research has made use of the \texttt{numpy} \citep{numpy}, \texttt{matplotlib} \citep{hunter07}, \texttt{pandas} \citep{mckinney10}, \texttt{scipy} \citep{scipy}, \texttt{astropy} \citep{astropy13, astropy18}, \texttt{scikit-learn} \citep{sklearn}, \texttt{PyTorch} \citep{pytorch} and \texttt{dustmaps} \citep{dustmaps} Python packages, and of \texttt{Topcat} \citep{topcat}.

\vspace{0.2cm}
\emph{Author contribution statement:} AAA designed the study and executed it, writing the full manuscript and doing all analyses. SGK provided the neural network from \citet{kane24} and helped adapt it for the present purpose. All other authors were involved in discussions and helped shape the manuscript. 

\section*{Data Availability}

This work uses public LAMOST and \Gaia data. The predictions for the \citetalias{lucey23} and \citetalias{andrae23} XP samples, as well as our LAMOST training sample with \texttt{FERRE} parameters and predictions, are available from \href{https://doi.org/10.5281/zenodo.14651677}{10.5281/zenodo.14651678} and at the CDS, with the data model described in the Appendix. The \texttt{FERRE} synthetic spectra grid used in this work is available upon reasonable request to Carlos Allende Prieto.



\bibliographystyle{mnras}
\bibliography{cemp}

\begin{thebibliography}{}
\makeatletter
\relax
\def\mn@urlcharsother{\let\do\@makeother \do\$\do\&\do\#\do\^\do\_\do\%\do\~}
\def\mn@doi{\begingroup\mn@urlcharsother \@ifnextchar [ {\mn@doi@}
  {\mn@doi@[]}}
\def\mn@doi@[#1]#2{\def\@tempa{#1}\ifx\@tempa\@empty \href
  {http://dx.doi.org/#2} {doi:#2}\else \href {http://dx.doi.org/#2} {#1}\fi
  \endgroup}
\def\mn@eprint#1#2{\mn@eprint@#1:#2::\@nil}
\def\mn@eprint@arXiv#1{\href {http://arxiv.org/abs/#1} {{\tt arXiv:#1}}}
\def\mn@eprint@dblp#1{\href {http://dblp.uni-trier.de/rec/bibtex/#1.xml}
  {dblp:#1}}
\def\mn@eprint@#1:#2:#3:#4\@nil{\def\@tempa {#1}\def\@tempb {#2}\def\@tempc
  {#3}\ifx \@tempc \@empty \let \@tempc \@tempb \let \@tempb \@tempa \fi \ifx
  \@tempb \@empty \def\@tempb {arXiv}\fi \@ifundefined
  {mn@eprint@\@tempb}{\@tempb:\@tempc}{\expandafter \expandafter \csname
  mn@eprint@\@tempb\endcsname \expandafter{\@tempc}}}

\bibitem[\protect\citeauthoryear{{Abate}, {Pols}, {Karakas}  \&
  {Izzard}}{{Abate} et~al.}{2015a}]{abate15_detailed}
{Abate} C.,  {Pols} O.~R.,  {Karakas} A.~I.,   {Izzard} R.~G.,  2015a, \mn@doi
  [\aap] {10.1051/0004-6361/201424739}, \href
  {http://adsabs.harvard.edu/abs/2015A%26A...576A.118A} {576, A118}

\bibitem[\protect\citeauthoryear{{Abate}, {Pols}, {Stancliffe}, {Izzard},
  {Karakas}, {Beers}  \& {Lee}}{{Abate} et~al.}{2015b}]{abate15_binarypop}
{Abate} C.,  {Pols} O.~R.,  {Stancliffe} R.~J.,  {Izzard} R.~G.,  {Karakas}
  A.~I.,  {Beers} T.~C.,   {Lee} Y.~S.,  2015b, \mn@doi [\aap]
  {10.1051/0004-6361/201526200}, \href
  {https://ui.adsabs.harvard.edu/abs/2015A&A...581A..62A} {581, A62}

\bibitem[\protect\citeauthoryear{{Abate}, {Pols}  \& {Stancliffe}}{{Abate}
  et~al.}{2018}]{abate18}
{Abate} C.,  {Pols} O.~R.,   {Stancliffe} R.~J.,  2018, \mn@doi [\aap]
  {10.1051/0004-6361/201833780}, \href
  {https://ui.adsabs.harvard.edu/abs/2018A&A...620A..63A} {620, A63}

\bibitem[\protect\citeauthoryear{{Abdurro'uf} et~al.,}{{Abdurro'uf}
  et~al.}{2022}]{apogeedr17}
{Abdurro'uf} et~al., 2022, \mn@doi [\apjs] {10.3847/1538-4365/ac4414}, \href
  {https://ui.adsabs.harvard.edu/abs/2022ApJS..259...35A} {259, 35}

\bibitem[\protect\citeauthoryear{{Abohalima} \& {Frebel}}{{Abohalima} \&
  {Frebel}}{2018}]{jinabase}
{Abohalima} A.,  {Frebel} A.,  2018, \mn@doi [\apjs]
  {10.3847/1538-4365/aadfe9}, \href
  {https://ui.adsabs.harvard.edu/abs/2018ApJS..238...36A} {238, 36}

\bibitem[\protect\citeauthoryear{{Aguado}, {Gonz{\'a}lez Hern{\'a}ndez},
  {Allende Prieto}  \& {Rebolo}}{{Aguado} et~al.}{2017}]{aguado17}
{Aguado} D.~S.,  {Gonz{\'a}lez Hern{\'a}ndez} J.~I.,  {Allende Prieto} C.,
  {Rebolo} R.,  2017, \mn@doi [\aap] {10.1051/0004-6361/201730654}, \href
  {https://ui.adsabs.harvard.edu/abs/2017A&A...605A..40A} {605, A40}

\bibitem[\protect\citeauthoryear{{Allende Prieto}, {Beers}, {Wilhelm},
  {Newberg}, {Rockosi}, {Yanny}  \& {Lee}}{{Allende Prieto}
  et~al.}{2006}]{allende06}
{Allende Prieto} C.,  {Beers} T.~C.,  {Wilhelm} R.,  {Newberg} H.~J.,
  {Rockosi} C.~M.,  {Yanny} B.,   {Lee} Y.~S.,  2006, \mn@doi [\apj]
  {10.1086/498131}, \href
  {https://ui.adsabs.harvard.edu/abs/2006ApJ...636..804A} {636, 804}

\bibitem[\protect\citeauthoryear{{Andrae}, {Rix}  \& {Chandra}}{{Andrae}
  et~al.}{2023}]{andrae23}
{Andrae} R.,  {Rix} H.-W.,   {Chandra} V.,  2023, \mn@doi [\apjs]
  {10.3847/1538-4365/acd53e}, \href
  {https://ui.adsabs.harvard.edu/abs/2023ApJS..267....8A} {267, 8}

\bibitem[\protect\citeauthoryear{{Aoki}, {Beers}, {Christlieb}, {Norris},
  {Ryan}  \& {Tsangarides}}{{Aoki} et~al.}{2007}]{aoki07}
{Aoki} W.,  {Beers} T.~C.,  {Christlieb} N.,  {Norris} J.~E.,  {Ryan} S.~G.,
  {Tsangarides} S.,  2007, \mn@doi [\apj] {10.1086/509817}, \href
  {https://ui.adsabs.harvard.edu/abs/2007ApJ...655..492A} {655, 492}

\bibitem[\protect\citeauthoryear{{Ardern-Arentsen} et~al.,}{{Ardern-Arentsen}
  et~al.}{2024}]{ardernarentsen24}
{Ardern-Arentsen} A.,  et~al., 2024, \mn@doi [\mnras] {10.1093/mnras/stae1049},
  \href {https://ui.adsabs.harvard.edu/abs/2024MNRAS.530.3391A} {530, 3391}

\bibitem[\protect\citeauthoryear{{Arentsen} et~al.,}{{Arentsen}
  et~al.}{2020}]{arentsen20b}
{Arentsen} A.,  et~al., 2020, \mn@doi [\mnras] {10.1093/mnras/staa1661}, \href
  {https://ui.adsabs.harvard.edu/abs/2020MNRAS.496.4964A} {496, 4964}

\bibitem[\protect\citeauthoryear{{Arentsen} et~al.,}{{Arentsen}
  et~al.}{2021}]{arentsen21}
{Arentsen} A.,  et~al., 2021, \mn@doi [\mnras] {10.1093/mnras/stab1343}, \href
  {https://ui.adsabs.harvard.edu/abs/2021MNRAS.505.1239A} {505, 1239}

\bibitem[\protect\citeauthoryear{{Arentsen}, {Placco}, {Lee}, {Aguado},
  {Martin}, {Starkenburg}  \& {Yoon}}{{Arentsen} et~al.}{2022}]{arentsen22}
{Arentsen} A.,  {Placco} V.~M.,  {Lee} Y.~S.,  {Aguado} D.~S.,  {Martin} N.~F.,
   {Starkenburg} E.,   {Yoon} J.,  2022, \mn@doi [\mnras]
  {10.1093/mnras/stac2062}, \href
  {https://ui.adsabs.harvard.edu/abs/2022MNRAS.515.4082A} {515, 4082}

\bibitem[\protect\citeauthoryear{{Arentsen}, {Aguado}, {Sestito}, {Gonz{\'a}lez
  Hern{\'a}ndez}, {Martin}, {Starkenburg}, {Jablonka}  \& {Yuan}}{{Arentsen}
  et~al.}{2023}]{arentsen23}
{Arentsen} A.,  {Aguado} D.~S.,  {Sestito} F.,  {Gonz{\'a}lez Hern{\'a}ndez}
  J.~I.,  {Martin} N.~F.,  {Starkenburg} E.,  {Jablonka} P.,   {Yuan} Z.,
  2023, \mn@doi [\mnras] {10.1093/mnras/stad043}, \href
  {https://ui.adsabs.harvard.edu/abs/2023MNRAS.519.5554A} {519, 5554}

\bibitem[\protect\citeauthoryear{{Astropy Collaboration} et~al.,}{{Astropy
  Collaboration} et~al.}{2013}]{astropy13}
{Astropy Collaboration} et~al., 2013, \mn@doi [\aap]
  {10.1051/0004-6361/201322068}, \href
  {http://adsabs.harvard.edu/abs/2013A%26A...558A..33A} {558, A33}

\bibitem[\protect\citeauthoryear{{Baumgardt} \& {Hilker}}{{Baumgardt} \&
  {Hilker}}{2018}]{baumgardt18}
{Baumgardt} H.,  {Hilker} M.,  2018, \mn@doi [\mnras] {10.1093/mnras/sty1057},
  \href {https://ui.adsabs.harvard.edu/abs/2018MNRAS.478.1520B} {478, 1520}

\bibitem[\protect\citeauthoryear{{Beers} \& {Christlieb}}{{Beers} \&
  {Christlieb}}{2005}]{beerschristlieb05}
{Beers} T.~C.,  {Christlieb} N.,  2005, \mn@doi [\araa]
  {10.1146/annurev.astro.42.053102.134057}, \href
  {https://ui.adsabs.harvard.edu/abs/2005ARA&A..43..531B} {43, 531}

\bibitem[\protect\citeauthoryear{{Beers}, {Preston}  \& {Shectman}}{{Beers}
  et~al.}{1992}]{beers92}
{Beers} T.~C.,  {Preston} G.~W.,   {Shectman} S.~A.,  1992, \mn@doi [\aj]
  {10.1086/116207}, \href
  {https://ui.adsabs.harvard.edu/abs/1992AJ....103.1987B} {103, 1987}

\bibitem[\protect\citeauthoryear{{Bekki} \& {Freeman}}{{Bekki} \&
  {Freeman}}{2003}]{bekki03}
{Bekki} K.,  {Freeman} K.~C.,  2003, \mn@doi [\mnras]
  {10.1046/j.1365-2966.2003.07275.x}, \href
  {https://ui.adsabs.harvard.edu/abs/2003MNRAS.346L..11B} {346, L11}

\bibitem[\protect\citeauthoryear{{Belokurov} \& {Kravtsov}}{{Belokurov} \&
  {Kravtsov}}{2023}]{belokurov23_Nrich}
{Belokurov} V.,  {Kravtsov} A.,  2023, \mn@doi [\mnras]
  {10.1093/mnras/stad2241}, \href
  {https://ui.adsabs.harvard.edu/abs/2023MNRAS.525.4456B} {525, 4456}

\bibitem[\protect\citeauthoryear{{Belokurov} et~al.,}{{Belokurov}
  et~al.}{2020}]{belokurov20}
{Belokurov} V.,  et~al., 2020, \mn@doi [\mnras] {10.1093/mnras/staa1522}, \href
  {https://ui.adsabs.harvard.edu/abs/2020MNRAS.496.1922B} {496, 1922}

\bibitem[\protect\citeauthoryear{{Bisterzo}, {Gallino}, {Straniero},
  {Cristallo}  \& {K{\"a}ppeler}}{{Bisterzo} et~al.}{2010}]{bisterzo10}
{Bisterzo} S.,  {Gallino} R.,  {Straniero} O.,  {Cristallo} S.,
  {K{\"a}ppeler} F.,  2010, \mn@doi [\mnras]
  {10.1111/j.1365-2966.2010.16369.x}, \href
  {https://ui.adsabs.harvard.edu/abs/2010MNRAS.404.1529B} {404, 1529}

\bibitem[\protect\citeauthoryear{{Bovy}}{{Bovy}}{2015}]{galpy}
{Bovy} J.,  2015, \mn@doi [\apjs] {10.1088/0067-0049/216/2/29}, \href
  {https://ui.adsabs.harvard.edu/abs/2015ApJS..216...29B} {216, 29}

\bibitem[\protect\citeauthoryear{{Bressan}, {Marigo}, {Girardi}, {Salasnich},
  {Dal Cero}, {Rubele}  \& {Nanni}}{{Bressan} et~al.}{2012}]{bressan12}
{Bressan} A.,  {Marigo} P.,  {Girardi} L.,  {Salasnich} B.,  {Dal Cero} C.,
  {Rubele} S.,   {Nanni} A.,  2012, \mn@doi [\mnras]
  {10.1111/j.1365-2966.2012.21948.x}, \href
  {https://ui.adsabs.harvard.edu/abs/2012MNRAS.427..127B} {427, 127}

\bibitem[\protect\citeauthoryear{{Buder} et~al.,}{{Buder}
  et~al.}{2024}]{galahdr4}
{Buder} S.,  et~al., 2024, \mn@doi [arXiv e-prints]
  {10.48550/arXiv.2409.19858}, \href
  {https://ui.adsabs.harvard.edu/abs/2024arXiv240919858B} {p. arXiv:2409.19858}

\bibitem[\protect\citeauthoryear{{Carrasco} et~al.,}{{Carrasco}
  et~al.}{2021}]{carrasco21}
{Carrasco} J.~M.,  et~al., 2021, \mn@doi [\aap] {10.1051/0004-6361/202141249},
  \href {https://ui.adsabs.harvard.edu/abs/2021A&A...652A..86C} {652, A86}

\bibitem[\protect\citeauthoryear{{Casagrande}, {Wolf}, {Mackey}, {Nordland er},
  {Yong}  \& {Bessell}}{{Casagrande} et~al.}{2019}]{casagrande19}
{Casagrande} L.,  {Wolf} C.,  {Mackey} A.~D.,  {Nordland er} T.,  {Yong} D.,
  {Bessell} M.,  2019, \mn@doi [\mnras] {10.1093/mnras/sty2878}, \href
  {https://ui.adsabs.harvard.edu/abs/2019MNRAS.482.2770C} {482, 2770}

\bibitem[\protect\citeauthoryear{Chen \& Guestrin}{Chen \&
  Guestrin}{2016}]{xgboost}
Chen T.,  Guestrin C.,  2016, in Proceedings of the 22nd ACM SIGKDD
  International Conference on Knowledge Discovery and Data Mining. KDD '16.
Association for Computing Machinery, New York, NY, USA, p. 785–794,
  \mn@doi{10.1145/2939672.2939785}, \url
  {https://doi.org/10.1145/2939672.2939785}

\bibitem[\protect\citeauthoryear{{Cohen}, {Christlieb}, {Thompson},
  {McWilliam}, {Shectman}, {Reimers}, {Wisotzki}  \& {Kirby}}{{Cohen}
  et~al.}{2013}]{cohen13}
{Cohen} J.~G.,  {Christlieb} N.,  {Thompson} I.,  {McWilliam} A.,  {Shectman}
  S.,  {Reimers} D.,  {Wisotzki} L.,   {Kirby} E.,  2013, \mn@doi [\apj]
  {10.1088/0004-637X/778/1/56}, \href
  {http://adsabs.harvard.edu/abs/2013ApJ...778...56C} {778, 56}

\bibitem[\protect\citeauthoryear{{D'Eugenio} et~al.,}{{D'Eugenio}
  et~al.}{2024}]{deugenio24}
{D'Eugenio} F.,  et~al., 2024, \mn@doi [\aap] {10.1051/0004-6361/202348636},
  \href {https://ui.adsabs.harvard.edu/abs/2024A&A...689A.152D} {689, A152}

\bibitem[\protect\citeauthoryear{{D'Orazi}, {Gratton}, {Lucatello}, {Carretta},
  {Bragaglia}  \& {Marino}}{{D'Orazi} et~al.}{2010}]{dorazi10}
{D'Orazi} V.,  {Gratton} R.,  {Lucatello} S.,  {Carretta} E.,  {Bragaglia} A.,
   {Marino} A.~F.,  2010, \mn@doi [\apjl] {10.1088/2041-8205/719/2/L213}, \href
  {https://ui.adsabs.harvard.edu/abs/2010ApJ...719L.213D} {719, L213}

\bibitem[\protect\citeauthoryear{{Da Costa} et~al.,}{{Da Costa}
  et~al.}{2019}]{dacosta19}
{Da Costa} G.~S.,  et~al., 2019, \mn@doi [\mnras] {10.1093/mnras/stz2550},
  \href {https://ui.adsabs.harvard.edu/abs/2019MNRAS.489.5900D} {489, 5900}

\bibitem[\protect\citeauthoryear{{Dalton} et~al.,}{{Dalton}
  et~al.}{2018}]{weave}
{Dalton} G.,  et~al., 2018, in \procspie. p. 107021B,
  \mn@doi{10.1117/12.2312031}

\bibitem[\protect\citeauthoryear{{De Angeli} et~al.,}{{De Angeli}
  et~al.}{2023}]{deangeli22}
{De Angeli} F.,  et~al., 2023, \mn@doi [\aap] {10.1051/0004-6361/202243680},
  \href {https://ui.adsabs.harvard.edu/abs/2023A&A...674A...2D} {674, A2}

\bibitem[\protect\citeauthoryear{{Ezzeddine} et~al.,}{{Ezzeddine}
  et~al.}{2020}]{ezzeddine20}
{Ezzeddine} R.,  et~al., 2020, \mn@doi [\apj] {10.3847/1538-4357/ab9d1a}, \href
  {https://ui.adsabs.harvard.edu/abs/2020ApJ...898..150E} {898, 150}

\bibitem[\protect\citeauthoryear{{Fallows} \& {Sanders}}{{Fallows} \&
  {Sanders}}{2024}]{fallows24}
{Fallows} C.~P.,  {Sanders} J.~L.,  2024, \mn@doi [\mnras]
  {10.1093/mnras/stae1303}, \href
  {https://ui.adsabs.harvard.edu/abs/2024MNRAS.531.2126F} {531, 2126}

\bibitem[\protect\citeauthoryear{{Frebel} et~al.,}{{Frebel}
  et~al.}{2006}]{frebel06}
{Frebel} A.,  et~al., 2006, \mn@doi [\apj] {10.1086/508506}, \href
  {https://ui.adsabs.harvard.edu/abs/2006ApJ...652.1585F} {652, 1585}

\bibitem[\protect\citeauthoryear{{Gaia Collaboration} et~al.,}{{Gaia
  Collaboration} et~al.}{2023}]{gaiadr3}
{Gaia Collaboration} et~al., 2023, \mn@doi [\aap]
  {10.1051/0004-6361/202243940}, \href
  {https://ui.adsabs.harvard.edu/abs/2023A&A...674A...1G} {674, A1}

\bibitem[\protect\citeauthoryear{{Gratton}, {Sneden}, {Carretta}  \&
  {Bragaglia}}{{Gratton} et~al.}{2000}]{gratton00}
{Gratton} R.~G.,  {Sneden} C.,  {Carretta} E.,   {Bragaglia} A.,  2000, \aap,
  \href {https://ui.adsabs.harvard.edu/abs/2000A&A...354..169G} {354, 169}

\bibitem[\protect\citeauthoryear{{Gratton}, {Sneden}  \& {Carretta}}{{Gratton}
  et~al.}{2004}]{gratton04}
{Gratton} R.,  {Sneden} C.,   {Carretta} E.,  2004, \mn@doi [\araa]
  {10.1146/annurev.astro.42.053102.133945}, \href
  {https://ui.adsabs.harvard.edu/abs/2004ARA&A..42..385G} {42, 385}

\bibitem[\protect\citeauthoryear{{Green}}{{Green}}{2018}]{dustmaps}
{Green} G.,  2018, \mn@doi [The Journal of Open Source Software]
  {10.21105/joss.00695}, \href
  {https://ui.adsabs.harvard.edu/abs/2018JOSS....3..695G} {3, 695}

\bibitem[\protect\citeauthoryear{{Hansen}, {Andersen}, {Nordstr{\"o}m},
  {Beers}, {Placco}, {Yoon}  \& {Buchhave}}{{Hansen} et~al.}{2016a}]{hansen16a}
{Hansen} T.~T.,  {Andersen} J.,  {Nordstr{\"o}m} B.,  {Beers} T.~C.,  {Placco}
  V.~M.,  {Yoon} J.,   {Buchhave} L.~A.,  2016a, \mn@doi [\aap]
  {10.1051/0004-6361/201527235}, \href
  {http://adsabs.harvard.edu/abs/2016A%26A...586A.160H} {586, A160}

\bibitem[\protect\citeauthoryear{{Hansen}, {Andersen}, {Nordstr{\"o}m},
  {Beers}, {Placco}, {Yoon}  \& {Buchhave}}{{Hansen} et~al.}{2016b}]{hansen16b}
{Hansen} T.~T.,  {Andersen} J.,  {Nordstr{\"o}m} B.,  {Beers} T.~C.,  {Placco}
  V.~M.,  {Yoon} J.,   {Buchhave} L.~A.,  2016b, \mn@doi [\aap]
  {10.1051/0004-6361/201527409}, \href
  {http://adsabs.harvard.edu/abs/2016A%26A...588A...3H} {588, A3}

\bibitem[\protect\citeauthoryear{{Hansen} et~al.,}{{Hansen}
  et~al.}{2018}]{hansen18}
{Hansen} T.~T.,  et~al., 2018, \mn@doi [\apj] {10.3847/1538-4357/aabacc}, \href
  {https://ui.adsabs.harvard.edu/abs/2018ApJ...858...92H} {858, 92}

\bibitem[\protect\citeauthoryear{Harris et~al.,}{Harris et~al.}{2020}]{numpy}
Harris C.~R.,  et~al., 2020, \mn@doi [Nature] {10.1038/s41586-020-2649-2}, 585,
  357

\bibitem[\protect\citeauthoryear{{Hattori}}{{Hattori}}{2024}]{hattori24}
{Hattori} K.,  2024, \mn@doi [arXiv e-prints] {10.48550/arXiv.2404.01269},
  \href {https://ui.adsabs.harvard.edu/abs/2024arXiv240401269H} {p.
  arXiv:2404.01269}

\bibitem[\protect\citeauthoryear{{Heger} \& {Woosley}}{{Heger} \&
  {Woosley}}{2010}]{hegerwoosley10}
{Heger} A.,  {Woosley} S.~E.,  2010, \mn@doi [\apj]
  {10.1088/0004-637X/724/1/341}, \href
  {https://ui.adsabs.harvard.edu/abs/2010ApJ...724..341H} {724, 341}

\bibitem[\protect\citeauthoryear{{Holmbeck} et~al.,}{{Holmbeck}
  et~al.}{2020}]{holmbeck20}
{Holmbeck} E.~M.,  et~al., 2020, \mn@doi [\apjs] {10.3847/1538-4365/ab9c19},
  \href {https://ui.adsabs.harvard.edu/abs/2020ApJS..249...30H} {249, 30}

\bibitem[\protect\citeauthoryear{{Horta} et~al.,}{{Horta}
  et~al.}{2021}]{horta21}
{Horta} D.,  et~al., 2021, \mn@doi [\mnras] {10.1093/mnras/staa3598}, \href
  {https://ui.adsabs.harvard.edu/abs/2021MNRAS.500.5462H} {500, 5462}

\bibitem[\protect\citeauthoryear{{Huang} et~al.,}{{Huang}
  et~al.}{2024}]{huang24}
{Huang} Y.,  et~al., 2024, \mn@doi [\apj] {10.3847/1538-4357/ad6b94}, \href
  {https://ui.adsabs.harvard.edu/abs/2024ApJ...974..192H} {974, 192}

\bibitem[\protect\citeauthoryear{Hunter}{Hunter}{2007}]{hunter07}
Hunter J.~D.,  2007, \mn@doi [Computing In Science \& Engineering]
  {10.1109/MCSE.2007.55}, 9, 90

\bibitem[\protect\citeauthoryear{{Ishigaki}, {Tominaga}, {Kobayashi}  \&
  {Nomoto}}{{Ishigaki} et~al.}{2018}]{ishigaki18}
{Ishigaki} M.~N.,  {Tominaga} N.,  {Kobayashi} C.,   {Nomoto} K.,  2018,
  \mn@doi [\apj] {10.3847/1538-4357/aab3de}, \href
  {https://ui.adsabs.harvard.edu/abs/2018ApJ...857...46I} {857, 46}

\bibitem[\protect\citeauthoryear{{Jorissen}, {Boffin}, {Karinkuzhi}, {Van Eck},
  {Escorza}, {Shetye}  \& {Van Winckel}}{{Jorissen} et~al.}{2019}]{jorissen19}
{Jorissen} A.,  {Boffin} H.~M.~J.,  {Karinkuzhi} D.,  {Van Eck} S.,  {Escorza}
  A.,  {Shetye} S.,   {Van Winckel} H.,  2019, \mn@doi [\aap]
  {10.1051/0004-6361/201834630}, \href
  {https://ui.adsabs.harvard.edu/abs/2019A&A...626A.127J} {626, A127}

\bibitem[\protect\citeauthoryear{{Kane}, {Belokurov}, {Cranmer}, {Monty},
  {Zhang}, {Ardern-Arentsen}  \& {Kane}}{{Kane} et~al.}{2024}]{kane24}
{Kane} S.~G.,  {Belokurov} V.,  {Cranmer} M.,  {Monty} S.,  {Zhang} H.,
  {Ardern-Arentsen} A.,   {Kane} E.,  2024, \mn@doi [arXiv e-prints]
  {10.48550/arXiv.2409.00197}, \href
  {https://ui.adsabs.harvard.edu/abs/2024arXiv240900197K} {p. arXiv:2409.00197}

\bibitem[\protect\citeauthoryear{{Kielty}, {Venn}, {Loewen}, {Shetrone},
  {Placco}, {Jahandar}, {M{\'e}sz{\'a}ros}  \& {Martell}}{{Kielty}
  et~al.}{2017}]{Kielty17}
{Kielty} C.~L.,  {Venn} K.~A.,  {Loewen} N.~B.,  {Shetrone} M.~D.,  {Placco}
  V.~M.,  {Jahandar} F.,  {M{\'e}sz{\'a}ros} S.,   {Martell} S.~L.,  2017,
  \mn@doi [\mnras] {10.1093/mnras/stx1594}, \href
  {http://adsabs.harvard.edu/abs/2017MNRAS.471..404K} {471, 404}

\bibitem[\protect\citeauthoryear{{Kim}, {Lee}, {Beers}  \& {Kim}}{{Kim}
  et~al.}{2023}]{kim23}
{Kim} C.,  {Lee} Y.~S.,  {Beers} T.~C.,   {Kim} Y.~K.,  2023, \mn@doi [Journal
  of Korean Astronomical Society] {10.5303/JKAS.2023.56.1.59}, \href
  {https://ui.adsabs.harvard.edu/abs/2023JKAS...56...59K} {56, 59}

\bibitem[\protect\citeauthoryear{{Kingma} \& {Ba}}{{Kingma} \&
  {Ba}}{2014}]{adam}
{Kingma} D.~P.,  {Ba} J.,  2014, \mn@doi [arXiv e-prints]
  {10.48550/arXiv.1412.6980}, \href
  {https://ui.adsabs.harvard.edu/abs/2014arXiv1412.6980K} {p. arXiv:1412.6980}

\bibitem[\protect\citeauthoryear{{Kirby} et~al.,}{{Kirby}
  et~al.}{2015}]{kirby15}
{Kirby} E.~N.,  et~al., 2015, \mn@doi [\apj] {10.1088/0004-637X/801/2/125},
  \href {https://ui.adsabs.harvard.edu/abs/2015ApJ...801..125K} {801, 125}

\bibitem[\protect\citeauthoryear{{Koesterke}, {Allende Prieto}  \&
  {Lambert}}{{Koesterke} et~al.}{2008}]{koesterke08}
{Koesterke} L.,  {Allende Prieto} C.,   {Lambert} D.~L.,  2008, \mn@doi [\apj]
  {10.1086/587471}, \href
  {https://ui.adsabs.harvard.edu/abs/2008ApJ...680..764K} {680, 764}

\bibitem[\protect\citeauthoryear{{Lee} et~al.,}{{Lee} et~al.}{2008}]{lee08}
{Lee} Y.~S.,  et~al., 2008, \mn@doi [\aj] {10.1088/0004-6256/136/5/2022}, \href
  {https://ui.adsabs.harvard.edu/abs/2008AJ....136.2022L} {136, 2022}

\bibitem[\protect\citeauthoryear{{Lee} et~al.,}{{Lee} et~al.}{2013}]{Lee13}
{Lee} Y.~S.,  et~al., 2013, \mn@doi [\aj] {10.1088/0004-6256/146/5/132}, \href
  {http://adsabs.harvard.edu/abs/2013AJ....146..132L} {146, 132}

\bibitem[\protect\citeauthoryear{{Li}, {Tan}  \& {Zhao}}{{Li}
  et~al.}{2018}]{li18}
{Li} H.,  {Tan} K.,   {Zhao} G.,  2018, \mn@doi [\apjs]
  {10.3847/1538-4365/aada4a}, \href
  {https://ui.adsabs.harvard.edu/abs/2018ApJS..238...16L} {238, 16}

\bibitem[\protect\citeauthoryear{{Li} et~al.,}{{Li} et~al.}{2022}]{li22}
{Li} H.,  et~al., 2022, \mn@doi [\apj] {10.3847/1538-4357/ac6514}, \href
  {https://ui.adsabs.harvard.edu/abs/2022ApJ...931..147L} {931, 147}

\bibitem[\protect\citeauthoryear{{Li}, {Wong}, {Hogg}, {Rix}  \&
  {Chandra}}{{Li} et~al.}{2024}]{ji24}
{Li} J.,  {Wong} K. W.~K.,  {Hogg} D.~W.,  {Rix} H.-W.,   {Chandra} V.,  2024,
  \mn@doi [\apjs] {10.3847/1538-4365/ad2b4d}, \href
  {https://ui.adsabs.harvard.edu/abs/2024ApJS..272....2L} {272, 2}

\bibitem[\protect\citeauthoryear{{Lucatello}, {Tsangarides}, {Beers},
  {Carretta}, {Gratton}  \& {Ryan}}{{Lucatello} et~al.}{2005a}]{lucatello05}
{Lucatello} S.,  {Tsangarides} S.,  {Beers} T.~C.,  {Carretta} E.,  {Gratton}
  R.~G.,   {Ryan} S.~G.,  2005a, \mn@doi [\apj] {10.1086/428104}, \href
  {http://adsabs.harvard.edu/abs/2005ApJ...625..825L} {625, 825}

\bibitem[\protect\citeauthoryear{{Lucatello}, {Gratton}, {Beers}  \&
  {Carretta}}{{Lucatello} et~al.}{2005b}]{lucatello05_imf}
{Lucatello} S.,  {Gratton} R.~G.,  {Beers} T.~C.,   {Carretta} E.,  2005b,
  \mn@doi [\apj] {10.1086/428105}, \href
  {https://ui.adsabs.harvard.edu/abs/2005ApJ...625..833L} {625, 833}

\bibitem[\protect\citeauthoryear{{Lucatello}, {Beers}, {Christlieb}, {Barklem},
  {Rossi}, {Marsteller}, {Sivarani}  \& {Lee}}{{Lucatello}
  et~al.}{2006}]{lucatello06}
{Lucatello} S.,  {Beers} T.~C.,  {Christlieb} N.,  {Barklem} P.~S.,  {Rossi}
  S.,  {Marsteller} B.,  {Sivarani} T.,   {Lee} Y.~S.,  2006, \mn@doi [\apjl]
  {10.1086/509780}, \href
  {https://ui.adsabs.harvard.edu/abs/2006ApJ...652L..37L} {652, L37}

\bibitem[\protect\citeauthoryear{{Lucatello}, {Sollima}, {Gratton},
  {Vesperini}, {D'Orazi}, {Carretta}  \& {Bragaglia}}{{Lucatello}
  et~al.}{2015}]{lucatello15}
{Lucatello} S.,  {Sollima} A.,  {Gratton} R.,  {Vesperini} E.,  {D'Orazi} V.,
  {Carretta} E.,   {Bragaglia} A.,  2015, \mn@doi [\aap]
  {10.1051/0004-6361/201526957}, \href
  {https://ui.adsabs.harvard.edu/abs/2015A&A...584A..52L} {584, A52}

\bibitem[\protect\citeauthoryear{{Lucchesi} et~al.,}{{Lucchesi}
  et~al.}{2024}]{lucchesi24}
{Lucchesi} R.,  et~al., 2024, \mn@doi [\aap] {10.1051/0004-6361/202348093},
  \href {https://ui.adsabs.harvard.edu/abs/2024A&A...686A.266L} {686, A266}

\bibitem[\protect\citeauthoryear{{Lucey} et~al.,}{{Lucey}
  et~al.}{2023}]{lucey23}
{Lucey} M.,  et~al., 2023, \mn@doi [\mnras] {10.1093/mnras/stad1675}, \href
  {https://ui.adsabs.harvard.edu/abs/2023MNRAS.523.4049L} {523, 4049}

\bibitem[\protect\citeauthoryear{{Martin} et~al.,}{{Martin}
  et~al.}{2024}]{martin24}
{Martin} N.~F.,  et~al., 2024, \mn@doi [\aap] {10.1051/0004-6361/202347633},
  \href {https://ui.adsabs.harvard.edu/abs/2024A&A...692A.115M} {692, A115}

\bibitem[\protect\citeauthoryear{{McClure} \& {Woodsworth}}{{McClure} \&
  {Woodsworth}}{1990}]{McClureWoodsworth90}
{McClure} R.~D.,  {Woodsworth} A.~W.,  1990, \mn@doi [\apj] {10.1086/168573},
  \href {http://adsabs.harvard.edu/abs/1990ApJ...352..709M} {352, 709}

\bibitem[\protect\citeauthoryear{McKinney}{McKinney}{2010}]{mckinney10}
McKinney W.,  2010, in van~der Walt S.,  Millman J.,  eds, Proceedings of the
  9th Python in Science Conference. pp 51 -- 56

\bibitem[\protect\citeauthoryear{{Mehta} et~al.,}{{Mehta}
  et~al.}{2024}]{mehta24}
{Mehta} V.~J.,  et~al., 2024, \mn@doi [arXiv e-prints]
  {10.48550/arXiv.2406.02755}, \href
  {https://ui.adsabs.harvard.edu/abs/2024arXiv240602755M} {p. arXiv:2406.02755}

\bibitem[\protect\citeauthoryear{{M{\'e}sz{\'a}ros} et~al.,}{{M{\'e}sz{\'a}ros}
  et~al.}{2012}]{meszaros12}
{M{\'e}sz{\'a}ros} S.,  et~al., 2012, \mn@doi [\aj]
  {10.1088/0004-6256/144/4/120}, \href
  {https://ui.adsabs.harvard.edu/abs/2012AJ....144..120M} {144, 120}

\bibitem[\protect\citeauthoryear{{Meynet}, {Ekstr{\"o}m}  \& {Maeder}}{{Meynet}
  et~al.}{2006}]{meynet06}
{Meynet} G.,  {Ekstr{\"o}m} S.,   {Maeder} A.,  2006, \mn@doi [\aap]
  {10.1051/0004-6361:20053070}, \href
  {http://adsabs.harvard.edu/abs/2006A%26A...447..623M} {447, 623}

\bibitem[\protect\citeauthoryear{{Milone} et~al.,}{{Milone}
  et~al.}{2012}]{milone12}
{Milone} A.~P.,  et~al., 2012, \mn@doi [\aap] {10.1051/0004-6361/201016384},
  \href {https://ui.adsabs.harvard.edu/abs/2012A&A...540A..16M} {540, A16}

\bibitem[\protect\citeauthoryear{{Montegriffo} et~al.,}{{Montegriffo}
  et~al.}{2023}]{montegriffo23}
{Montegriffo} P.,  et~al., 2023, \mn@doi [\aap] {10.1051/0004-6361/202243880},
  \href {https://ui.adsabs.harvard.edu/abs/2023A&A...674A...3M} {674, A3}

\bibitem[\protect\citeauthoryear{{Nomoto}, {Kobayashi}  \& {Tominaga}}{{Nomoto}
  et~al.}{2013}]{nomoto13}
{Nomoto} K.,  {Kobayashi} C.,   {Tominaga} N.,  2013, \mn@doi [\araa]
  {10.1146/annurev-astro-082812-140956}, \href
  {https://ui.adsabs.harvard.edu/abs/2013ARA&A..51..457N} {51, 457}

\bibitem[\protect\citeauthoryear{{Norris} et~al.,}{{Norris}
  et~al.}{2013}]{norris13b}
{Norris} J.~E.,  et~al., 2013, \mn@doi [\apj] {10.1088/0004-637X/762/1/28},
  \href {http://adsabs.harvard.edu/abs/2013ApJ...762...28N} {762, 28}

\bibitem[\protect\citeauthoryear{{Paszke} et~al.,}{{Paszke}
  et~al.}{2019}]{pytorch}
{Paszke} A.,  et~al., 2019, \mn@doi [arXiv e-prints]
  {10.48550/arXiv.1912.01703}, \href
  {https://ui.adsabs.harvard.edu/abs/2019arXiv191201703P} {p. arXiv:1912.01703}

\bibitem[\protect\citeauthoryear{Pedregosa et~al.,}{Pedregosa
  et~al.}{2011}]{sklearn}
Pedregosa F.,  et~al., 2011, Journal of machine learning research, 12, 2825

\bibitem[\protect\citeauthoryear{{Placco}, {Frebel}, {Beers}  \&
  {Stancliffe}}{{Placco} et~al.}{2014}]{placco14}
{Placco} V.~M.,  {Frebel} A.,  {Beers} T.~C.,   {Stancliffe} R.~J.,  2014,
  \mn@doi [\apj] {10.1088/0004-637X/797/1/21}, \href
  {https://ui.adsabs.harvard.edu/abs/2014ApJ...797...21P} {797, 21}

\bibitem[\protect\citeauthoryear{Price-Whelan}{Price-Whelan}{2017}]{gala}
Price-Whelan A.~M.,  2017, \mn@doi [The Journal of Open Source Software]
  {10.21105/joss.00388}, 2

\bibitem[\protect\citeauthoryear{{Price-Whelan} et~al.,}{{Price-Whelan}
  et~al.}{2018}]{astropy18}
{Price-Whelan} A.~M.,  et~al., 2018, \mn@doi [\aj] {10.3847/1538-3881/aabc4f},
  \href {https://ui.adsabs.harvard.edu/#abs/2018AJ....156..123T} {156, 123}

\bibitem[\protect\citeauthoryear{Price-Whelan et~al.,}{Price-Whelan
  et~al.}{2020}]{pricewhelan20_gala}
Price-Whelan A.,  et~al., 2020, adrn/gala: v1.3,
  \mn@doi{10.5281/zenodo.4159870}, \url
  {https://doi.org/10.5281/zenodo.4159870}

\bibitem[\protect\citeauthoryear{{Rix} et~al.,}{{Rix} et~al.}{2022}]{rix22}
{Rix} H.-W.,  et~al., 2022, \mn@doi [\apj] {10.3847/1538-4357/ac9e01}, \href
  {https://ui.adsabs.harvard.edu/abs/2022ApJ...941...45R} {941, 45}

\bibitem[\protect\citeauthoryear{{Roederer}, {Preston}, {Thompson}, {Shectman},
  {Sneden}, {Burley}  \& {Kelson}}{{Roederer} et~al.}{2014}]{roederer14}
{Roederer} I.~U.,  {Preston} G.~W.,  {Thompson} I.~B.,  {Shectman} S.~A.,
  {Sneden} C.,  {Burley} G.~S.,   {Kelson} D.~D.,  2014, \mn@doi [\aj]
  {10.1088/0004-6256/147/6/136}, \href
  {http://adsabs.harvard.edu/abs/2014AJ....147..136R} {147, 136}

\bibitem[\protect\citeauthoryear{{Saccardi} et~al.,}{{Saccardi}
  et~al.}{2023}]{saccardi23}
{Saccardi} A.,  et~al., 2023, \mn@doi [\apj] {10.3847/1538-4357/acc39f}, \href
  {https://ui.adsabs.harvard.edu/abs/2023ApJ...948...35S} {948, 35}

\bibitem[\protect\citeauthoryear{{Salvadori}, {Sk{\'u}lad{\'o}ttir}  \&
  {Tolstoy}}{{Salvadori} et~al.}{2015}]{salvadori15}
{Salvadori} S.,  {Sk{\'u}lad{\'o}ttir} {\'A}.,   {Tolstoy} E.,  2015, \mn@doi
  [\mnras] {10.1093/mnras/stv1969}, \href
  {https://ui.adsabs.harvard.edu/abs/2015MNRAS.454.1320S} {454, 1320}

\bibitem[\protect\citeauthoryear{{Sanders} \& {Matsunaga}}{{Sanders} \&
  {Matsunaga}}{2023}]{sanders23}
{Sanders} J.~L.,  {Matsunaga} N.,  2023, \mn@doi [\mnras]
  {10.1093/mnras/stad574}, \href
  {https://ui.adsabs.harvard.edu/abs/2023MNRAS.521.2745S} {521, 2745}

\bibitem[\protect\citeauthoryear{{Schiavon} et~al.,}{{Schiavon}
  et~al.}{2017}]{schiavon17}
{Schiavon} R.~P.,  et~al., 2017, \mn@doi [\mnras] {10.1093/mnras/stw2162},
  \href {https://ui.adsabs.harvard.edu/abs/2017MNRAS.465..501S} {465, 501}

\bibitem[\protect\citeauthoryear{{Schlafly} \& {Finkbeiner}}{{Schlafly} \&
  {Finkbeiner}}{2011}]{schlafly11}
{Schlafly} E.~F.,  {Finkbeiner} D.~P.,  2011, \mn@doi [\apj]
  {10.1088/0004-637X/737/2/103}, \href
  {https://ui.adsabs.harvard.edu/abs/2011ApJ...737..103S} {737, 103}

\bibitem[\protect\citeauthoryear{{Schlegel}, {Finkbeiner}  \&
  {Davis}}{{Schlegel} et~al.}{1998}]{schlegel98}
{Schlegel} D.~J.,  {Finkbeiner} D.~P.,   {Davis} M.,  1998, \mn@doi [\apj]
  {10.1086/305772}, \href
  {https://ui.adsabs.harvard.edu/abs/1998ApJ...500..525S} {500, 525}

\bibitem[\protect\citeauthoryear{{Sestito} et~al.,}{{Sestito}
  et~al.}{2024}]{sestito24_sgrcarbon}
{Sestito} F.,  et~al., 2024, \mn@doi [\aap] {10.1051/0004-6361/202451258},
  \href {https://ui.adsabs.harvard.edu/abs/2024A&A...690A.333S} {690, A333}

\bibitem[\protect\citeauthoryear{{Stancliffe}, {Church}, {Angelou}  \&
  {Lattanzio}}{{Stancliffe} et~al.}{2009}]{stancliffe09}
{Stancliffe} R.~J.,  {Church} R.~P.,  {Angelou} G.~C.,   {Lattanzio} J.~C.,
  2009, \mn@doi [\mnras] {10.1111/j.1365-2966.2009.14900.x}, \href
  {https://ui.adsabs.harvard.edu/abs/2009MNRAS.396.2313S} {396, 2313}

\bibitem[\protect\citeauthoryear{{Taylor}}{{Taylor}}{2005}]{topcat}
{Taylor} M.~B.,  2005, in {Shopbell} P.,  {Britton} M.,   {Ebert} R.,  eds,
  Astronomical Society of the Pacific Conference Series Vol. 347, Astronomical
  Data Analysis Software and Systems XIV. p.~29

\bibitem[\protect\citeauthoryear{{Tominaga}, {Iwamoto}  \& {Nomoto}}{{Tominaga}
  et~al.}{2014}]{tominaga14}
{Tominaga} N.,  {Iwamoto} N.,   {Nomoto} K.,  2014, \mn@doi [\apj]
  {10.1088/0004-637X/785/2/98}, \href
  {https://ui.adsabs.harvard.edu/abs/2014ApJ...785...98T} {785, 98}

\bibitem[\protect\citeauthoryear{{Tumlinson}}{{Tumlinson}}{2007}]{tumlinson07}
{Tumlinson} J.,  2007, \mn@doi [\apjl] {10.1086/520930}, \href
  {https://ui.adsabs.harvard.edu/abs/2007ApJ...664L..63T} {664, L63}

\bibitem[\protect\citeauthoryear{{Umeda} \& {Nomoto}}{{Umeda} \&
  {Nomoto}}{2003}]{umedanomoto03}
{Umeda} H.,  {Nomoto} K.,  2003, \mn@doi [\nat] {10.1038/nature01571}, \href
  {http://adsabs.harvard.edu/abs/2003Natur.422..871U} {422, 871}

\bibitem[\protect\citeauthoryear{{Vasiliev} \& {Baumgardt}}{{Vasiliev} \&
  {Baumgardt}}{2021}]{vasiliev21}
{Vasiliev} E.,  {Baumgardt} H.,  2021, \mn@doi [\mnras]
  {10.1093/mnras/stab1475}, \href
  {https://ui.adsabs.harvard.edu/abs/2021MNRAS.505.5978V} {505, 5978}

\bibitem[\protect\citeauthoryear{Virtanen et~al.,}{Virtanen
  et~al.}{2020}]{scipy}
Virtanen P.,  et~al., 2020, \mn@doi [Nature Methods]
  {10.1038/s41592-019-0686-2}, \href {https://rdcu.be/b08Wh} {17, 261}

\bibitem[\protect\citeauthoryear{{Whitten} et~al.,}{{Whitten}
  et~al.}{2021}]{whitten21}
{Whitten} D.~D.,  et~al., 2021, \mn@doi [\apj] {10.3847/1538-4357/abee7e},
  \href {https://ui.adsabs.harvard.edu/abs/2021ApJ...912..147W} {912, 147}

\bibitem[\protect\citeauthoryear{{Witten} et~al.,}{{Witten}
  et~al.}{2022}]{witten22}
{Witten} C. E.~C.,  et~al., 2022, \mn@doi [\mnras] {10.1093/mnras/stac2273},
  \href {https://ui.adsabs.harvard.edu/abs/2022MNRAS.516.3254W} {516, 3254}

\bibitem[\protect\citeauthoryear{{Ye}, {Cui}, {Li}, {Luo}  \& {Jones}}{{Ye}
  et~al.}{2024}]{ye24}
{Ye} S.,  {Cui} W.-Y.,  {Li} Y.-B.,  {Luo} A.-L.,   {Jones} R.~A.~H.,  2024,
  \mn@doi [arXiv e-prints] {10.48550/arXiv.2407.18754}, \href
  {https://ui.adsabs.harvard.edu/abs/2024arXiv240718754Y} {p. arXiv:2407.18754}

\bibitem[\protect\citeauthoryear{{Yong} et~al.,}{{Yong}
  et~al.}{2013a}]{yong13a}
{Yong} D.,  et~al., 2013a, \mn@doi [\apj] {10.1088/0004-637X/762/1/26}, \href
  {https://ui.adsabs.harvard.edu/abs/2013ApJ...762...26Y} {762, 26}

\bibitem[\protect\citeauthoryear{{Yong} et~al.,}{{Yong}
  et~al.}{2013b}]{yong13b}
{Yong} D.,  et~al., 2013b, \mn@doi [\apj] {10.1088/0004-637X/762/1/27}, \href
  {https://ui.adsabs.harvard.edu/abs/2013ApJ...762...27Y} {762, 27}

\bibitem[\protect\citeauthoryear{{Yoon} et~al.,}{{Yoon} et~al.}{2016}]{yoon16}
{Yoon} J.,  et~al., 2016, \mn@doi [\apj] {10.3847/0004-637X/833/1/20}, \href
  {http://adsabs.harvard.edu/abs/2016ApJ...833...20Y} {833, 20}

\bibitem[\protect\citeauthoryear{{Yoon} et~al.,}{{Yoon} et~al.}{2018}]{yoon18}
{Yoon} J.,  et~al., 2018, \mn@doi [\apj] {10.3847/1538-4357/aaccea}, \href
  {https://ui.adsabs.harvard.edu/abs/2018ApJ...861..146Y} {861, 146}

\bibitem[\protect\citeauthoryear{{York} et~al.,}{{York} et~al.}{2000}]{york00}
{York} D.~G.,  et~al., 2000, \mn@doi [\aj] {10.1086/301513}, \href
  {https://ui.adsabs.harvard.edu/abs/2000AJ....120.1579Y} {120, 1579}

\bibitem[\protect\citeauthoryear{{Zhang}, {Green}  \& {Rix}}{{Zhang}
  et~al.}{2023}]{zhang23}
{Zhang} X.,  {Green} G.~M.,   {Rix} H.-W.,  2023, \mn@doi [\mnras]
  {10.1093/mnras/stad1941}, \href
  {https://ui.adsabs.harvard.edu/abs/2023MNRAS.524.1855Z} {524, 1855}

\makeatother
\end{thebibliography}



\newpage

\appendix

\section{Comparison to other XP/photometric predictions}\label{sec:photcomp}

\subsection{Metallicity and carbon from J-PLUS by Huang et al. (2024)}

\begin{figure*}
\centering
\includegraphics[width=0.45\hsize,trim={0.0cm 0.0cm 0.0cm 0.0cm}]{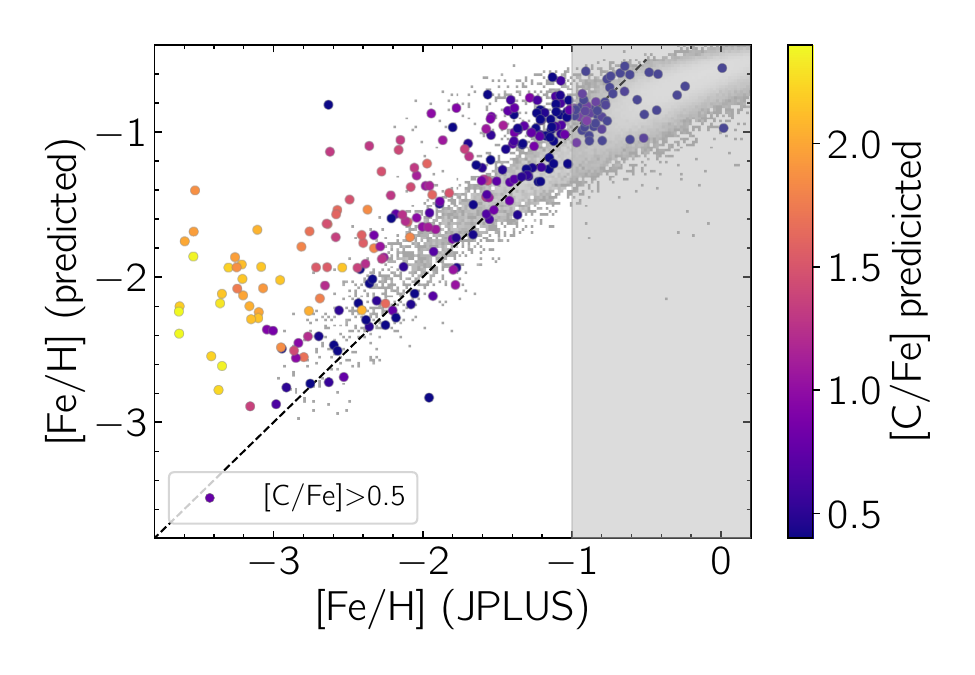}
\includegraphics[width=0.35\hsize,trim={0.0cm 0.0cm 0.0cm 0.0cm}]{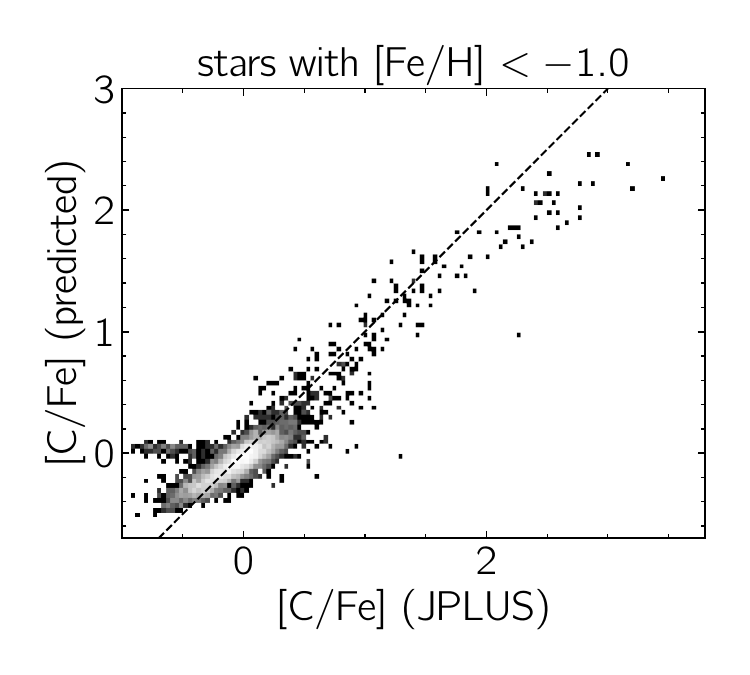}
\caption{Comparison between our predictions and those from J-PLUS by \citet{huang24}, in similar style to Figure~\ref{fig:apogalahcomp_a23}, with slightly different axis limits.}
\label{fig:jplus} 
\end{figure*}

Recently, \citet{huang24} derived \teff, \logg, \feh, \cfe, \mgfe and \alphafe from J-PLUS narrow-band photometry using Kernel Principal Component Analysis. Their training sample is based on a literature compilation of high-resolution spectroscopy (which defines their \feh scale) and (n-)SSPP analyses of SDSS and LAMOST spectra (which defines their \cfe scale). We compare our predictions for stars in common in Figure~\ref{fig:jplus}, showing \feh on the left, with C-rich stars highlighted, and \cfe for metal-poor stars on the right. 

For C-normal stars, the \feh comparison looks similar to our comparisons with high-resolution samples in Section~\ref{sec:speccomp}. For C-rich stars, however, there is a strong discrepancy with rising \cfe; \citet{huang24} are estimating significantly lower metallicities than we are. As can be appreciated from the comparison with our test sample in Figures~\ref{fig:testpred} and \ref{fig:pred}, we are indeed likely \emph{slightly} overestimating the metallicity for the most metal-poor CEMP stars with $\feh < -2.5$, but they seem to pile up around $-2.5$ and do not get assigned \emph{much} higher metallicities as suggested by the J-PLUS comparison. More work is necessary to investigate how metal-poor these stars truly are, for example through spectroscopic follow-up, but a direct comparison between the training samples used in our work and in \citet{huang24} might also be useful. Differences between \texttt{FERRE} and (n-)SSPP results for carbon in VMP stars were explored by \citet{arentsen21}, but the focus there was more on stars with $\cfe < +1$ rather than these extreme ones.

The right-hand panel of the Figure shows that up to $\cfe = +2$, our \cfe predictions are in very good agreement for the bulk of the stars (even if \feh is biased). Above that carbon abundance, \citet{huang24} predict systematically higher \cfe, which could partly be connected to the strong difference in predicted \feh.

\begin{figure*}
\centering
\includegraphics[width=0.45\hsize,trim={0.0cm 0.0cm 0.0cm 0.5cm}]{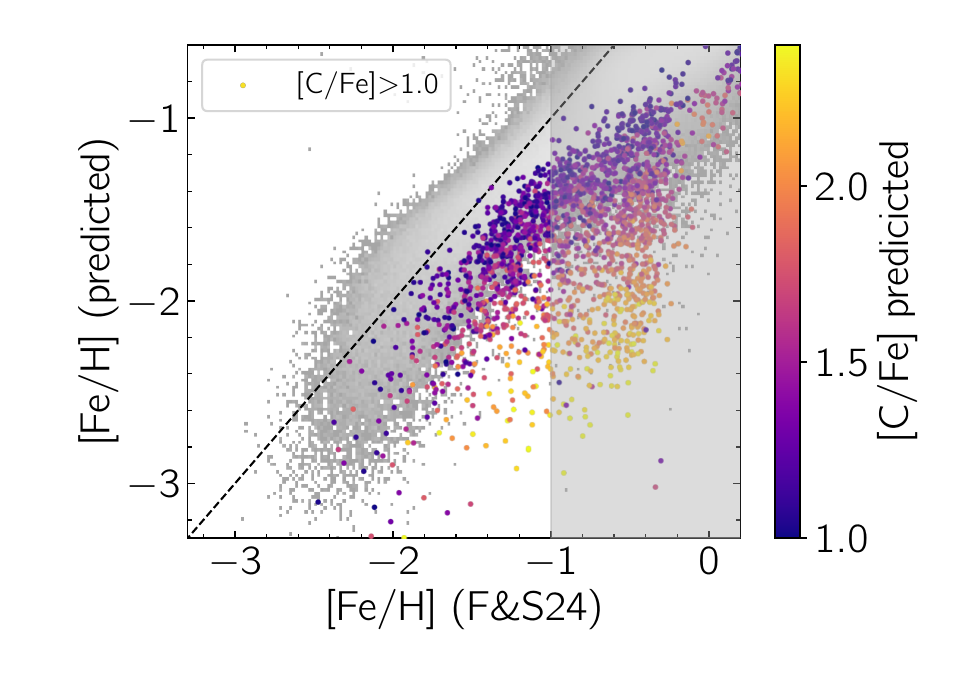}
\includegraphics[width=0.35\hsize,trim={0.0cm 0.0cm 0.0cm 0.5cm}]{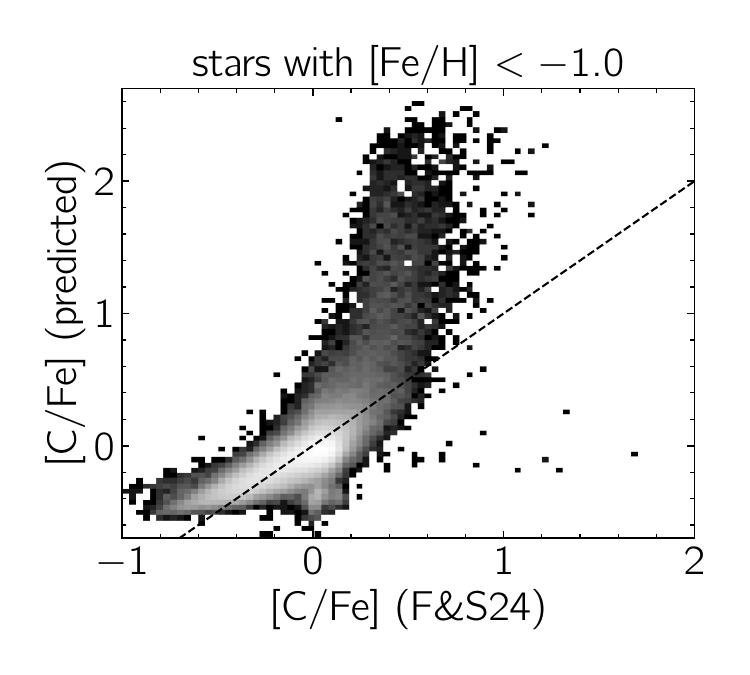}
\caption{Comparison between our predictions and those from \citet{fallows24}, in similar style to Figure~\ref{fig:apogalahcomp_a23}. Note that the axis limits are the same as in that Figure and slightly different from Figure~\ref{fig:jplus}, and the colour bar has a different lower limit. Metal-poor stars are selected based on our predicted \feh.}
\label{fig:fs24} 
\end{figure*}

\subsection{Metallicity and carbon from XP by Fallows \& Sanders (2024)}

In their recent work, \citet{fallows24} used a neural network to derive \teff, \logg, \feh, \cfe, \nfe and [$\alpha$/M] from the \Gaia XP spectra, paying special attention to estimating robust uncertainties of their predictions. They trained on APOGEE, so they are expected to have similar limitations as APOGEE at very low metallicity and/or high carbon (see Section~\ref{sec:apogee}). We present a comparison of our predictions for stars in the \citetalias{andrae23} sample (consisting of RGB stars only) in Figure~\ref{fig:fs24}. The left panel shows that our metallicities agree reasonably well for C-normal stars with $\feh \gtrsim -2$, but for C-rich stars (with $\cfe > +1.0$) the \citet{fallows24} predictions are biased -- the higher the \cfe, the more biased they are. This is not surprising since stars with the right labels were lacking from their training sample, so very C-rich, metal-poor stars are classified as metal-rich instead. The \cfe comparison in the right panel looks similar to that for APOGEE (middle panels of Figure~\ref{fig:apogalahcomp_a23}), saturating strongly at high \cfe but still showing a roughly monotonic relationship. 

\begin{figure}
\centering
\includegraphics[width=0.9\hsize,trim={0.5cm 0.0cm 0.5cm 0.5cm}]{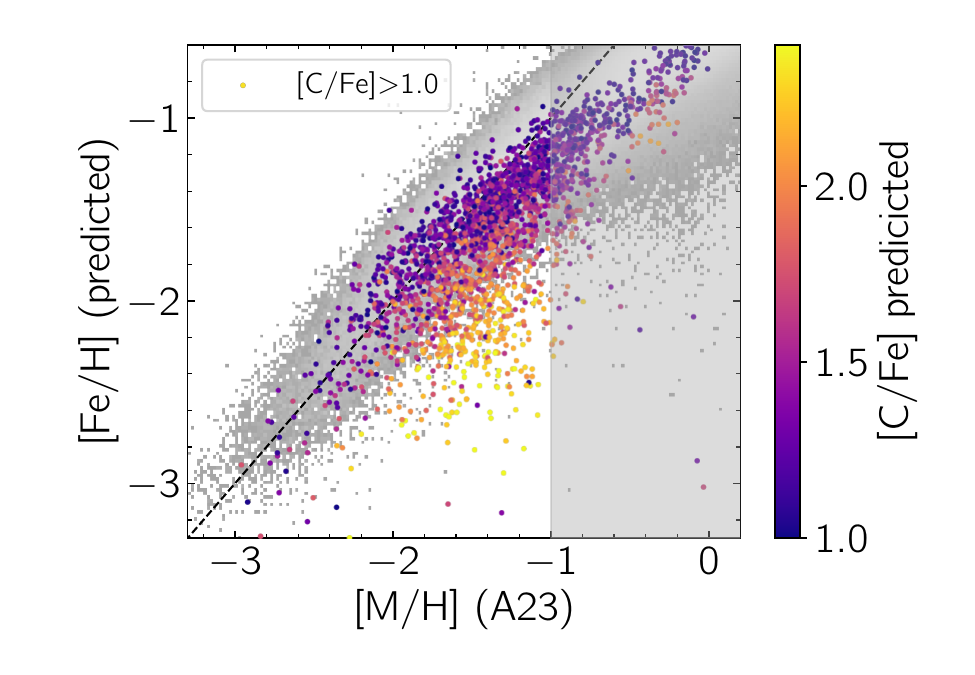}
\caption{Comparison between our metallicity predictions and those from \citetalias{andrae23}, in similar style to Figure~\ref{fig:apogalahcomp_a23} but only for metallicity since \citetalias{andrae23} do not predict carbon. The colour bar is the same as in Figure~\ref{fig:fs24}.}
\label{fig:a23} 
\end{figure}

\subsection{Metallicity from XP by A23}\label{sec:a23met}

\citetalias{andrae23} do not predict carbon abundances, but we can investigate the effect of carbon abundances on the \citetalias{andrae23} metallicities. For C-normal stars, our metallicities agree very well. For $\feh < -2.0$, the median difference is $0.01$~dex, with a dispersion of $0.19$~dex. This might not be too surprising since \citetalias{andrae23} also included the \citet{li22} high-resolution VMP sample in their training set. \citet{martin24} compare the \citetalias{andrae23} with other VMP spectroscopic samples as well, which show very good agreement. For very C-rich stars, however, our comparison shows that the \citetalias{andrae23} [M/H] is higher than our \feh. This  is again expected: since the molecular carbon features are large, they raise the overall metallicity estimate. 

\section{Tables}

This section gives descriptions for the tables for the reference sample (Table~\ref{table:ref}), the predictions for the \citetalias{lucey23} sample (Table~\ref{table:L23}) and the predictions for the \citetalias{andrae23} sample (Table~\ref{table:A23}). The full tables are available at \href{https://doi.org/10.5281/zenodo.14651677}{10.5281/zenodo.14651678} and at the CDS.  

\newpage

\textcolor{white}{.}

\newpage

\begin{table}
\centering
\caption{\label{table:ref}Table description for the reference sample}
\begin{tabular}{lc}
 \hline
  Column name & Description \\
 \hline
 source\_id &  Gaia DR3 identifier\\
 ra & Gaia DR3 Right ascension [deg] \\
 dec & Gaia DR3 Declination [deg] \\
 phot\_g\_mean\_mag & Gaia DR3 G-band magnitude \\
 parallax & Gaia DR3 parallax [mas]  \\
 parallax\_error & Gaia DR3 parallax error [mas] \\
 pvar & Photometric variability probability$^1$ \\
 ebv & E(B$-$V) from SFD$^2$\\
 BPRP\_0 & Dereddened Gaia (BP$-$RP) \\
 G\_0 & Dereddened Gaia G \\
 teff\_spec &  Spectroscopic \teff [K]\\
 feh\_spec &  Spectroscopic \feh \\
 cfe\_spec & Spectroscopic \cfe \\
 logg\_spec &  Spectroscopic \logg \\
 teff & Predicted \teff [K]\\
 feh & Predicted \feh \\
 cfe & Predicted \cfe \\
 logg & Predicted \logg \\
 err\_teff & Predicted \teff uncertainty [K]\\
 err\_feh & Predicted \feh uncertainty  \\
 err\_cfe & Predicted \cfe uncertainty  \\
 err\_logg & Predicted \logg uncertainty  \\
 std\_teff & Predicted \teff standard deviation [K] \\
 std\_feh & Predicted \feh standard deviation   \\
 std\_cfe & Predicted \cfe standard deviation   \\
 std\_logg & Predicted \logg standard deviation   \\
 cat & Source of spectroscopic parameters$^1$ \\
 train & 1 = in training sample, 0 = in test sample \\
 \hline
\end{tabular} \\
\vspace{0.1cm}
 \footnotesize{$^1$ Possible values: \citet{yoon16}, \citet{li22}, this work (normal LAMOST), this work (\citetalias{lucey23} LAMOST)} \\
\end{table}

\begin{table}
\centering
\caption{\label{table:L23}Table description for the \citetalias{lucey23} sample}
\begin{tabular}{lc}
 \hline
  Column name & Description \\
 \hline
 source\_id &  Gaia DR3 identifier\\
 ra & Gaia DR3 Right ascension [deg] \\
 dec & Gaia DR3 Declination [deg] \\
 phot\_g\_mean\_mag & Gaia DR3 G-band magnitude \\
 parallax & Gaia DR3 parallax [mas]  \\
 parallax\_error & Gaia DR3 parallax error [mas] \\
 pvar & Photometric variability probability$^1$ \\
 ebv & E(B$-$V) from SFD$^2$\\
 BPRP\_0 & Dereddened Gaia (BP$-$RP) \\
 G\_0 & Dereddened Gaia G \\
 teff & Predicted \teff [K] \\
 feh & Predicted \feh \\
 cfe & Predicted \cfe \\
 logg & Predicted \logg \\
 err\_teff & Predicted \teff uncertainty [K] \\
 err\_feh & Predicted \feh uncertainty  \\
 err\_cfe & Predicted \cfe uncertainty  \\
 err\_logg & Predicted \logg uncertainty  \\
 std\_teff & Predicted \teff standard deviation [K] \\
 std\_feh & Predicted \feh standard deviation   \\
 std\_cfe & Predicted \cfe standard deviation   \\
 std\_logg & Predicted \logg standard deviation   \\
 \hline 
\end{tabular} \\
\vspace{0.1cm}
 \footnotesize{$^1$ \citet{martin24}} \\
 \footnotesize{$^2$ \citet{schlafly11} scale}
\end{table}

\begin{table}
\centering
\caption{\label{table:A23}Table description for the \citetalias{andrae23} sample (only for the ``vetted RGB sample'' with radial velocities)}
\begin{tabular}{lc}
 \hline
  Column name & Description \\
 \hline
 source\_id &  Gaia DR3 identifier\\
 ra & Gaia DR3 Right ascension [deg] \\
 dec & Gaia DR3 Declination [deg] \\
 phot\_g\_mean\_mag & Gaia DR3 G-band magnitude \\
 parallax & Gaia DR3 parallax [mas] \\
 parallax\_error & Gaia DR3 parallax error [mas]  \\
 pvar & Photometric variability probability$^1$ \\
 ebv & E(B$-$V) from SFD$^2$\\
 BPRP\_0 & Dereddened Gaia (BP$-$RP) \\
 G\_0 & Dereddened Gaia G \\
 teff & Predicted \teff [K]\\
 feh & Predicted \feh \\
 cfe & Predicted \cfe \\
 logg & Predicted \logg \\
 err\_teff & Predicted \teff uncertainty [K] \\
 err\_feh & Predicted \feh uncertainty  \\
 err\_cfe & Predicted \cfe uncertainty  \\
 err\_logg & Predicted \logg uncertainty  \\
 std\_teff & Predicted \teff standard deviation [K]  \\
 std\_feh & Predicted \feh standard deviation   \\
 std\_cfe & Predicted \cfe standard deviation   \\
 std\_logg & Predicted \logg standard deviation   \\
 Ccor & Evolutionary carbon correction$^3$ \\
 energy & Orbital energy [km$^2$ s$^{-2}$]\\
 Lz & Orbital angular momentum [km kpc s$^-1$] \\
 \hline 
\end{tabular} \\
\vspace{0.1cm}
 \footnotesize{$^1$ \citet{martin24}} \\ 
 \footnotesize{$^2$ \citet{schlafly11} scale} \\
 \footnotesize{$^3$ \citet{placco14}}
\end{table}


\bsp	
\label{lastpage}
\end{document}